\documentclass[prd,reprint,showpacs,superscriptaddress,bibnotes,floatfix]{revtex4-1}
\usepackage{natbib}
\usepackage{graphics}
\usepackage{graphicx}
\usepackage{multirow}
\usepackage{amsbsy}
\usepackage{mathrsfs}
\usepackage{footmisc}
\usepackage{hyperref}
\usepackage{amsmath}
\hypersetup{
  colorlinks = true,
  urlcolor = black,
  linkcolor=black,
  citecolor=black
}
\def\rrange{0.020^{+0.021}_{-0.018}}
\def\rul{0.072}
\def\rztop{0.66}
\def\rztopps{18}

\def\Adrange{4.6^{+1.1}_{-0.9}}
\def\Adcentval{4.6}

\def\Asrange{1.0^{+1.2}_{-0.8}}
\def\Ascentval{1.0}
\def\Asul{3.7}

\def\rmlm{0.020}
\def\Asmlm{1.5}
\def\Admlm{4.7}
\def\Bsmlm{-3.0}
\def\asmlm{-0.27}
\def\Bdmlm{1.6}
\def\admlm{-0.58}
\def\emlm{-0.38}

\def\chitwo{760}
\def\chitwodc{759}
\def\chitwoptenom{0.06}
\def\chitwoptesim{0.19}
\def\chioneptesim{0.23}

\def\Ad{A_\mathrm{d}}
\def\Adf{A_\mathrm{d,353}}
\def\As{A_\mathrm{sync}}
\def\Asf{A_\mathrm{sync,23}}

\def\Bd{\beta_\mathrm{d}}
\def\Bs{\beta_\mathrm{s}}

\def\ad{\alpha_\mathrm{d}}
\def\as{\alpha_\mathrm{s}}

\def\dd{\Delta_\mathrm{d}}
\def\ddp{\Delta'_\mathrm{d}}

\def\bicep{BICEP}

\def\bicepone{{\sc BICEP1}}
\def\biceptwo{{\sc BICEP2}}
\def\bicepthree{{\sc BICEP3}}
\def\biceparray{{\sc BICEP} Array}

\def\keck{{\it Keck}}
\def\keckarray{{\it Keck Array}}
\def\bk{\bicep/\keck}
\def\BKfourteen{{BK14}}

\def\planck{{\it Planck}} % They DO italicize

\def\wmap{WMAP}

\def\spass{S-PASS}

\def\uk{$\mu{\mathrm K}$}
\def\uksq{$\mu{\mathrm K^2}$}
\def\ukcmb{$\mu{\mathrm K}_{\mathrm{\mbox{\tiny\sc cmb}}}$}

\def\deg{^\circ}
\def\emode{$E$-mode}
\def\bmode{$B$-mode}
\newcommand{\cl}{$C_\ell$ }
\def\clstar{\ell \left( \ell + 1 \right) C_\ell / 2 \pi}
\def\lcdm{$\Lambda$CDM}

\def\aap{Astron.\ Astrophys.}

\def\apj{Astrophys.\ J.}

\def\apjs{Astrophys.\ J.\ Suppl.\ Ser.}

\def\jcap{J.\ Cosmol.\ Astropart.\ Phys.}

\def\mnras{Mon.\ Not.\ R.\ Astron.\ Soc.}

\def\prd{Phys.\ Rev.\ D}

\def\prl{Phys.\ Rev.\ Lett.}

\begin{document}

\title{\biceptwo\ / \keckarray\ X:
Constraints on Primordial Gravitational Waves using \planck, \wmap, and New
\biceptwo/\keck\ Observations through the 2015 Season}

\author{\keckarray\ and \biceptwo\ Collaborations: P.~A.~R.~Ade}
\affiliation{School of Physics and Astronomy, Cardiff University, Cardiff, CF24 3AA, United Kingdom}
\author{Z.~Ahmed}
\affiliation{Kavli Institute for Particle Astrophysics and Cosmology, SLAC National Accelerator Laboratory, 2575 Sand Hill Rd, Menlo Park, California 94025, USA}
\author{R.~W.~Aikin}
\affiliation{Department of Physics, California Institute of Technology, Pasadena, California 91125, USA}
\author{K.~D.~Alexander}
\affiliation{Harvard-Smithsonian Center for Astrophysics, 60 Garden Street MS 42, Cambridge, Massachusetts 02138, USA}
\author{D.~Barkats}
\affiliation{Harvard-Smithsonian Center for Astrophysics, 60 Garden Street MS 42, Cambridge, Massachusetts 02138, USA}
\author{S.~J.~Benton}
\affiliation{Department of Physics, Princeton University, Princeton, NJ 08544, USA}
\author{C.~A.~Bischoff}
\affiliation{Department of Physics, University of Cincinnati, Cincinnati, Ohio 45221, USA}
\author{J.~J.~Bock}
\affiliation{Department of Physics, California Institute of Technology, Pasadena, California 91125, USA}
\affiliation{Jet Propulsion Laboratory, Pasadena, California 91109, USA}
\author{R.~Bowens-Rubin}
\affiliation{Harvard-Smithsonian Center for Astrophysics, 60 Garden Street MS 42, Cambridge, Massachusetts 02138, USA}
\author{J.~A.~Brevik}
\affiliation{Department of Physics, California Institute of Technology, Pasadena, California 91125, USA}
\author{I.~Buder}
\affiliation{Harvard-Smithsonian Center for Astrophysics, 60 Garden Street MS 42, Cambridge, Massachusetts 02138, USA}
\author{E.~Bullock}
\affiliation{Minnesota Institute for Astrophysics, University of Minnesota, Minneapolis, Minnesota 55455, USA}
\author{V.~Buza}
\affiliation{Harvard-Smithsonian Center for Astrophysics, 60 Garden Street MS 42, Cambridge, Massachusetts 02138, USA}
\affiliation{Department of Physics, Harvard University, Cambridge, MA 02138, USA}
\author{J.~Connors}
\affiliation{Harvard-Smithsonian Center for Astrophysics, 60 Garden Street MS 42, Cambridge, Massachusetts 02138, USA}
\author{J.~Cornelison}
\affiliation{Harvard-Smithsonian Center for Astrophysics, 60 Garden Street MS 42, Cambridge, Massachusetts 02138, USA}
\author{B.~P.~Crill}
\affiliation{Jet Propulsion Laboratory, Pasadena, California 91109, USA}
\author{M.~Crumrine}
\affiliation{School of Physics and Astronomy, University of Minnesota, Minneapolis, Minnesota 55455, USA}
\author{M.~Dierickx}
\affiliation{Harvard-Smithsonian Center for Astrophysics, 60 Garden Street MS 42, Cambridge, Massachusetts 02138, USA}
\author{L.~Duband}
\affiliation{Service des Basses Temp\'{e}ratures, Commissariat \`{a} l'Energie Atomique, 38054 Grenoble, France}
\author{C.~Dvorkin}
\affiliation{Department of Physics, Harvard University, Cambridge, MA 02138, USA}
\author{J.~P.~Filippini}
\affiliation{Department of Physics, University of Illinois at Urbana-Champaign, Urbana, Illinois 61801, USA}
\affiliation{Department of Astronomy, University of Illinois at Urbana-Champaign, Urbana, Illinois 61801, USA}
\author{S.~Fliescher}
\affiliation{School of Physics and Astronomy, University of Minnesota, Minneapolis, Minnesota 55455, USA}
\author{J.~Grayson}
\affiliation{Department of Physics, Stanford University, Stanford, California 94305, USA}
\author{G.~Hall}
\affiliation{School of Physics and Astronomy, University of Minnesota, Minneapolis, Minnesota 55455, USA}
\author{M.~Halpern}
\affiliation{Department of Physics and Astronomy, University of British Columbia, Vancouver, British Columbia, V6T 1Z1, Canada}
\author{S.~Harrison}
\affiliation{Harvard-Smithsonian Center for Astrophysics, 60 Garden Street MS 42, Cambridge, Massachusetts 02138, USA}
\author{S.~R.~Hildebrandt}
\affiliation{Department of Physics, California Institute of Technology, Pasadena, California 91125, USA}
\affiliation{Jet Propulsion Laboratory, Pasadena, California 91109, USA}
\author{G.~C.~Hilton}
\affiliation{National Institute of Standards and Technology, Boulder, Colorado 80305, USA}
\author{H.~Hui}
\affiliation{Department of Physics, California Institute of Technology, Pasadena, California 91125, USA}
\author{K.~D.~Irwin}
\affiliation{Department of Physics, Stanford University, Stanford, California 94305, USA}
\affiliation{Kavli Institute for Particle Astrophysics and Cosmology, SLAC National Accelerator Laboratory, 2575 Sand Hill Rd, Menlo Park, California 94025, USA}
\affiliation{National Institute of Standards and Technology, Boulder, Colorado 80305, USA}
\author{J.~Kang}
\affiliation{Department of Physics, Stanford University, Stanford, California 94305, USA}
\author{K.~S.~Karkare}
\affiliation{Harvard-Smithsonian Center for Astrophysics, 60 Garden Street MS 42, Cambridge, Massachusetts 02138, USA}
\affiliation{Kavli Institute for Cosmological Physics, University of Chicago, Chicago, IL 60637, USA}
\author{E.~Karpel}
\affiliation{Department of Physics, Stanford University, Stanford, California 94305, USA}
\author{J.~P.~Kaufman}
\affiliation{Department of Physics, University of California at San Diego, La Jolla, California 92093, USA}
\author{B.~G.~Keating}
\affiliation{Department of Physics, University of California at San Diego, La Jolla, California 92093, USA}
\author{S.~Kefeli}
\affiliation{Department of Physics, California Institute of Technology, Pasadena, California 91125, USA}
\author{S.~A.~Kernasovskiy}
\affiliation{Department of Physics, Stanford University, Stanford, California 94305, USA}
\author{J.~M.~Kovac}
\affiliation{Harvard-Smithsonian Center for Astrophysics, 60 Garden Street MS 42, Cambridge, Massachusetts 02138, USA}
\affiliation{Department of Physics, Harvard University, Cambridge, MA 02138, USA}
\author{C.~L.~Kuo}
\affiliation{Department of Physics, Stanford University, Stanford, California 94305, USA}
\affiliation{Kavli Institute for Particle Astrophysics and Cosmology, SLAC National Accelerator Laboratory, 2575 Sand Hill Rd, Menlo Park, California 94025, USA}
\author{N.~A.~Larsen}
\affiliation{Kavli Institute for Cosmological Physics, University of Chicago, Chicago, IL 60637, USA}
\author{K.~Lau}
\affiliation{School of Physics and Astronomy, University of Minnesota, Minneapolis, Minnesota 55455, USA}
\author{E.~M.~Leitch}
\affiliation{Kavli Institute for Cosmological Physics, University of Chicago, Chicago, IL 60637, USA}
\author{M.~Lueker}
\affiliation{Department of Physics, California Institute of Technology, Pasadena, California 91125, USA}
\author{K.~G.~Megerian}
\affiliation{Jet Propulsion Laboratory, Pasadena, California 91109, USA}
\author{L.~Moncelsi}
\affiliation{Department of Physics, California Institute of Technology, Pasadena, California 91125, USA}
\author{T.~Namikawa}
\affiliation{Leung Center for Cosmology and Particle Astrophysics, National Taiwan University, Taipei 10617, Taiwan}
\author{C.~B.~Netterfield}
\affiliation{Department of Physics, University of Toronto, Toronto, Ontario, M5S 1A7, Canada}
\affiliation{Canadian Institute for Advanced Research, Toronto, Ontario, M5G 1Z8, Canada}
\author{H.~T.~Nguyen}
\affiliation{Jet Propulsion Laboratory, Pasadena, California 91109, USA}
\author{R.~O'Brient}
\affiliation{Department of Physics, California Institute of Technology, Pasadena, California 91125, USA}
\affiliation{Jet Propulsion Laboratory, Pasadena, California 91109, USA}
\author{R.~W.~Ogburn~IV}
\affiliation{Department of Physics, Stanford University, Stanford, California 94305, USA}
\affiliation{Kavli Institute for Particle Astrophysics and Cosmology, SLAC National Accelerator Laboratory, 2575 Sand Hill Rd, Menlo Park, California 94025, USA}
\author{S.~Palladino}
\affiliation{Department of Physics, University of Cincinnati, Cincinnati, Ohio 45221, USA}
\author{C.~Pryke}
\email{pryke@physics.umn.edu}
\affiliation{School of Physics and Astronomy, University of Minnesota, Minneapolis, Minnesota 55455, USA}
\affiliation{Minnesota Institute for Astrophysics, University of Minnesota, Minneapolis, Minnesota 55455, USA}
\author{B.~Racine}
\affiliation{Harvard-Smithsonian Center for Astrophysics, 60 Garden Street MS 42, Cambridge, Massachusetts 02138, USA}
\author{S.~Richter}
\affiliation{Harvard-Smithsonian Center for Astrophysics, 60 Garden Street MS 42, Cambridge, Massachusetts 02138, USA}
\author{A.~Schillaci}
\affiliation{Department of Physics, California Institute of Technology, Pasadena, California 91125, USA}
\author{R.~Schwarz}
\affiliation{School of Physics and Astronomy, University of Minnesota, Minneapolis, Minnesota 55455, USA}
\author{C.~D.~Sheehy}
\affiliation{Physics Department, Brookhaven National Laboratory, Upton, NY 11973}
\author{A.~Soliman}
\affiliation{Department of Physics, California Institute of Technology, Pasadena, California 91125, USA}
\author{T.~St.~Germaine}
\affiliation{Harvard-Smithsonian Center for Astrophysics, 60 Garden Street MS 42, Cambridge, Massachusetts 02138, USA}
\author{Z.~K.~Staniszewski}
\affiliation{Department of Physics, California Institute of Technology, Pasadena, California 91125, USA}
\affiliation{Jet Propulsion Laboratory, Pasadena, California 91109, USA}
\author{B.~Steinbach}
\affiliation{Department of Physics, California Institute of Technology, Pasadena, California 91125, USA}
\author{R.~V.~Sudiwala}
\affiliation{School of Physics and Astronomy, Cardiff University, Cardiff, CF24 3AA, United Kingdom}
\author{G.~P.~Teply}
\affiliation{Department of Physics, California Institute of Technology, Pasadena, California 91125, USA}
\affiliation{Department of Physics, University of California at San Diego, La Jolla, California 92093, USA}
\author{K.~L.~Thompson}
\affiliation{Department of Physics, Stanford University, Stanford, California 94305, USA}
\affiliation{Kavli Institute for Particle Astrophysics and Cosmology, SLAC National Accelerator Laboratory, 2575 Sand Hill Rd, Menlo Park, California 94025, USA}
\author{J.~E.~Tolan}
\affiliation{Department of Physics, Stanford University, Stanford, California 94305, USA}
\author{C.~Tucker}
\affiliation{School of Physics and Astronomy, Cardiff University, Cardiff, CF24 3AA, United Kingdom}
\author{A.~D.~Turner}
\affiliation{Jet Propulsion Laboratory, Pasadena, California 91109, USA}
\author{C.~Umilt\`{a}}
\affiliation{Department of Physics, University of Cincinnati, Cincinnati, Ohio 45221, USA}
\author{A.~G.~Vieregg}
\affiliation{Department of Physics, Enrico Fermi Institute, University of Chicago, Chicago, IL 60637, USA}
\affiliation{Kavli Institute for Cosmological Physics, University of Chicago, Chicago, IL 60637, USA}
\author{A.~Wandui}
\affiliation{Department of Physics, California Institute of Technology, Pasadena, California 91125, USA}
\author{A.~C.~Weber}
\affiliation{Jet Propulsion Laboratory, Pasadena, California 91109, USA}
\author{D.~V.~Wiebe}
\affiliation{Department of Physics and Astronomy, University of British Columbia, Vancouver, British Columbia, V6T 1Z1, Canada}
\author{J.~Willmert}
\affiliation{School of Physics and Astronomy, University of Minnesota, Minneapolis, Minnesota 55455, USA}
\author{C.~L.~Wong}
\affiliation{Harvard-Smithsonian Center for Astrophysics, 60 Garden Street MS 42, Cambridge, Massachusetts 02138, USA}
\affiliation{Department of Physics, Harvard University, Cambridge, MA 02138, USA}
\author{W.~L.~K.~Wu}
\affiliation{Kavli Institute for Cosmological Physics, University of Chicago, Chicago, IL 60637, USA}
\author{H.~Yang}
\affiliation{Department of Physics, Stanford University, Stanford, California 94305, USA}
\author{K.~W.~Yoon}
\affiliation{Department of Physics, Stanford University, Stanford, California 94305, USA}
\affiliation{Kavli Institute for Particle Astrophysics and Cosmology, SLAC National Accelerator Laboratory, 2575 Sand Hill Rd, Menlo Park, California 94025, USA}
\author{C.~Zhang}
\affiliation{Department of Physics, California Institute of Technology, Pasadena, California 91125, USA}

\date[Draft~]{As accepted by PRL}

\begin{abstract}
We present results from an analysis of all data taken by the 
\biceptwo/\keck\ CMB polarization experiments
up to and including the 2015 observing season.
This includes the first \keckarray\ observations at 220\,GHz
and additional observations at 95 \& 150\,GHz. % REFB
The $Q/U$ maps reach depths of 5.2, 2.9 and 26\,\ukcmb\,arcmin at
95, 150 and 220\,GHz respectively over an effective area
of $\approx 400$ square degrees.
The 220\,GHz maps achieve a signal-to-noise on polarized dust emission % REFB
approximately equal to that of \planck\ at 353\,GHz.
We take auto- and cross-spectra between these maps and publicly
available \wmap\ and \planck\ maps at frequencies from 23 to 353\,GHz.
We evaluate the joint likelihood of the spectra versus
a multicomponent model of lensed-\lcdm+$r$+dust+synchrotron+noise.
The foreground model has seven parameters, and we impose priors on
some of these using external information from \planck\ and
\wmap\ derived from larger regions of sky.
The model is shown to be an adequate description of the data
at the current noise levels.
The likelihood analysis yields the constraint $r_{0.05}<0.07$ at
95\% confidence, which tightens to $r_{0.05}<0.06$ in conjunction
with \planck\ temperature measurements and other data.
The lensing signal is detected at $8.8 \sigma$ significance. % REFB
Running maximum likelihood search on simulations we 
obtain unbiased results and find that $\sigma(r)=0.020$.
These are the strongest constraints to date on primordial gravitational waves.
\end{abstract}

\doi{xyz}

\maketitle

{\it Introduction.}---It is remarkable that our 
standard model of cosmology, known as \lcdm,
is able to statistically describe the observable universe with
only six parameters
(tensions between high and low redshift probes notwithstanding~\cite{planck2018VI}). % REFC  
Observations of the cosmic microwave background (CMB)~\cite{penzias65} 
have played a central role in establishing this model and 
now constrain these parameters 
to percent-level precision
(see most recently Ref.~\cite{planck2015XIII}).

The success of this model focuses our attention on the deep physical
mysteries it exposes.
Dark matter and dark energy dominate the present-day universe, but
we lack understanding of both their nature and abundance.
Perhaps most fundamentally, the standard model offers no explanation for the
observed initial conditions of the universe: highly 
uniform and flat with small, nearly scale-invariant,
adiabatic density perturbations.  
Inflation is an extension to the standard model that
addresses initial conditions by postulating
that the observable universe arose from a tiny, causally-connected
volume in a period of accelerated expansion 
within the first fraction of a nanosecond,
during which quantum fluctuations of the spacetime metric
gave rise to both the observed primordial density
perturbations and a potentially-observable
background of gravitational waves
(see Ref.~\cite{kamionkowski2015} for a recent review and 
citations to the original literature).

Probing for these primordial gravitational waves through the
faint \bmode\ polarization patterns that they would imprint on
the CMB is recognized as one of the most important goals in
cosmology today, with the potential to either confirm inflation,
and establish its energy scale, or to powerfully
limit the space of allowed inflationary models~\cite{S4_sciencebook2016}.
Multiple groups are making measurements of CMB polarization, % REFB
some focused on the gravitational wave goal at larger angular
scales, and others focused on other science at smaller angular
scales---examples include~\cite{keisler15,actpol16,polarbear17,abs18}.

In principle \bmode\ polarization patterns offer a unique 
probe of primordial gravitational waves because
they cannot be sourced by primordial density 
perturbations~\cite{seljak97b,kamionkowski97,seljak97a}.
However, in practice there are two sources of foreground: % REFB
gravitational deflections of the CMB photons in flight
leads to a lensing \bmode\ component~\cite{zaldarriaga98}, and 
polarized emission from our own galaxy can also produce {\bmode}s.
The latter can be separated out
through their differing frequency spectral behavior, so
extremely sensitive multi-frequency observations are needed
to advance the leading constraints on primordial gravitational waves.

Our \bicep/\keck\ program first reported detection of an
excess over the lensing \bmode\ expectation at % REFB
150\,GHz in Ref.~\cite{biceptwoI}.
In a joint analysis using multi-frequency data from 
the \planck\ experiment it was shown that most
or all of this is due to polarized emission from dust in our own
galaxy~\cite[hereafter BKP]{bkp}.
We first started to diversify our own frequency coverage by
adding data taken in 2014 with \keckarray\ at 95\,GHz,
yielding the tightest previous constraints on primordial
gravitational waves~\cite[hereafter BK14]{biceptwoVI}.

In this letter [hereafter BK15], we advance these constraints using
new data taken by \keckarray\ in the 2015 season including
two 95\,GHz receivers, a single 150\,GHz receiver,
and, for the first time, two 220\,GHz receivers.
This analysis thus doubles the 95\,GHz dataset from two receiver-years
to four, while adding a new higher frequency band
that significantly improves the constraints on the dust contribution
over what is possible using the \planck\ 353\,GHz data alone.
The constraint on primordial gravitational waves parametrized by
tensor to scalar ratio $r$ is improved
to $r_{0.05}<0.062$ (95\%), disfavoring the important class of inflationary models
represented by a $\phi$ potential\cite{kamionkowski2015,S4_sciencebook2016}.

{\it Instrument and observations.}---\keckarray\ consists of a set
of five microwave receivers similar in design to the precursor \biceptwo\ % REFB
instrument~\cite{biceptwoII,biceptwoV}.
Each receiver employs a $\approx 0.25$\,m aperture all cold
refracting telescope focusing microwave radiation onto
a focal plane of polarized antenna-coupled bolometric detectors~\cite{bkdets}.
The receivers are mounted on a movable platform (or mount) which
scans their pointing direction across the sky in a controlled manner.
The detectors are read out through a time-domain multiplexed
SQUID readout system.
Orthogonally-polarized detectors are arranged as coincident pairs in the
focal plane, and the pair-difference timestream data thus traces out
changes in the polarization signal from place to place on
the sky.
The telescopes are located at the South Pole in Antarctica
where the atmosphere is extremely stable and transparent
at the relevant frequencies.
The data are recorded to disk and transmitted back to the
US daily for analysis.

To date we have mapped a single region of sky centered at
RA 0h, Dec.\ $-57.5\deg$.
From 2010 to 2013, \biceptwo\ and \keckarray\ jointly
recorded a total of 13 receiver-years of data in a band
centered on 150\,GHz.
Two of the \keck\ receivers were switched to 95\,GHz before the 2014 season,
and two more were switched to 220\,GHz before the 2015 season.
The BK15 data set thus consists of 4/17/2 receiver-years
at 95/150/220\,GHz respectively.

{\it Maps and Power Spectra}---We make maps and power spectra
using the same procedures as used for BK14
and previous analyses~\cite{biceptwoI}.
Briefly: the telescope timestream data are filtered
and then binned into sky pixels with the multiple
detector pairs being co-added together using knowledge
of their individual pointing directions as the telescope
scans across the sky.
Maps of the polarization Stokes parameters $Q$ and $U$ are
constructed by also knowing the polarization
sensitivity angle of each pair as projected onto the sky.

After apodizing to downweight the noisy regions around
the edge of the observed area, the $Q/U$ maps are Fourier
transformed and converted to the $E/B$ basis in which the
primordial gravitational wave signal is expected to be
maximally distinct from the standard \lcdm\ signal.

Two details worth noting are the deprojection of
leading order temperature to polarization leakage
terms, and the adjustment of the absolute polarization angle
to minimize the $EB$ cross spectrum.
See Ref.~\cite{biceptwoI} for more information.

\begin{figure}
\resizebox{\columnwidth}{!}{\includegraphics{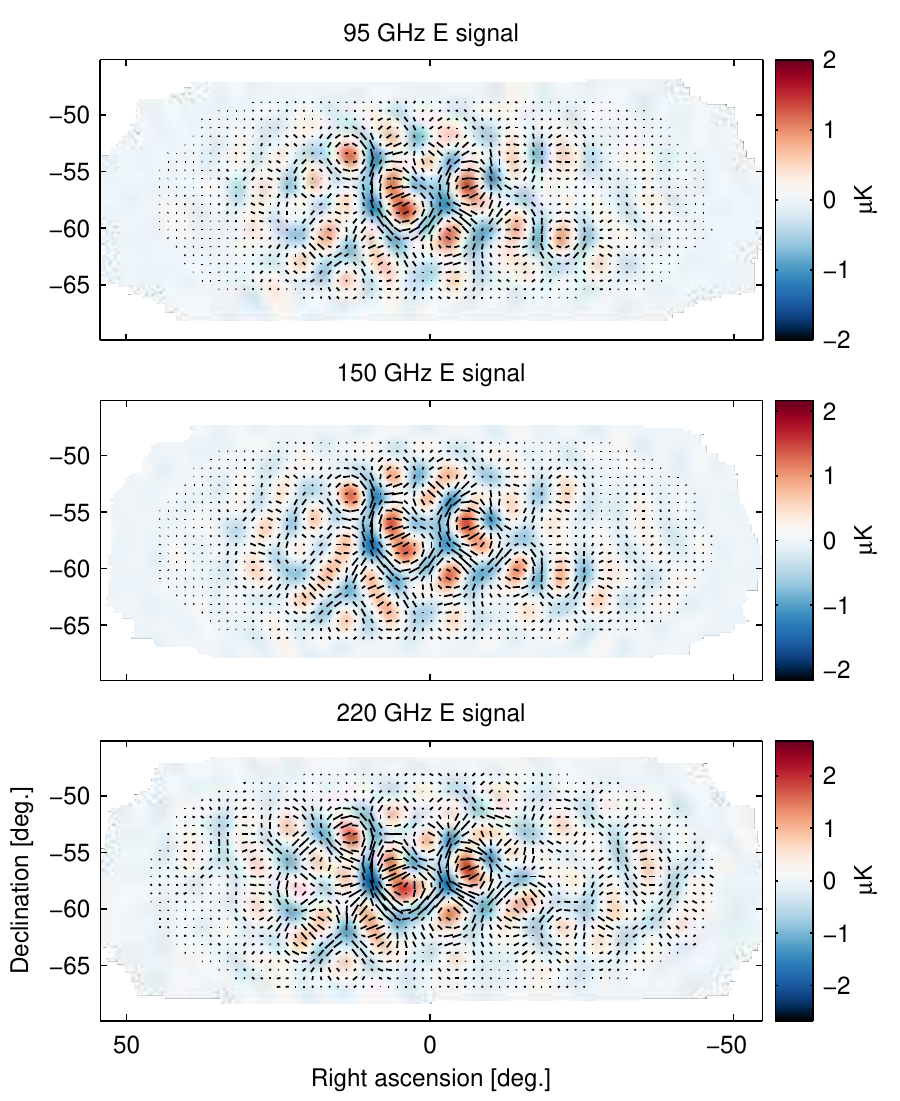}}
\caption{Maps of degree angular scale {\emode}s ($50<\ell<120$)
at three frequencies made using \keckarray\ data from the 2015
season only.
The similarity of the pattern indicates that \lcdm\ {\emode}s
dominate at all three frequencies (and that the signal-to-noise
is high).
The color scale is in \uk, and the range is allowed to vary
slightly to (partially) compensate for the decrease in beam size
with increasing frequency.}
\label{fig:k2015_emaps}
\end{figure}

For illustration purposes we can inverse Fourier transform to
form $E/B$ maps.
Fig.~\ref{fig:k2015_emaps} shows
\emode\ maps formed from the 2015 data alone---the data
which is being added to the previous data in this analysis.
The similarity of the pattern at all three frequencies
indicates that \lcdm\ {\emode}s
dominate, and that the signal-to-noise is high.
The effective area of these maps is $\sim 1$\% of the full
sky.
(See Appendix~\ref{app:maps} for the full set of $T/Q/U$ maps.)

To suppress $E$ to $B$ leakage we use the matrix purification
technique which we have developed~\cite{biceptwoI,biceptwoVII}.
We then take the variance within annuli of the Fourier plane
to estimate the angular power spectra.
To test for systematic contamination we carry out our usual
``jackknife'' internal consistency tests on the new 95\,GHz and 220\,GHz
data as described in Appendices~\ref{app:mapjack}
and~\ref{app:specjack}---the
distributions of $\chi$ and $\chi^2$ PTE values are consistent with % REFB
uniform showing no evidence for problems.

In this paper we use the three bands of \biceptwo/\keck\
plus the 23 \& 33\,GHz bands of
\wmap~\footnote{See \url{http://lambda.gsfc.nasa.gov/product/map/dr5/m_products.cfm}}\citep{bennett13}
and all seven polarized bands of
\planck~\footnote{Public Release 2 ``full mission'' maps as available
at \url{http://www.cosmos.esa.int/web/planck/pla}.
We will update to PR3 in our next analysis.}\citep{planck2015I}. % REFA
We take all possible auto- and cross-power spectra between
these twelve bands---the full set of spectra are shown
in Appendix~\ref{app:allspec}.

Fig.~\ref{fig:powspecres_bkbands} shows the $EE$ and $BB$
auto- and cross-spectra for the \biceptwo/\keck\ bands plus the
\planck\ 353~GHz band which is important for constraining the
polarized dust contribution.
The spectra are compared to the ``baseline'' lensed-\lcdm+dust model
from our previous BK14 analysis.
Note that the $BB$ spectra involving 220\,GHz were not used to derive
this model but agree well with it.
The $EE$ spectra were also not used to derive the model but
agree well with it under the assumption that $EE/BB=2$ for dust,
as it is shown to be close to in \planck\ analysis of larger regions
of sky~\cite{planckiXXX,planckiLIV}.
(Note that many of the \bicep/\keck\ spectra are sample variance % REFB
dominated.)

\begin{figure*}
\resizebox{0.8\textwidth}{!}{\includegraphics{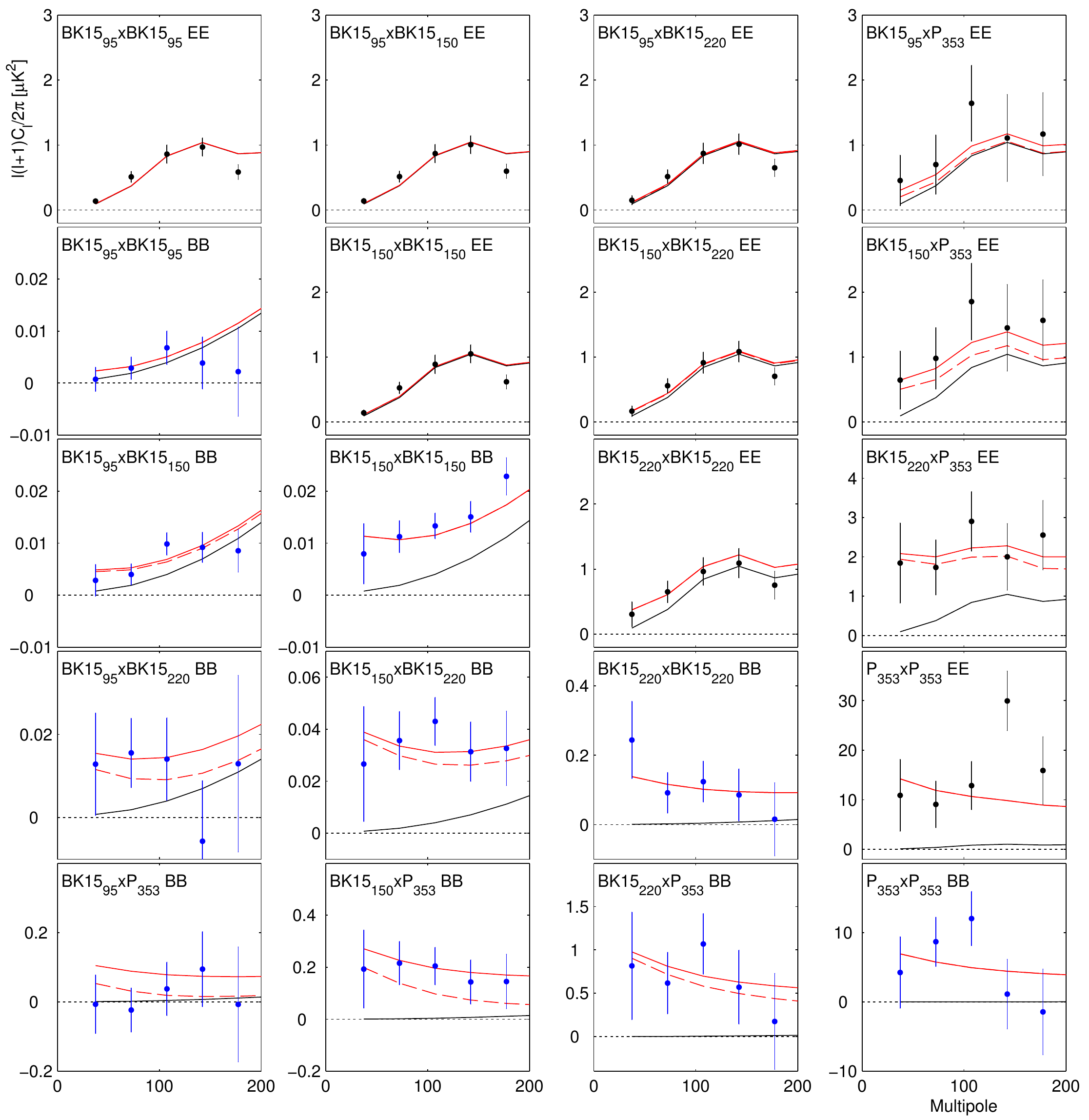}}
\caption{$EE$ and $BB$ auto- and cross-spectra calculated using
\biceptwo/\keck\ 95, 150 \& 220\,GHz maps and the \planck\ 353\,GHz
map.
The \biceptwo/\keck\ maps use all data taken up to and including
the 2015 observing season---we refer to these as BK15.
The black lines show the model expectation values
for lensed-\lcdm, while the red lines show the expectation values of the
baseline lensed-\lcdm+dust model from our previous BK14 analysis
($r=0$, $\Adf=4.3$\,\uksq, $\Bd=1.6$, $\ad=-0.4$),
and the error bars are scaled to that model.
Note that the model shown was fit to $BB$ only and did not
use the 220\,GHz points (which are entirely new).
The agreement with the spectra involving 220\,GHz and all the
$EE$ spectra (under the assumption that $EE/BB=2$ for dust)
is therefore a validation of the model.
(The dashed red lines show the expectation values of the
lensed-\lcdm+dust model when adding strong spectral decorrelation
of the dust pattern---see Appendix~\ref{app:decorr} for further information.)
}
\label{fig:powspecres_bkbands}
\end{figure*}

\begin{figure}
\resizebox{\columnwidth}{!}{\includegraphics{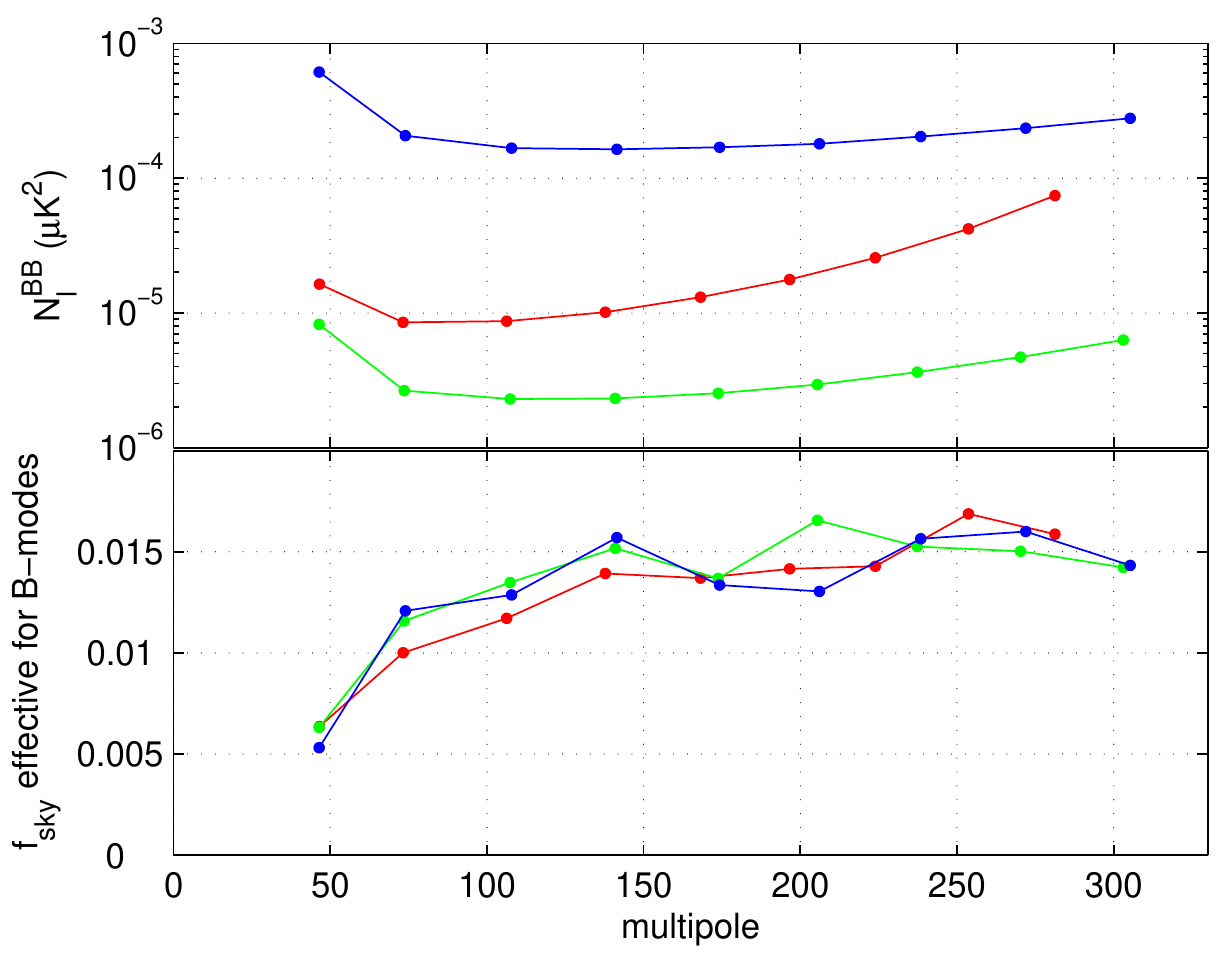}}
\caption{{\it Upper:} The noise spectra of the BK15 maps
for 95\,GHz (red), 150\,GHz (green) and 220\,GHz (blue)
after correction for the filtering of signal which occurs
due to the beam roll-off and timestream filtering.
(Note that no $\ell^2$ scaling is applied.)
{\it Lower:} The effective sky fraction as calculated from
the ratio of the mean noise realization bandpowers to their fluctuation
$f_\mathrm{sky}(\ell)=\frac{1}{2\ell \Delta \ell} \left( \frac{\sqrt{2}\bar{N_b}}{\sigma(N_b)} \right)^2$,
i.e.\ the observed number of {\bmode} degrees of freedom divided by the
nominal full-sky number.
The turn-down at low $\ell$ is due to mode loss to the timestream filtering and
matrix purification.
}
\label{fig:nl_fsky}
\end{figure}

Fig.~\ref{fig:nl_fsky} upper shows the noise spectra
(derived using the sign-flip technique~\cite{biceptwoI,vanengelen12}) % REFA
for the three BK15 bands after correction for the filter and beam
suppression.
The turn up at low-$\ell$ is partially due to residual
atmospheric $1/f$ in the pair-difference data and hence is
weakest in the 95\,GHz band where water vapor emission is
weakest.
In an auto-spectrum the quantity which determines the ability to constrain % REFC
$r$ is the fluctuation of the noise bandpowers rather than their mean.
The lower panel therefore shows the effective sky fraction observed
as inferred from the fractional noise fluctuation.
Together, these panels provide a useful synoptic measure of the % REFB
loss of information due to noise, filtering, and $EE/BB$
separation in the lowest bandpowers.
We suggest that other experiments reproduce this plot
for comparison purposes.

{\it Likelihood Analysis.}---We perform likelihood analysis using
the methods introduced in BKP and refined in BK14.
We use the Hamimeche-Lewis approximation~\citep[hereafter HL]{hamimeche08} to the joint
likelihood of the ensemble of 78 $BB$ auto- and cross-spectra taken % REFB
between the \biceptwo/\keck\ and \wmap/\planck\
maps.
We compare the observed bandpower values for ${20<\ell<330}$
(9 bandpowers per spectrum) % REFB
to an eight parameter model of lensed-\lcdm+$r$+dust+synchrotron+noise
and explore the parameter space using \texttt{COSMOMC}~\citep{cosmomc}
(which implements a Markov chain Monte Carlo). % REFB
As in our previous analyses the bandpower covariance matrix is derived % REFA+
from 499 simulations of signal and noise, explicitly setting to zero terms
such as the covariance of signal-only bandpowers with noise-only bandpowers
or covariance of \bicep/\keck\ noise bandpowers with \wmap/\planck\ noise bandpowers~\cite{bkp}.
The tensor/scalar power ratio $r$ is evaluated at a pivot scale of
0.05~Mpc$^{-1}$, and we fix the tensor spectral index $n_t=0$.
The \texttt{COSMOMC} module containing the data and model
is available for download at \url{http://bicepkeck.org}.
We make only one change to the ``baseline'' analysis choices of BK14,
expanding the prior on the dust/sync correlation parameter.
The following paragraphs briefly summarize.

We include dust with amplitude $\Adf$ evaluated
at 353\,GHz and ${\ell=80}$.
The frequency spectral behavior is taken as a modified black body spectrum
with ${T_\mathrm{d}=19.6}$\,K and ${\Bd=1.59 \pm 0.11}$, using a Gaussian prior
with the given $1\sigma$ width, this being an upper limit
on the patch-to-patch variation~\cite{planckiXXII,bkp}.
We note that the latest \planck\ analysis finds a slightly
lower central value of ${\Bd=1.53}$~\cite{planckiLIV} (well within our
prior range) with no detected trends with galactic latitude, angular scale or $EE$ vs.\ $BB$.
The spatial power spectrum is taken as a power law
${\mathcal{D}_\ell \propto \ell^{\ad}}$ marginalizing
uniformly over the (generous) range ${-1<\ad<0}$
(where ${\mathcal{D}_\ell \equiv \clstar}$).
\planck\ analysis consistently finds approximate power
law behavior of both the $EE$ and $BB$ dust spectra with
exponents $\approx -0.4$ ~\cite{planckiXXX,planckiLIV}. % REFB

We include synchrotron with amplitude $\Asf$
evaluated at 23\,GHz (the lowest \wmap\ band) and $\ell=80$,
assuming a simple power law for the frequency spectral behavior
${\As \propto \nu^{\Bs}}$ with a Gaussian prior
${\Bs=-3.1\pm0.3}$~\citep{fuskeland14}.
We note that recent analysis of 2.3\,GHz data from \spass\ in
conjunction with \wmap\ and \planck\ finds $\Bs=-3.2$ with no detected
trends with galactic latitude or angular scale~\cite{krachmalnicoff18}.
The spatial power spectrum is taken as a power law
${\mathcal{D}_\ell \propto \ell^{\as}}$ marginalizing
over the range ${-1<\as<0}$~\cite{dunkley08}.
The recent \spass\ analysis finds a value at the
bottom end of this range ($\approx -1$) for $BB$ at high galactic latitude. % REFB

Finally we include sync/dust correlation parameter $\epsilon$
(called $\rho$ in some other papers~\cite{choi15,planckiLIV,krachmalnicoff18}).
In BK14 we marginalized over the range ${0<\epsilon<1}$ but
in this paper we extend to the full possible range
${-1<\epsilon<1}$.
The latest \planck\ analysis does not detect sync/dust
correlation at high galactic latitude and the $\ell$
range of interest~\cite{planckiLIV}.

\begin{figure*}
\begin{center}
\resizebox{0.7\textwidth}{!}{\includegraphics{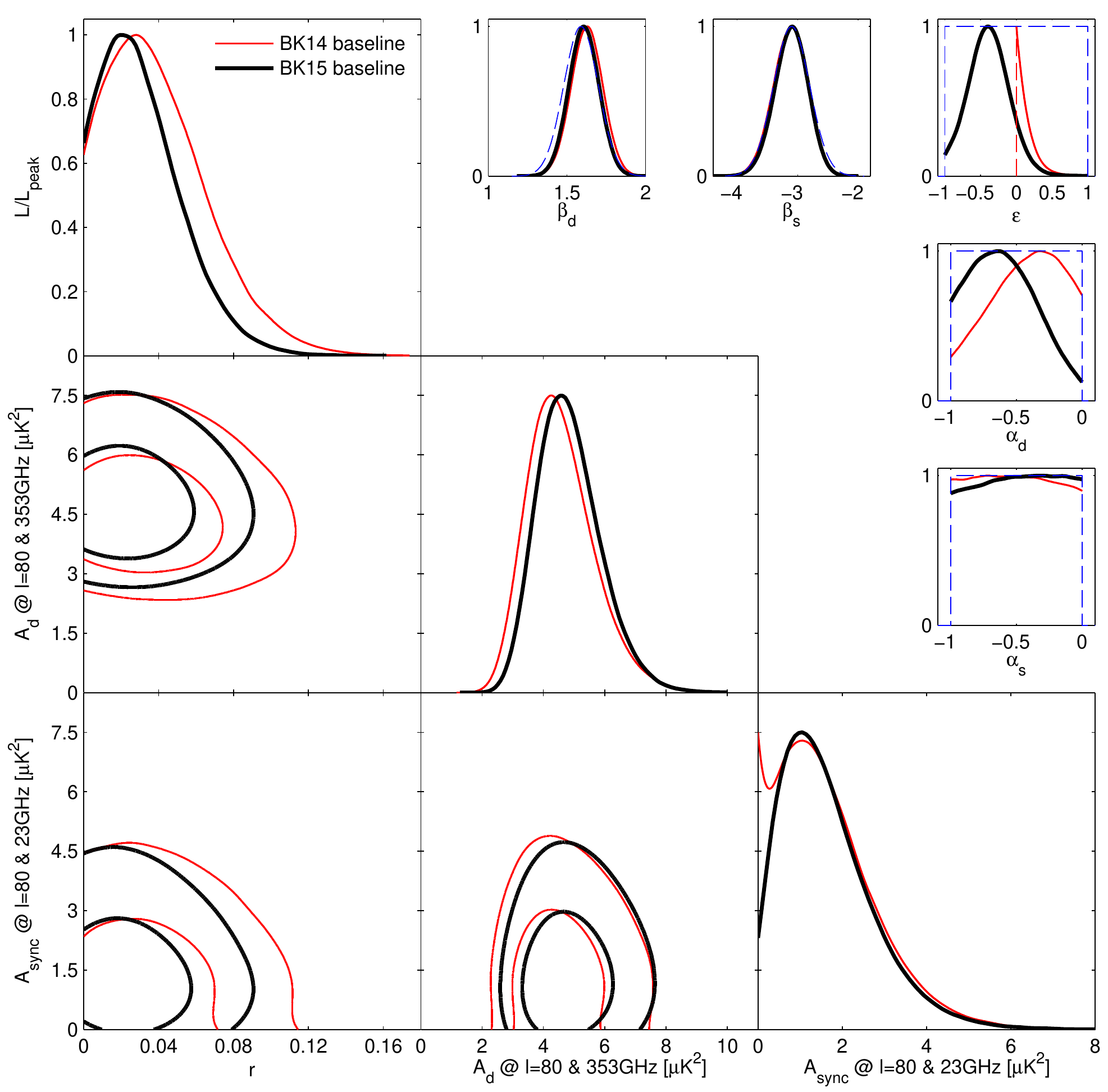}}
\end{center}
\caption{Results of a multicomponent multi-spectral likelihood
analysis of \biceptwo/\keck+\wmap/\planck\ data.
The red faint curves are the baseline result from the previous
BK14 paper (the black curves from Fig.~4 of that paper).
The bold black curves are the new baseline BK15 results.
Differences between these analyses include adding
\keckarray\ data taken during the 2015 observing season,
in particular doubling the 95\,GHz sensitivity
and adding, for the first time, a 220\,GHz channel.
(In addition the $\epsilon$ prior is modified.)
The upper limit on the tensor-to-scalar ratio tightens
to $r_{0.05}<\rul$ at 95\protect\% confidence.
The parameters $\Ad$ and $\As$ are the
amplitudes of the dust and synchrotron $B$-mode power spectra, where
$\beta$ and $\alpha$ are the respective frequency and spatial
spectral indices.
The correlation coefficient between the dust and synchrotron patterns is
$\epsilon$.
In the $\beta$, $\alpha$ and $\epsilon$ panels the dashed lines
show the priors placed on these parameters (either Gaussian or uniform).
Broadening or tightening the uniform prior range on $\alpha_s$
and $\alpha_d$ results in very small changes, and negligible changes to
the $r$ constraint.} % REFA+
\label{fig:likebase}
\end{figure*}

Results of the baseline analysis are shown in
Fig.~\ref{fig:likebase} and yield the following
statistics:
$r_{0.05}=\rrange$ ($r_{0.05}<\rul$ at 95\% confidence),
$\Adf=\Adrange$\,\uksq, and
$\Asf=\Asrange$\,\uksq, ($\Asf<\Asul$\,\uksq at 95\% confidence).
For $r$, the zero-to-peak likelihood ratio is \rztop.
Taking
${\frac{1}{2} \left( 1-f \left( -2\log{L_0/L_{\rm peak}} \right) \right)}$,
where $f$ is the $\chi^2$ CDF (for one degree of freedom),
we estimate that the probability to get a likelihood ratio smaller than this is
\rztopps\% if, in fact, $r=0$.
As compared to the previous analysis, the likelihood curve
for $r$ shifts down slightly and tightens.
The $\Ad$ curve shifts up very slightly but remains about the
same width (presumably saturated at sample variance),
and the $\As$ curve loses the second bump at zero.

The maximum likelihood model (including priors) has parameters
$r_{0.05}=\rmlm$, $\Adf=\Admlm$\,\uksq,
$\Asf=\Asmlm$\,\uksq, 
$\Bd=\Bdmlm$, $\Bs=\Bsmlm$,
$\ad=\admlm$, $\as=\asmlm$,
and $\epsilon=\emlm$.
This model is an acceptable fit to the data with the probability to
exceed (PTE) the observed value of $\chi^2$ being \chitwoptesim.
Thus, while fluctuation about the assumed power law
behavior of the dust component is in general expected to be
``super-Gaussian''~\cite{planckiLIV}, we find no evidence for this at the
present noise level---see Appendix~\ref{app:allspec}
for further details.

We have explored several variations from the baseline analysis choices
and data selection and find that these do not significantly alter
the results.
Removing the prior on $\Bd$ makes the $r$ constraint curve % REFA
slightly broader resulting in $r_{0.05}<0.079$ (95\protect\%),
while using the \bicep/\keck\ data only shifts the peak
position down to zero resulting in $r_{0.05}<0.063$.
Concerns have been raised that the known problems with the LFI % REFB+
maps~\cite{planck2015II} might affect the analysis---excluding LFI the $r$
constraint curve peak position shifts down to $r=0.012^{+0.022}_{-0.012}$ ($r_{0.05}<0.065$,
with zero-to-peak likelihood ratio of 0.90, and 32\% probability to get a
smaller value if $r=0$),
while the constraint on $\Asf$ becomes $2.4^{+1.9}_{-1.4}$\,\uksq.
The shifts when varying the data set selection (e.g. omitting \planck)
are not statistically significant when compared to shifts of lensed-\lcdm+dust+noise
simulations---see Appendices~\ref{app:likeevol} and~\ref{app:likevar} for further details.
Freeing the amplitude of the lensing power we obtain % REFA
$A_{\rm L}= 1.15^{+0.16}_{-0.14}$, and detect lensing
at $8.8 \sigma$ significance.

The results of likelihood analysis where the parameters
are restricted to, and marginalized over, physical values only
can potentially be biased.
Running the baseline analysis on an ensemble of
lensed-\lcdm+dust+noise simulations with simple Gaussian dust
we do not detect bias.
Half of the $r$ constraint curves peak at zero and the CDF of the
zero-to-peak likelihood ratios closely follows the idealized analytic
expectation.
When running maximum likelihood searches on the simulations
with the parameters unrestricted we again obtain unbiased results
and find that $\sigma(r)=0.020$.
See Appendix~\ref{app:likevalid} for further details.

We extend the maximum likelihood validation study to a suite
of third-party foreground models~\citep{thorne17,hensley2015,kritsuk17}.
These models do not necessarily conform to the foreground
parameterization which we are using, and when fit to
it are in general expected to produce bias on $r$.
However, for the models considered we find that such bias
is small compared to the instrumental noise---see
Appendix~\ref{app:altfore}.

Spatial variation of the frequency spectral behavior of dust
will lead to a decorrelation of the dust patterns
as observed in different frequency bands.
Since the baseline parametric model assumes a fixed dust
pattern as a function of frequency such variation will lead
to bias on $r$.
Dust decorrelation surely exists at some level---the
question is whether it is relevant as compared to the
current experimental noise.
For the third-party foreground models mentioned above,
decorrelation is very small.
Since our previous BK14 paper \planck\ Intermediate Paper
L~\cite{planckiL} appeared claiming a detection of relatively
strong dust decorrelation between 217 and 353\,GHz.
This was followed up by Ref.~\citep{sheehy17}, which analyzed the
same data and found no evidence for dust decorrelation, and
\planck\ Intermediate Paper~LIV~\citep{planckiLIV}, which performed
a more sophisticated multi-frequency analysis and again
found no evidence.
In the meantime we added a decorrelation parameter
to our analysis framework.
Including it only increases $\sigma(r)$ from 0.020 to 0.021, but % REFA
for the present data set this parameter is partially
degenerate with $r$ and including it results in a downward bias
on $r$ in simulations---see Appendix~\ref{app:decorr} for more details.

By cross correlating against the \planck\ CO map we find
that the contamination of our 220\,GHz map by CO is equivalent
to $r \sim 10^{-4}$.

{\it Conclusions.}---The previous BK14 analysis yielded the
constraint $r_{0.05}<0.090$ (95\%).
Adding the \keckarray\ data taken during 2015 we obtain
the BK15 result $r_{0.05}<\rul$.
The distributions of maximum likelihood $r$ values
in simulations where the true value of $r$ is zero give
$\sigma(r_{0.05})=0.024$ and $\sigma(r_{0.05})=0.020$ for BK14 and BK15
respectively.
The BK15 simulations have a median 95\% upper limit
of $r_{0.05}<0.046$.

\begin{figure}
\begin{center}
\resizebox{\columnwidth}{!}{\includegraphics{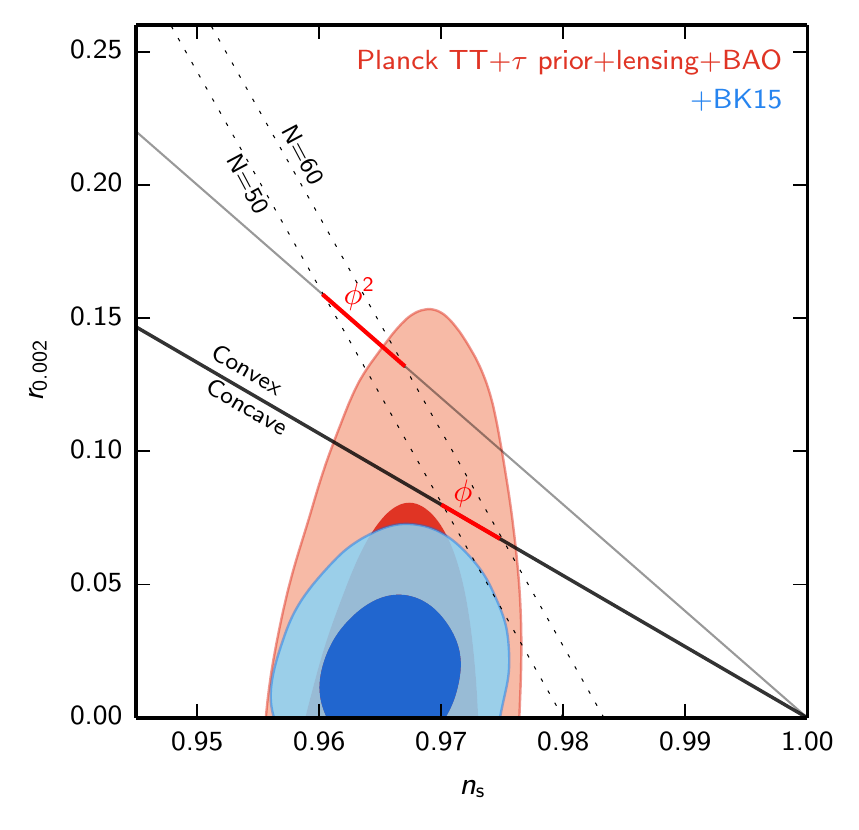}}
\end{center}
\caption{
Constraints in the $r$ vs.\ $n_s$ plane when using \planck\ 2015
plus additional data,
and when also adding \biceptwo/\keck\ data through the end of the 2015
season---the constraint on $r$ tightens
from $r_{0.05}<0.12$ to $r_{0.05}<0.06$.
This figure is adapted from Fig.~21 of Ref.~\cite{planck2015XIII}, with two
notable differences: switching \textit{lowP} to \textit{lowT} plus a $\tau$
prior of $0.055\pm 0.009$ Ref.~\cite{planck2016XLVI}, and the exclusion of JLA
data and the $H_0$ prior.}
\label{fig:rns}
\end{figure}

Fig.~\ref{fig:rns} shows the constraints in the
$r$ vs. $n_s$ plane for \planck\ 2015 plus additional data ($r_{0.05}<0.12$)
and when adding in also BK15 ($r_{0.05}<0.062$).
In contrast to the BK14 result the $\phi$ model now lies entirely % REFA+
outside of the 95\% contour.

\begin{figure}
\resizebox{\columnwidth}{!}{\includegraphics{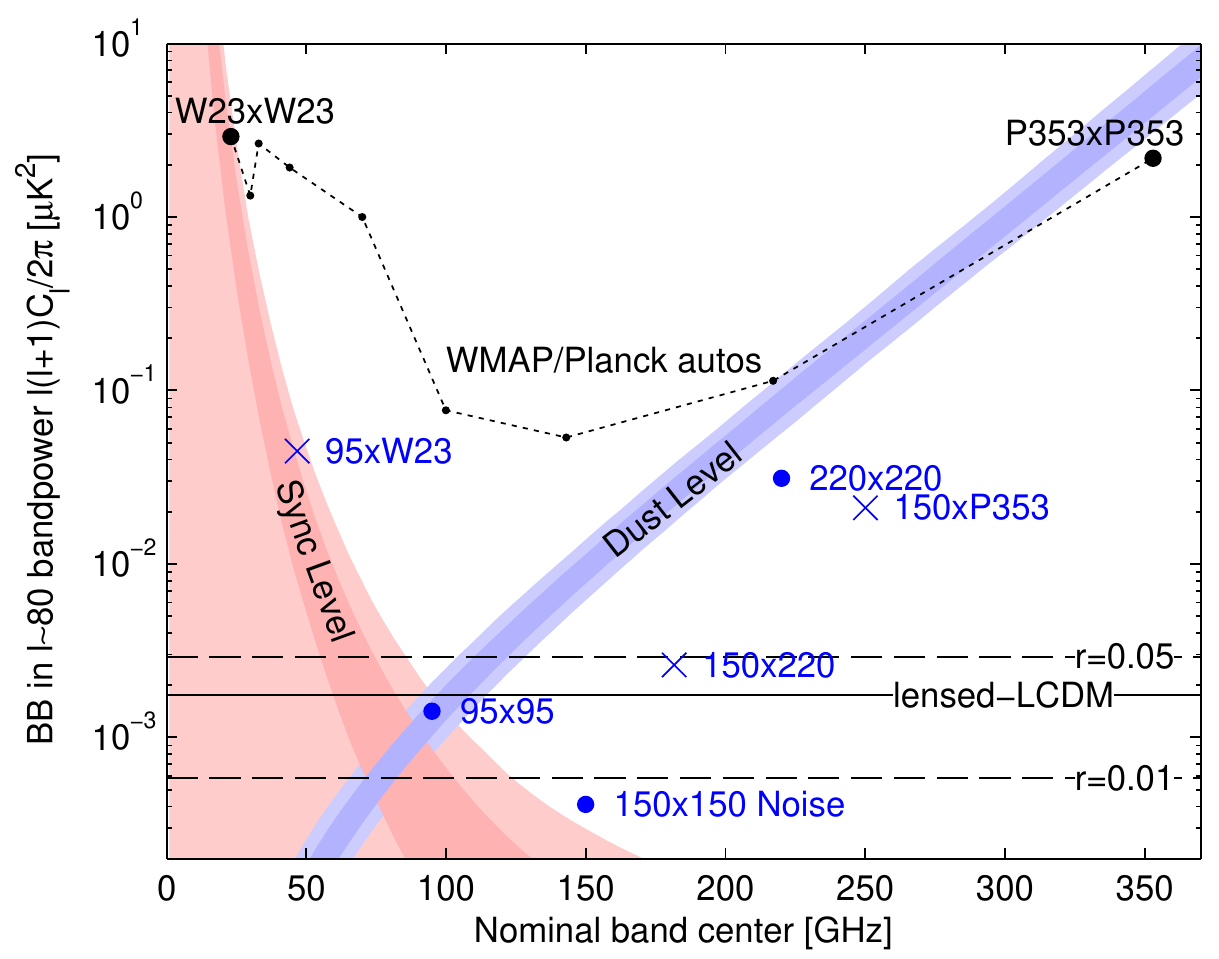}}
\caption{
Expectation values and noise uncertainties
for the $\ell\sim80$ $BB$ bandpower in the \biceptwo/\keck\ field.
The solid and dashed black lines show the expected signal power of
lensed-\lcdm\ and $r_{0.05}=0.05$ \& 0.01. 
Since CMB units are used, the levels corresponding
to these are flat with frequency.
The blue/red bands show the 1 and $2\sigma$ ranges of dust and
synchrotron in the baseline analysis including the uncertainties
in the amplitude and frequency spectral index parameters
($\Asf,\Bs$ and $\Adf,\Bd$).
The \biceptwo/\keck\ auto-spectrum noise uncertainties are shown as large blue
circles, and the noise uncertainties of the \wmap/\planck\ single-frequency
spectra evaluated in the \biceptwo/\keck\ field are shown in black.
The blue crosses show the noise uncertainty of selected cross-spectra,
and are plotted at horizontal positions such that they
can be compared vertically with the dust and sync curves.}
\label{fig:noilev}
\end{figure}

Fig.~\ref{fig:noilev} shows the BK15 noise uncertainties in the $\ell\approx 80$
bandpowers as compared to the signal levels.
Note that the new \keck\ 220\,GHz band has approximately the same
signal-to-noise on dust as \planck\ 353\,GHz with two receiver-years
of operation.
In 2016 and 2017 we recorded an additional eight receiver-years
of data which will reduce the noise by a factor of 5 \& $\sqrt{5}$
for $220\times220$ \& $150\times220$ respectively.

As seen in the lower right panel of Fig.~\ref{fig:likebase}
with four \keck\ receiver-years of data, our 95\,GHz data starts
to weakly prefer a non-zero value for the synchrotron amplitude
for the first time.
In 2017 alone \bicepthree\ recorded nearly twice as much
data in the 95\,GHz band as is included in the current result.
We plan to proceed directly to a BK17 result which can be expected
to improve substantially on the current results.

Dust decorrelation, and foreground complexity more generally,
will remain a serious concern.
With higher quality data we will be able to constrain the
foreground behavior ever better, but of course we will also
need to constrain it ever better.
The \biceparray\ experiment which is under construction will
provide \bicepthree\ class receivers in the 30/40, 95, 150 and
220/270\,GHz bands and is projected to reach $\sigma(r)<0.005$
within five years.

\acknowledgments

The \biceptwo/\keckarray\ projects have been made possible through
a series of grants from the National Science Foundation
including 0742818, 0742592, 1044978, 1110087, 1145172, 1145143, 1145248,
1639040, 1638957, 1638978, \& 1638970, and by the Keck Foundation.
The development of antenna-coupled detector technology was supported
by the JPL Research and Technology Development Fund, and by NASA Grants
06-ARPA206-0040, 10-SAT10-0017, 12-SAT12-0031, 14-SAT14-0009
\& 16-SAT-16-0002.
The development and testing of focal planes were supported
by the Gordon and Betty Moore Foundation at Caltech.
Readout electronics were supported by a Canada Foundation
for Innovation grant to UBC.
Support for quasi-optical filtering was provided by UK STFC grant ST/N000706/1.
The computations in this paper were run on the Odyssey cluster
supported by the FAS Science Division Research Computing Group at
Harvard University.
The analysis effort at Stanford and SLAC is partially supported by
the U.S. DOE Office of Science.
We thank the staff of the U.S. Antarctic Program and in particular
the South Pole Station without whose help this research would not
have been possible.
Most special thanks go to our heroic winter-overs Robert Schwarz
and Steffen Richter.
We thank all those who have contributed past efforts to the \bicep--\keckarray\
series of experiments, including the \bicepone\ team.
We also thank the \planck\ and \wmap\ teams for the use of their
data, and are grateful to the \planck\ team for helpful discussions
including on the use of the $\tau$ prior in Figure~\ref{fig:rns}.

\bibliography{ms}

\begin{thebibliography}{48}%
\makeatletter
\providecommand \@ifxundefined [1]{%
 \@ifx{#1\undefined}
}%
\providecommand \@ifnum [1]{%
 \ifnum #1\expandafter \@firstoftwo
 \else \expandafter \@secondoftwo
 \fi
}%
\providecommand \@ifx [1]{%
 \ifx #1\expandafter \@firstoftwo
 \else \expandafter \@secondoftwo
 \fi
}%
\providecommand \natexlab [1]{#1}%
\providecommand \enquote  [1]{``#1''}%
\providecommand \bibnamefont  [1]{#1}%
\providecommand \bibfnamefont [1]{#1}%
\providecommand \citenamefont [1]{#1}%
\providecommand \href@noop [0]{\@secondoftwo}%
\providecommand \href [0]{\begingroup \@sanitize@url \@href}%
\providecommand \@href[1]{\@@startlink{#1}\@@href}%
\providecommand \@@href[1]{\endgroup#1\@@endlink}%
\providecommand \@sanitize@url [0]{\catcode `\\12\catcode `\$12\catcode
  `\&12\catcode `\#12\catcode `\^12\catcode `\_12\catcode `\%12\relax}%
\providecommand \@@startlink[1]{}%
\providecommand \@@endlink[0]{}%
\providecommand \url  [0]{\begingroup\@sanitize@url \@url }%
\providecommand \@url [1]{\endgroup\@href {#1}{\urlprefix }}%
\providecommand \urlprefix  [0]{URL }%
\providecommand \Eprint [0]{\href }%
\providecommand \doibase [0]{http://dx.doi.org/}%
\providecommand \selectlanguage [0]{\@gobble}%
\providecommand \bibinfo  [0]{\@secondoftwo}%
\providecommand \bibfield  [0]{\@secondoftwo}%
\providecommand \translation [1]{[#1]}%
\providecommand \BibitemOpen [0]{}%
\providecommand \bibitemStop [0]{}%
\providecommand \bibitemNoStop [0]{.\EOS\space}%
\providecommand \EOS [0]{\spacefactor3000\relax}%
\providecommand \BibitemShut  [1]{\csname bibitem#1\endcsname}%
\let\auto@bib@innerbib\@empty
%</preamble>
\bibitem [{\citenamefont {{Planck Collaboration 2018
  VI}}(2018)}]{planck2018VI}%
  \BibitemOpen
  \bibfield  {author} {\bibinfo {author} {\bibnamefont {{Planck Collaboration
  2018 VI}}},\ }\href@noop {} {\bibfield  {journal} {\bibinfo  {journal} {ArXiv
  e-prints}\ } (\bibinfo {year} {2018})},\ \Eprint
  {http://arxiv.org/abs/1807.06209} {arXiv:1807.06209} \BibitemShut {NoStop}%
\bibitem [{\citenamefont {{Penzias}}\ and\ \citenamefont
  {{Wilson}}(1965)}]{penzias65}%
  \BibitemOpen
  \bibfield  {author} {\bibinfo {author} {\bibfnamefont {A.~A.}\ \bibnamefont
  {{Penzias}}}\ and\ \bibinfo {author} {\bibfnamefont {R.~W.}\ \bibnamefont
  {{Wilson}}},\ }\href {\doibase 10.1086/148307} {\bibfield  {journal}
  {\bibinfo  {journal} {\apj}\ }\textbf {\bibinfo {volume} {142}},\ \bibinfo
  {pages} {419} (\bibinfo {year} {1965})}\BibitemShut {NoStop}%
\bibitem [{\citenamefont {{Planck Collaboration 2015
  XIII}}(2016)}]{planck2015XIII}%
  \BibitemOpen
  \bibfield  {author} {\bibinfo {author} {\bibnamefont {{Planck Collaboration
  2015 XIII}}},\ }\href {\doibase 10.1051/0004-6361/201525830} {\bibfield
  {journal} {\bibinfo  {journal} {\aap}\ }\textbf {\bibinfo {volume} {594}},\
  \bibinfo {eid} {A13} (\bibinfo {year} {2016})},\ \Eprint
  {http://arxiv.org/abs/1502.01589} {arXiv:1502.01589} \BibitemShut {NoStop}%
\bibitem [{\citenamefont {{Kamionkowski}}\ and\ \citenamefont
  {{Kovetz}}(2016)}]{kamionkowski2015}%
  \BibitemOpen
  \bibfield  {author} {\bibinfo {author} {\bibfnamefont {M.}~\bibnamefont
  {{Kamionkowski}}}\ and\ \bibinfo {author} {\bibfnamefont {E.~D.}\
  \bibnamefont {{Kovetz}}},\ }\href {\doibase
  10.1146/annurev-astro-081915-023433} {\bibfield  {journal} {\bibinfo
  {journal} {Annual Review of Astronomy and Astrophysics}\ }\textbf {\bibinfo
  {volume} {54}},\ \bibinfo {pages} {227} (\bibinfo {year} {2016})},\ \Eprint
  {http://arxiv.org/abs/1510.06042} {arXiv:1510.06042} \BibitemShut {NoStop}%
\bibitem [{\citenamefont {{Abazajian}}\ \emph {et~al.}(2016)\citenamefont
  {{Abazajian}}, \citenamefont {{Adshead}}, \citenamefont {{Ahmed}},
  \citenamefont {{Allen}}, \citenamefont {{Alonso}}, \citenamefont {{Arnold}},
  \citenamefont {{Baccigalupi}}, \citenamefont {{Bartlett}}, \citenamefont
  {{Battaglia}}, \citenamefont {{Benson}}, \citenamefont {{Bischoff}},
  \citenamefont {{Borrill}}, \citenamefont {{Buza}}, \citenamefont
  {{Calabrese}}, \citenamefont {{Caldwell}}, \citenamefont {{Carlstrom}},
  \citenamefont {{Chang}}, \citenamefont {{Crawford}}, \citenamefont
  {{Cyr-Racine}}, \citenamefont {{De Bernardis}}, \citenamefont {{de Haan}},
  \citenamefont {{di Serego Alighieri}}, \citenamefont {{Dunkley}},
  \citenamefont {{Dvorkin}}, \citenamefont {{Errard}}, \citenamefont
  {{Fabbian}}, \citenamefont {{Feeney}}, \citenamefont {{Ferraro}},
  \citenamefont {{Filippini}}, \citenamefont {{Flauger}}, \citenamefont
  {{Fuller}}, \citenamefont {{Gluscevic}}, \citenamefont {{Green}},
  \citenamefont {{Grin}}, \citenamefont {{Grohs}}, \citenamefont {{Henning}},
  \citenamefont {{Hill}}, \citenamefont {{Hlozek}}, \citenamefont {{Holder}},
  \citenamefont {{Holzapfel}}, \citenamefont {{Hu}}, \citenamefont
  {{Huffenberger}}, \citenamefont {{Keskitalo}}, \citenamefont {{Knox}},
  \citenamefont {{Kosowsky}}, \citenamefont {{Kovac}}, \citenamefont
  {{Kovetz}}, \citenamefont {{Kuo}}, \citenamefont {{Kusaka}}, \citenamefont
  {{Le Jeune}}, \citenamefont {{Lee}}, \citenamefont {{Lilley}}, \citenamefont
  {{Loverde}}, \citenamefont {{Madhavacheril}}, \citenamefont {{Mantz}},
  \citenamefont {{Marsh}}, \citenamefont {{McMahon}}, \citenamefont
  {{Meerburg}}, \citenamefont {{Meyers}}, \citenamefont {{Miller}},
  \citenamefont {{Munoz}}, \citenamefont {{Nguyen}}, \citenamefont {{Niemack}},
  \citenamefont {{Peloso}}, \citenamefont {{Peloton}}, \citenamefont
  {{Pogosian}}, \citenamefont {{Pryke}}, \citenamefont {{Raveri}},
  \citenamefont {{Reichardt}}, \citenamefont {{Rocha}}, \citenamefont
  {{Rotti}}, \citenamefont {{Schaan}}, \citenamefont {{Schmittfull}},
  \citenamefont {{Scott}}, \citenamefont {{Sehgal}}, \citenamefont
  {{Shandera}}, \citenamefont {{Sherwin}}, \citenamefont {{Smith}},
  \citenamefont {{Sorbo}}, \citenamefont {{Starkman}}, \citenamefont {{Story}},
  \citenamefont {{van Engelen}}, \citenamefont {{Vieira}}, \citenamefont
  {{Watson}}, \citenamefont {{Whitehorn}},\ and\ \citenamefont {{Kimmy
  Wu}}}]{S4_sciencebook2016}%
  \BibitemOpen
  \bibfield  {author} {\bibinfo {author} {\bibfnamefont {K.~N.}\ \bibnamefont
  {{Abazajian}}}, \bibinfo {author} {\bibfnamefont {P.}~\bibnamefont
  {{Adshead}}}, \bibinfo {author} {\bibfnamefont {Z.}~\bibnamefont {{Ahmed}}},
  \bibinfo {author} {\bibfnamefont {S.~W.}\ \bibnamefont {{Allen}}}, \bibinfo
  {author} {\bibfnamefont {D.}~\bibnamefont {{Alonso}}}, \bibinfo {author}
  {\bibfnamefont {K.~S.}\ \bibnamefont {{Arnold}}}, \bibinfo {author}
  {\bibfnamefont {C.}~\bibnamefont {{Baccigalupi}}}, \bibinfo {author}
  {\bibfnamefont {J.~G.}\ \bibnamefont {{Bartlett}}}, \bibinfo {author}
  {\bibfnamefont {N.}~\bibnamefont {{Battaglia}}}, \bibinfo {author}
  {\bibfnamefont {B.~A.}\ \bibnamefont {{Benson}}}, \bibinfo {author}
  {\bibfnamefont {C.~A.}\ \bibnamefont {{Bischoff}}}, \bibinfo {author}
  {\bibfnamefont {J.}~\bibnamefont {{Borrill}}}, \bibinfo {author}
  {\bibfnamefont {V.}~\bibnamefont {{Buza}}}, \bibinfo {author} {\bibfnamefont
  {E.}~\bibnamefont {{Calabrese}}}, \bibinfo {author} {\bibfnamefont
  {R.}~\bibnamefont {{Caldwell}}}, \bibinfo {author} {\bibfnamefont {J.~E.}\
  \bibnamefont {{Carlstrom}}}, \bibinfo {author} {\bibfnamefont {C.~L.}\
  \bibnamefont {{Chang}}}, \bibinfo {author} {\bibfnamefont {T.~M.}\
  \bibnamefont {{Crawford}}}, \bibinfo {author} {\bibfnamefont {F.-Y.}\
  \bibnamefont {{Cyr-Racine}}}, \bibinfo {author} {\bibfnamefont
  {F.}~\bibnamefont {{De Bernardis}}}, \bibinfo {author} {\bibfnamefont
  {T.}~\bibnamefont {{de Haan}}}, \bibinfo {author} {\bibfnamefont
  {S.}~\bibnamefont {{di Serego Alighieri}}}, \bibinfo {author} {\bibfnamefont
  {J.}~\bibnamefont {{Dunkley}}}, \bibinfo {author} {\bibfnamefont
  {C.}~\bibnamefont {{Dvorkin}}}, \bibinfo {author} {\bibfnamefont
  {J.}~\bibnamefont {{Errard}}}, \bibinfo {author} {\bibfnamefont
  {G.}~\bibnamefont {{Fabbian}}}, \bibinfo {author} {\bibfnamefont
  {S.}~\bibnamefont {{Feeney}}}, \bibinfo {author} {\bibfnamefont
  {S.}~\bibnamefont {{Ferraro}}}, \bibinfo {author} {\bibfnamefont {J.~P.}\
  \bibnamefont {{Filippini}}}, \bibinfo {author} {\bibfnamefont
  {R.}~\bibnamefont {{Flauger}}}, \bibinfo {author} {\bibfnamefont {G.~M.}\
  \bibnamefont {{Fuller}}}, \bibinfo {author} {\bibfnamefont {V.}~\bibnamefont
  {{Gluscevic}}}, \bibinfo {author} {\bibfnamefont {D.}~\bibnamefont
  {{Green}}}, \bibinfo {author} {\bibfnamefont {D.}~\bibnamefont {{Grin}}},
  \bibinfo {author} {\bibfnamefont {E.}~\bibnamefont {{Grohs}}}, \bibinfo
  {author} {\bibfnamefont {J.~W.}\ \bibnamefont {{Henning}}}, \bibinfo {author}
  {\bibfnamefont {J.~C.}\ \bibnamefont {{Hill}}}, \bibinfo {author}
  {\bibfnamefont {R.}~\bibnamefont {{Hlozek}}}, \bibinfo {author}
  {\bibfnamefont {G.}~\bibnamefont {{Holder}}}, \bibinfo {author}
  {\bibfnamefont {W.}~\bibnamefont {{Holzapfel}}}, \bibinfo {author}
  {\bibfnamefont {W.}~\bibnamefont {{Hu}}}, \bibinfo {author} {\bibfnamefont
  {K.~M.}\ \bibnamefont {{Huffenberger}}}, \bibinfo {author} {\bibfnamefont
  {R.}~\bibnamefont {{Keskitalo}}}, \bibinfo {author} {\bibfnamefont
  {L.}~\bibnamefont {{Knox}}}, \bibinfo {author} {\bibfnamefont
  {A.}~\bibnamefont {{Kosowsky}}}, \bibinfo {author} {\bibfnamefont
  {J.}~\bibnamefont {{Kovac}}}, \bibinfo {author} {\bibfnamefont {E.~D.}\
  \bibnamefont {{Kovetz}}}, \bibinfo {author} {\bibfnamefont {C.-L.}\
  \bibnamefont {{Kuo}}}, \bibinfo {author} {\bibfnamefont {A.}~\bibnamefont
  {{Kusaka}}}, \bibinfo {author} {\bibfnamefont {M.}~\bibnamefont {{Le
  Jeune}}}, \bibinfo {author} {\bibfnamefont {A.~T.}\ \bibnamefont {{Lee}}},
  \bibinfo {author} {\bibfnamefont {M.}~\bibnamefont {{Lilley}}}, \bibinfo
  {author} {\bibfnamefont {M.}~\bibnamefont {{Loverde}}}, \bibinfo {author}
  {\bibfnamefont {M.~S.}\ \bibnamefont {{Madhavacheril}}}, \bibinfo {author}
  {\bibfnamefont {A.}~\bibnamefont {{Mantz}}}, \bibinfo {author} {\bibfnamefont
  {D.~J.~E.}\ \bibnamefont {{Marsh}}}, \bibinfo {author} {\bibfnamefont
  {J.}~\bibnamefont {{McMahon}}}, \bibinfo {author} {\bibfnamefont {P.~D.}\
  \bibnamefont {{Meerburg}}}, \bibinfo {author} {\bibfnamefont
  {J.}~\bibnamefont {{Meyers}}}, \bibinfo {author} {\bibfnamefont {A.~D.}\
  \bibnamefont {{Miller}}}, \bibinfo {author} {\bibfnamefont {J.~B.}\
  \bibnamefont {{Munoz}}}, \bibinfo {author} {\bibfnamefont {H.~N.}\
  \bibnamefont {{Nguyen}}}, \bibinfo {author} {\bibfnamefont {M.~D.}\
  \bibnamefont {{Niemack}}}, \bibinfo {author} {\bibfnamefont {M.}~\bibnamefont
  {{Peloso}}}, \bibinfo {author} {\bibfnamefont {J.}~\bibnamefont {{Peloton}}},
  \bibinfo {author} {\bibfnamefont {L.}~\bibnamefont {{Pogosian}}}, \bibinfo
  {author} {\bibfnamefont {C.}~\bibnamefont {{Pryke}}}, \bibinfo {author}
  {\bibfnamefont {M.}~\bibnamefont {{Raveri}}}, \bibinfo {author}
  {\bibfnamefont {C.~L.}\ \bibnamefont {{Reichardt}}}, \bibinfo {author}
  {\bibfnamefont {G.}~\bibnamefont {{Rocha}}}, \bibinfo {author} {\bibfnamefont
  {A.}~\bibnamefont {{Rotti}}}, \bibinfo {author} {\bibfnamefont
  {E.}~\bibnamefont {{Schaan}}}, \bibinfo {author} {\bibfnamefont {M.~M.}\
  \bibnamefont {{Schmittfull}}}, \bibinfo {author} {\bibfnamefont
  {D.}~\bibnamefont {{Scott}}}, \bibinfo {author} {\bibfnamefont
  {N.}~\bibnamefont {{Sehgal}}}, \bibinfo {author} {\bibfnamefont
  {S.}~\bibnamefont {{Shandera}}}, \bibinfo {author} {\bibfnamefont {B.~D.}\
  \bibnamefont {{Sherwin}}}, \bibinfo {author} {\bibfnamefont {T.~L.}\
  \bibnamefont {{Smith}}}, \bibinfo {author} {\bibfnamefont {L.}~\bibnamefont
  {{Sorbo}}}, \bibinfo {author} {\bibfnamefont {G.~D.}\ \bibnamefont
  {{Starkman}}}, \bibinfo {author} {\bibfnamefont {K.~T.}\ \bibnamefont
  {{Story}}}, \bibinfo {author} {\bibfnamefont {A.}~\bibnamefont {{van
  Engelen}}}, \bibinfo {author} {\bibfnamefont {J.~D.}\ \bibnamefont
  {{Vieira}}}, \bibinfo {author} {\bibfnamefont {S.}~\bibnamefont {{Watson}}},
  \bibinfo {author} {\bibfnamefont {N.}~\bibnamefont {{Whitehorn}}}, \ and\
  \bibinfo {author} {\bibfnamefont {W.~L.}\ \bibnamefont {{Kimmy Wu}}},\
  }\href@noop {} {\bibfield  {journal} {\bibinfo  {journal} {ArXiv e-prints}\ }
  (\bibinfo {year} {2016})},\ \Eprint {http://arxiv.org/abs/1610.02743}
  {arXiv:1610.02743} \BibitemShut {NoStop}%
\bibitem [{\citenamefont {{Keisler}}\ \emph {et~al.}(2015)\citenamefont
  {{Keisler}}, \citenamefont {{Hoover}}, \citenamefont {{Harrington}},
  \citenamefont {{Henning}}, \citenamefont {{Ade}}, \citenamefont {{Aird}},
  \citenamefont {{Austermann}}, \citenamefont {{Beall}}, \citenamefont
  {{Bender}}, \citenamefont {{Benson}}, \citenamefont {{Bleem}}, \citenamefont
  {{Carlstrom}}, \citenamefont {{Chang}}, \citenamefont {{Chiang}},
  \citenamefont {{Cho}}, \citenamefont {{Citron}}, \citenamefont {{Crawford}},
  \citenamefont {{Crites}}, \citenamefont {{de Haan}}, \citenamefont {{Dobbs}},
  \citenamefont {{Everett}}, \citenamefont {{Gallicchio}}, \citenamefont
  {{Gao}}, \citenamefont {{George}}, \citenamefont {{Gilbert}}, \citenamefont
  {{Halverson}}, \citenamefont {{Hanson}}, \citenamefont {{Hilton}},
  \citenamefont {{Holder}}, \citenamefont {{Holzapfel}}, \citenamefont {{Hou}},
  \citenamefont {{Hrubes}}, \citenamefont {{Huang}}, \citenamefont {{Hubmayr}},
  \citenamefont {{Irwin}}, \citenamefont {{Knox}}, \citenamefont {{Lee}},
  \citenamefont {{Leitch}}, \citenamefont {{Li}}, \citenamefont {{Luong-Van}},
  \citenamefont {{Marrone}}, \citenamefont {{McMahon}}, \citenamefont {{Mehl}},
  \citenamefont {{Meyer}}, \citenamefont {{Mocanu}}, \citenamefont {{Natoli}},
  \citenamefont {{Nibarger}}, \citenamefont {{Novosad}}, \citenamefont
  {{Padin}}, \citenamefont {{Pryke}}, \citenamefont {{Reichardt}},
  \citenamefont {{Ruhl}}, \citenamefont {{Saliwanchik}}, \citenamefont
  {{Sayre}}, \citenamefont {{Schaffer}}, \citenamefont {{Shirokoff}},
  \citenamefont {{Smecher}}, \citenamefont {{Stark}}, \citenamefont {{Story}},
  \citenamefont {{Tucker}}, \citenamefont {{Vanderlinde}}, \citenamefont
  {{Vieira}}, \citenamefont {{Wang}}, \citenamefont {{Whitehorn}},
  \citenamefont {{Yefremenko}},\ and\ \citenamefont {{Zahn}}}]{keisler15}%
  \BibitemOpen
  \bibfield  {author} {\bibinfo {author} {\bibfnamefont {R.}~\bibnamefont
  {{Keisler}}}, \bibinfo {author} {\bibfnamefont {S.}~\bibnamefont {{Hoover}}},
  \bibinfo {author} {\bibfnamefont {N.}~\bibnamefont {{Harrington}}}, \bibinfo
  {author} {\bibfnamefont {J.~W.}\ \bibnamefont {{Henning}}}, \bibinfo {author}
  {\bibfnamefont {P.~A.~R.}\ \bibnamefont {{Ade}}}, \bibinfo {author}
  {\bibfnamefont {K.~A.}\ \bibnamefont {{Aird}}}, \bibinfo {author}
  {\bibfnamefont {J.~E.}\ \bibnamefont {{Austermann}}}, \bibinfo {author}
  {\bibfnamefont {J.~A.}\ \bibnamefont {{Beall}}}, \bibinfo {author}
  {\bibfnamefont {A.~N.}\ \bibnamefont {{Bender}}}, \bibinfo {author}
  {\bibfnamefont {B.~A.}\ \bibnamefont {{Benson}}}, \bibinfo {author}
  {\bibfnamefont {L.~E.}\ \bibnamefont {{Bleem}}}, \bibinfo {author}
  {\bibfnamefont {J.~E.}\ \bibnamefont {{Carlstrom}}}, \bibinfo {author}
  {\bibfnamefont {C.~L.}\ \bibnamefont {{Chang}}}, \bibinfo {author}
  {\bibfnamefont {H.~C.}\ \bibnamefont {{Chiang}}}, \bibinfo {author}
  {\bibfnamefont {H.-M.}\ \bibnamefont {{Cho}}}, \bibinfo {author}
  {\bibfnamefont {R.}~\bibnamefont {{Citron}}}, \bibinfo {author}
  {\bibfnamefont {T.~M.}\ \bibnamefont {{Crawford}}}, \bibinfo {author}
  {\bibfnamefont {A.~T.}\ \bibnamefont {{Crites}}}, \bibinfo {author}
  {\bibfnamefont {T.}~\bibnamefont {{de Haan}}}, \bibinfo {author}
  {\bibfnamefont {M.~A.}\ \bibnamefont {{Dobbs}}}, \bibinfo {author}
  {\bibfnamefont {W.}~\bibnamefont {{Everett}}}, \bibinfo {author}
  {\bibfnamefont {J.}~\bibnamefont {{Gallicchio}}}, \bibinfo {author}
  {\bibfnamefont {J.}~\bibnamefont {{Gao}}}, \bibinfo {author} {\bibfnamefont
  {E.~M.}\ \bibnamefont {{George}}}, \bibinfo {author} {\bibfnamefont
  {A.}~\bibnamefont {{Gilbert}}}, \bibinfo {author} {\bibfnamefont {N.~W.}\
  \bibnamefont {{Halverson}}}, \bibinfo {author} {\bibfnamefont
  {D.}~\bibnamefont {{Hanson}}}, \bibinfo {author} {\bibfnamefont {G.~C.}\
  \bibnamefont {{Hilton}}}, \bibinfo {author} {\bibfnamefont {G.~P.}\
  \bibnamefont {{Holder}}}, \bibinfo {author} {\bibfnamefont {W.~L.}\
  \bibnamefont {{Holzapfel}}}, \bibinfo {author} {\bibfnamefont
  {Z.}~\bibnamefont {{Hou}}}, \bibinfo {author} {\bibfnamefont {J.~D.}\
  \bibnamefont {{Hrubes}}}, \bibinfo {author} {\bibfnamefont {N.}~\bibnamefont
  {{Huang}}}, \bibinfo {author} {\bibfnamefont {J.}~\bibnamefont {{Hubmayr}}},
  \bibinfo {author} {\bibfnamefont {K.~D.}\ \bibnamefont {{Irwin}}}, \bibinfo
  {author} {\bibfnamefont {L.}~\bibnamefont {{Knox}}}, \bibinfo {author}
  {\bibfnamefont {A.~T.}\ \bibnamefont {{Lee}}}, \bibinfo {author}
  {\bibfnamefont {E.~M.}\ \bibnamefont {{Leitch}}}, \bibinfo {author}
  {\bibfnamefont {D.}~\bibnamefont {{Li}}}, \bibinfo {author} {\bibfnamefont
  {D.}~\bibnamefont {{Luong-Van}}}, \bibinfo {author} {\bibfnamefont {D.~P.}\
  \bibnamefont {{Marrone}}}, \bibinfo {author} {\bibfnamefont {J.~J.}\
  \bibnamefont {{McMahon}}}, \bibinfo {author} {\bibfnamefont {J.}~\bibnamefont
  {{Mehl}}}, \bibinfo {author} {\bibfnamefont {S.~S.}\ \bibnamefont {{Meyer}}},
  \bibinfo {author} {\bibfnamefont {L.}~\bibnamefont {{Mocanu}}}, \bibinfo
  {author} {\bibfnamefont {T.}~\bibnamefont {{Natoli}}}, \bibinfo {author}
  {\bibfnamefont {J.~P.}\ \bibnamefont {{Nibarger}}}, \bibinfo {author}
  {\bibfnamefont {V.}~\bibnamefont {{Novosad}}}, \bibinfo {author}
  {\bibfnamefont {S.}~\bibnamefont {{Padin}}}, \bibinfo {author} {\bibfnamefont
  {C.}~\bibnamefont {{Pryke}}}, \bibinfo {author} {\bibfnamefont {C.~L.}\
  \bibnamefont {{Reichardt}}}, \bibinfo {author} {\bibfnamefont {J.~E.}\
  \bibnamefont {{Ruhl}}}, \bibinfo {author} {\bibfnamefont {B.~R.}\
  \bibnamefont {{Saliwanchik}}}, \bibinfo {author} {\bibfnamefont {J.~T.}\
  \bibnamefont {{Sayre}}}, \bibinfo {author} {\bibfnamefont {K.~K.}\
  \bibnamefont {{Schaffer}}}, \bibinfo {author} {\bibfnamefont
  {E.}~\bibnamefont {{Shirokoff}}}, \bibinfo {author} {\bibfnamefont
  {G.}~\bibnamefont {{Smecher}}}, \bibinfo {author} {\bibfnamefont {A.~A.}\
  \bibnamefont {{Stark}}}, \bibinfo {author} {\bibfnamefont {K.~T.}\
  \bibnamefont {{Story}}}, \bibinfo {author} {\bibfnamefont {C.}~\bibnamefont
  {{Tucker}}}, \bibinfo {author} {\bibfnamefont {K.}~\bibnamefont
  {{Vanderlinde}}}, \bibinfo {author} {\bibfnamefont {J.~D.}\ \bibnamefont
  {{Vieira}}}, \bibinfo {author} {\bibfnamefont {G.}~\bibnamefont {{Wang}}},
  \bibinfo {author} {\bibfnamefont {N.}~\bibnamefont {{Whitehorn}}}, \bibinfo
  {author} {\bibfnamefont {V.}~\bibnamefont {{Yefremenko}}}, \ and\ \bibinfo
  {author} {\bibfnamefont {O.}~\bibnamefont {{Zahn}}},\ }\href {\doibase
  10.1088/0004-637X/807/2/151} {\bibfield  {journal} {\bibinfo  {journal}
  {\apj}\ }\textbf {\bibinfo {volume} {807}},\ \bibinfo {eid} {151} (\bibinfo
  {year} {2015})},\ \Eprint {http://arxiv.org/abs/1503.02315}
  {arXiv:1503.02315} \BibitemShut {NoStop}%
\bibitem [{\citenamefont {{Louis}}\ \emph {et~al.}(2017)\citenamefont
  {{Louis}}, \citenamefont {{Grace}}, \citenamefont {{Hasselfield}},
  \citenamefont {{Lungu}}, \citenamefont {{Maurin}}, \citenamefont {{Addison}},
  \citenamefont {{Ade}}, \citenamefont {{Aiola}}, \citenamefont {{Allison}},
  \citenamefont {{Amiri}}, \citenamefont {{Angile}}, \citenamefont
  {{Battaglia}}, \citenamefont {{Beall}}, \citenamefont {{de Bernardis}},
  \citenamefont {{Bond}}, \citenamefont {{Britton}}, \citenamefont
  {{Calabrese}}, \citenamefont {{Cho}}, \citenamefont {{Choi}}, \citenamefont
  {{Coughlin}}, \citenamefont {{Crichton}}, \citenamefont {{Crowley}},
  \citenamefont {{Datta}}, \citenamefont {{Devlin}}, \citenamefont {{Dicker}},
  \citenamefont {{Dunkley}}, \citenamefont {{D{\"u}nner}}, \citenamefont
  {{Ferraro}}, \citenamefont {{Fox}}, \citenamefont {{Gallardo}}, \citenamefont
  {{Gralla}}, \citenamefont {{Halpern}}, \citenamefont {{Henderson}},
  \citenamefont {{Hill}}, \citenamefont {{Hilton}}, \citenamefont {{Hilton}},
  \citenamefont {{Hincks}}, \citenamefont {{Hlozek}}, \citenamefont {{Ho}},
  \citenamefont {{Huang}}, \citenamefont {{Hubmayr}}, \citenamefont
  {{Huffenberger}}, \citenamefont {{Hughes}}, \citenamefont {{Infante}},
  \citenamefont {{Irwin}}, \citenamefont {{Muya Kasanda}}, \citenamefont
  {{Klein}}, \citenamefont {{Koopman}}, \citenamefont {{Kosowsky}},
  \citenamefont {{Li}}, \citenamefont {{Madhavacheril}}, \citenamefont
  {{Marriage}}, \citenamefont {{McMahon}}, \citenamefont {{Menanteau}},
  \citenamefont {{Moodley}}, \citenamefont {{Munson}}, \citenamefont {{Naess}},
  \citenamefont {{Nati}}, \citenamefont {{Newburgh}}, \citenamefont
  {{Nibarger}}, \citenamefont {{Niemack}}, \citenamefont {{Nolta}},
  \citenamefont {{Nu{\~n}ez}}, \citenamefont {{Page}}, \citenamefont
  {{Pappas}}, \citenamefont {{Partridge}}, \citenamefont {{Rojas}},
  \citenamefont {{Schaan}}, \citenamefont {{Schmitt}}, \citenamefont
  {{Sehgal}}, \citenamefont {{Sherwin}}, \citenamefont {{Sievers}},
  \citenamefont {{Simon}}, \citenamefont {{Spergel}}, \citenamefont {{Staggs}},
  \citenamefont {{Switzer}}, \citenamefont {{Thornton}}, \citenamefont
  {{Trac}}, \citenamefont {{Treu}}, \citenamefont {{Tucker}}, \citenamefont
  {{Van Engelen}}, \citenamefont {{Ward}},\ and\ \citenamefont
  {{Wollack}}}]{actpol16}%
  \BibitemOpen
  \bibfield  {author} {\bibinfo {author} {\bibfnamefont {T.}~\bibnamefont
  {{Louis}}}, \bibinfo {author} {\bibfnamefont {E.}~\bibnamefont {{Grace}}},
  \bibinfo {author} {\bibfnamefont {M.}~\bibnamefont {{Hasselfield}}}, \bibinfo
  {author} {\bibfnamefont {M.}~\bibnamefont {{Lungu}}}, \bibinfo {author}
  {\bibfnamefont {L.}~\bibnamefont {{Maurin}}}, \bibinfo {author}
  {\bibfnamefont {G.~E.}\ \bibnamefont {{Addison}}}, \bibinfo {author}
  {\bibfnamefont {P.~A.~R.}\ \bibnamefont {{Ade}}}, \bibinfo {author}
  {\bibfnamefont {S.}~\bibnamefont {{Aiola}}}, \bibinfo {author} {\bibfnamefont
  {R.}~\bibnamefont {{Allison}}}, \bibinfo {author} {\bibfnamefont
  {M.}~\bibnamefont {{Amiri}}}, \bibinfo {author} {\bibfnamefont
  {E.}~\bibnamefont {{Angile}}}, \bibinfo {author} {\bibfnamefont
  {N.}~\bibnamefont {{Battaglia}}}, \bibinfo {author} {\bibfnamefont {J.~A.}\
  \bibnamefont {{Beall}}}, \bibinfo {author} {\bibfnamefont {F.}~\bibnamefont
  {{de Bernardis}}}, \bibinfo {author} {\bibfnamefont {J.~R.}\ \bibnamefont
  {{Bond}}}, \bibinfo {author} {\bibfnamefont {J.}~\bibnamefont {{Britton}}},
  \bibinfo {author} {\bibfnamefont {E.}~\bibnamefont {{Calabrese}}}, \bibinfo
  {author} {\bibfnamefont {H.-m.}\ \bibnamefont {{Cho}}}, \bibinfo {author}
  {\bibfnamefont {S.~K.}\ \bibnamefont {{Choi}}}, \bibinfo {author}
  {\bibfnamefont {K.}~\bibnamefont {{Coughlin}}}, \bibinfo {author}
  {\bibfnamefont {D.}~\bibnamefont {{Crichton}}}, \bibinfo {author}
  {\bibfnamefont {K.}~\bibnamefont {{Crowley}}}, \bibinfo {author}
  {\bibfnamefont {R.}~\bibnamefont {{Datta}}}, \bibinfo {author} {\bibfnamefont
  {M.~J.}\ \bibnamefont {{Devlin}}}, \bibinfo {author} {\bibfnamefont {S.~R.}\
  \bibnamefont {{Dicker}}}, \bibinfo {author} {\bibfnamefont {J.}~\bibnamefont
  {{Dunkley}}}, \bibinfo {author} {\bibfnamefont {R.}~\bibnamefont
  {{D{\"u}nner}}}, \bibinfo {author} {\bibfnamefont {S.}~\bibnamefont
  {{Ferraro}}}, \bibinfo {author} {\bibfnamefont {A.~E.}\ \bibnamefont
  {{Fox}}}, \bibinfo {author} {\bibfnamefont {P.}~\bibnamefont {{Gallardo}}},
  \bibinfo {author} {\bibfnamefont {M.}~\bibnamefont {{Gralla}}}, \bibinfo
  {author} {\bibfnamefont {M.}~\bibnamefont {{Halpern}}}, \bibinfo {author}
  {\bibfnamefont {S.}~\bibnamefont {{Henderson}}}, \bibinfo {author}
  {\bibfnamefont {J.~C.}\ \bibnamefont {{Hill}}}, \bibinfo {author}
  {\bibfnamefont {G.~C.}\ \bibnamefont {{Hilton}}}, \bibinfo {author}
  {\bibfnamefont {M.}~\bibnamefont {{Hilton}}}, \bibinfo {author}
  {\bibfnamefont {A.~D.}\ \bibnamefont {{Hincks}}}, \bibinfo {author}
  {\bibfnamefont {R.}~\bibnamefont {{Hlozek}}}, \bibinfo {author}
  {\bibfnamefont {S.~P.~P.}\ \bibnamefont {{Ho}}}, \bibinfo {author}
  {\bibfnamefont {Z.}~\bibnamefont {{Huang}}}, \bibinfo {author} {\bibfnamefont
  {J.}~\bibnamefont {{Hubmayr}}}, \bibinfo {author} {\bibfnamefont {K.~M.}\
  \bibnamefont {{Huffenberger}}}, \bibinfo {author} {\bibfnamefont {J.~P.}\
  \bibnamefont {{Hughes}}}, \bibinfo {author} {\bibfnamefont {L.}~\bibnamefont
  {{Infante}}}, \bibinfo {author} {\bibfnamefont {K.}~\bibnamefont {{Irwin}}},
  \bibinfo {author} {\bibfnamefont {S.}~\bibnamefont {{Muya Kasanda}}},
  \bibinfo {author} {\bibfnamefont {J.}~\bibnamefont {{Klein}}}, \bibinfo
  {author} {\bibfnamefont {B.}~\bibnamefont {{Koopman}}}, \bibinfo {author}
  {\bibfnamefont {A.}~\bibnamefont {{Kosowsky}}}, \bibinfo {author}
  {\bibfnamefont {D.}~\bibnamefont {{Li}}}, \bibinfo {author} {\bibfnamefont
  {M.}~\bibnamefont {{Madhavacheril}}}, \bibinfo {author} {\bibfnamefont
  {T.~A.}\ \bibnamefont {{Marriage}}}, \bibinfo {author} {\bibfnamefont
  {J.}~\bibnamefont {{McMahon}}}, \bibinfo {author} {\bibfnamefont
  {F.}~\bibnamefont {{Menanteau}}}, \bibinfo {author} {\bibfnamefont
  {K.}~\bibnamefont {{Moodley}}}, \bibinfo {author} {\bibfnamefont
  {C.}~\bibnamefont {{Munson}}}, \bibinfo {author} {\bibfnamefont
  {S.}~\bibnamefont {{Naess}}}, \bibinfo {author} {\bibfnamefont
  {F.}~\bibnamefont {{Nati}}}, \bibinfo {author} {\bibfnamefont
  {L.}~\bibnamefont {{Newburgh}}}, \bibinfo {author} {\bibfnamefont
  {J.}~\bibnamefont {{Nibarger}}}, \bibinfo {author} {\bibfnamefont {M.~D.}\
  \bibnamefont {{Niemack}}}, \bibinfo {author} {\bibfnamefont {M.~R.}\
  \bibnamefont {{Nolta}}}, \bibinfo {author} {\bibfnamefont {C.}~\bibnamefont
  {{Nu{\~n}ez}}}, \bibinfo {author} {\bibfnamefont {L.~A.}\ \bibnamefont
  {{Page}}}, \bibinfo {author} {\bibfnamefont {C.}~\bibnamefont {{Pappas}}},
  \bibinfo {author} {\bibfnamefont {B.}~\bibnamefont {{Partridge}}}, \bibinfo
  {author} {\bibfnamefont {F.}~\bibnamefont {{Rojas}}}, \bibinfo {author}
  {\bibfnamefont {E.}~\bibnamefont {{Schaan}}}, \bibinfo {author}
  {\bibfnamefont {B.~L.}\ \bibnamefont {{Schmitt}}}, \bibinfo {author}
  {\bibfnamefont {N.}~\bibnamefont {{Sehgal}}}, \bibinfo {author}
  {\bibfnamefont {B.~D.}\ \bibnamefont {{Sherwin}}}, \bibinfo {author}
  {\bibfnamefont {J.}~\bibnamefont {{Sievers}}}, \bibinfo {author}
  {\bibfnamefont {S.}~\bibnamefont {{Simon}}}, \bibinfo {author} {\bibfnamefont
  {D.~N.}\ \bibnamefont {{Spergel}}}, \bibinfo {author} {\bibfnamefont {S.~T.}\
  \bibnamefont {{Staggs}}}, \bibinfo {author} {\bibfnamefont {E.~R.}\
  \bibnamefont {{Switzer}}}, \bibinfo {author} {\bibfnamefont {R.}~\bibnamefont
  {{Thornton}}}, \bibinfo {author} {\bibfnamefont {H.}~\bibnamefont {{Trac}}},
  \bibinfo {author} {\bibfnamefont {J.}~\bibnamefont {{Treu}}}, \bibinfo
  {author} {\bibfnamefont {C.}~\bibnamefont {{Tucker}}}, \bibinfo {author}
  {\bibfnamefont {A.}~\bibnamefont {{Van Engelen}}}, \bibinfo {author}
  {\bibfnamefont {J.~T.}\ \bibnamefont {{Ward}}}, \ and\ \bibinfo {author}
  {\bibfnamefont {E.~J.}\ \bibnamefont {{Wollack}}},\ }\href {\doibase
  10.1088/1475-7516/2017/06/031} {\bibfield  {journal} {\bibinfo  {journal}
  {\jcap}\ }\textbf {\bibinfo {volume} {6}},\ \bibinfo {eid} {031} (\bibinfo
  {year} {2017})},\ \Eprint {http://arxiv.org/abs/1610.02360}
  {arXiv:1610.02360} \BibitemShut {NoStop}%
\bibitem [{\citenamefont {{POLARBEAR Collaboration}}\ \emph
  {et~al.}(2017)\citenamefont {{POLARBEAR Collaboration}}, \citenamefont
  {{Ade}}, \citenamefont {{Aguilar}}, \citenamefont {{Akiba}}, \citenamefont
  {{Arnold}}, \citenamefont {{Baccigalupi}}, \citenamefont {{Barron}},
  \citenamefont {{Beck}}, \citenamefont {{Bianchini}}, \citenamefont
  {{Boettger}}, \citenamefont {{Borrill}}, \citenamefont {{Chapman}},
  \citenamefont {{Chinone}}, \citenamefont {{Crowley}}, \citenamefont
  {{Cukierman}}, \citenamefont {{D{\"u}nner}}, \citenamefont {{Dobbs}},
  \citenamefont {{Ducout}}, \citenamefont {{Elleflot}}, \citenamefont
  {{Errard}}, \citenamefont {{Fabbian}}, \citenamefont {{Feeney}},
  \citenamefont {{Feng}}, \citenamefont {{Fujino}}, \citenamefont {{Galitzki}},
  \citenamefont {{Gilbert}}, \citenamefont {{Goeckner-Wald}}, \citenamefont
  {{Groh}}, \citenamefont {{Hall}}, \citenamefont {{Halverson}}, \citenamefont
  {{Hamada}}, \citenamefont {{Hasegawa}}, \citenamefont {{Hazumi}},
  \citenamefont {{Hill}}, \citenamefont {{Howe}}, \citenamefont {{Inoue}},
  \citenamefont {{Jaehnig}}, \citenamefont {{Jaffe}}, \citenamefont {{Jeong}},
  \citenamefont {{Kaneko}}, \citenamefont {{Katayama}}, \citenamefont
  {{Keating}}, \citenamefont {{Keskitalo}}, \citenamefont {{Kisner}},
  \citenamefont {{Krachmalnicoff}}, \citenamefont {{Kusaka}}, \citenamefont
  {{Le Jeune}}, \citenamefont {{Lee}}, \citenamefont {{Leitch}}, \citenamefont
  {{Leon}}, \citenamefont {{Linder}}, \citenamefont {{Lowry}}, \citenamefont
  {{Matsuda}}, \citenamefont {{Matsumura}}, \citenamefont {{Minami}},
  \citenamefont {{Montgomery}}, \citenamefont {{Navaroli}}, \citenamefont
  {{Nishino}}, \citenamefont {{Paar}}, \citenamefont {{Peloton}}, \citenamefont
  {{Pham}}, \citenamefont {{Poletti}}, \citenamefont {{Puglisi}}, \citenamefont
  {{Reichardt}}, \citenamefont {{Richards}}, \citenamefont {{Ross}},
  \citenamefont {{Segawa}}, \citenamefont {{Sherwin}}, \citenamefont
  {{Silva-Feaver}}, \citenamefont {{Siritanasak}}, \citenamefont {{Stebor}},
  \citenamefont {{Stompor}}, \citenamefont {{Suzuki}}, \citenamefont
  {{Tajima}}, \citenamefont {{Takakura}}, \citenamefont {{Takatori}},
  \citenamefont {{Tanabe}}, \citenamefont {{Teply}}, \citenamefont {{Tomaru}},
  \citenamefont {{Tucker}}, \citenamefont {{Whitehorn}},\ and\ \citenamefont
  {{Zahn}}}]{polarbear17}%
  \BibitemOpen
  \bibfield  {author} {\bibinfo {author} {\bibnamefont {{POLARBEAR
  Collaboration}}}, \bibinfo {author} {\bibfnamefont {P.~A.~R.}\ \bibnamefont
  {{Ade}}}, \bibinfo {author} {\bibfnamefont {M.}~\bibnamefont {{Aguilar}}},
  \bibinfo {author} {\bibfnamefont {Y.}~\bibnamefont {{Akiba}}}, \bibinfo
  {author} {\bibfnamefont {K.}~\bibnamefont {{Arnold}}}, \bibinfo {author}
  {\bibfnamefont {C.}~\bibnamefont {{Baccigalupi}}}, \bibinfo {author}
  {\bibfnamefont {D.}~\bibnamefont {{Barron}}}, \bibinfo {author}
  {\bibfnamefont {D.}~\bibnamefont {{Beck}}}, \bibinfo {author} {\bibfnamefont
  {F.}~\bibnamefont {{Bianchini}}}, \bibinfo {author} {\bibfnamefont
  {D.}~\bibnamefont {{Boettger}}}, \bibinfo {author} {\bibfnamefont
  {J.}~\bibnamefont {{Borrill}}}, \bibinfo {author} {\bibfnamefont
  {S.}~\bibnamefont {{Chapman}}}, \bibinfo {author} {\bibfnamefont
  {Y.}~\bibnamefont {{Chinone}}}, \bibinfo {author} {\bibfnamefont
  {K.}~\bibnamefont {{Crowley}}}, \bibinfo {author} {\bibfnamefont
  {A.}~\bibnamefont {{Cukierman}}}, \bibinfo {author} {\bibfnamefont
  {R.}~\bibnamefont {{D{\"u}nner}}}, \bibinfo {author} {\bibfnamefont
  {M.}~\bibnamefont {{Dobbs}}}, \bibinfo {author} {\bibfnamefont
  {A.}~\bibnamefont {{Ducout}}}, \bibinfo {author} {\bibfnamefont
  {T.}~\bibnamefont {{Elleflot}}}, \bibinfo {author} {\bibfnamefont
  {J.}~\bibnamefont {{Errard}}}, \bibinfo {author} {\bibfnamefont
  {G.}~\bibnamefont {{Fabbian}}}, \bibinfo {author} {\bibfnamefont {S.~M.}\
  \bibnamefont {{Feeney}}}, \bibinfo {author} {\bibfnamefont {C.}~\bibnamefont
  {{Feng}}}, \bibinfo {author} {\bibfnamefont {T.}~\bibnamefont {{Fujino}}},
  \bibinfo {author} {\bibfnamefont {N.}~\bibnamefont {{Galitzki}}}, \bibinfo
  {author} {\bibfnamefont {A.}~\bibnamefont {{Gilbert}}}, \bibinfo {author}
  {\bibfnamefont {N.}~\bibnamefont {{Goeckner-Wald}}}, \bibinfo {author}
  {\bibfnamefont {J.~C.}\ \bibnamefont {{Groh}}}, \bibinfo {author}
  {\bibfnamefont {G.}~\bibnamefont {{Hall}}}, \bibinfo {author} {\bibfnamefont
  {N.}~\bibnamefont {{Halverson}}}, \bibinfo {author} {\bibfnamefont
  {T.}~\bibnamefont {{Hamada}}}, \bibinfo {author} {\bibfnamefont
  {M.}~\bibnamefont {{Hasegawa}}}, \bibinfo {author} {\bibfnamefont
  {M.}~\bibnamefont {{Hazumi}}}, \bibinfo {author} {\bibfnamefont {C.~A.}\
  \bibnamefont {{Hill}}}, \bibinfo {author} {\bibfnamefont {L.}~\bibnamefont
  {{Howe}}}, \bibinfo {author} {\bibfnamefont {Y.}~\bibnamefont {{Inoue}}},
  \bibinfo {author} {\bibfnamefont {G.}~\bibnamefont {{Jaehnig}}}, \bibinfo
  {author} {\bibfnamefont {A.~H.}\ \bibnamefont {{Jaffe}}}, \bibinfo {author}
  {\bibfnamefont {O.}~\bibnamefont {{Jeong}}}, \bibinfo {author} {\bibfnamefont
  {D.}~\bibnamefont {{Kaneko}}}, \bibinfo {author} {\bibfnamefont
  {N.}~\bibnamefont {{Katayama}}}, \bibinfo {author} {\bibfnamefont
  {B.}~\bibnamefont {{Keating}}}, \bibinfo {author} {\bibfnamefont
  {R.}~\bibnamefont {{Keskitalo}}}, \bibinfo {author} {\bibfnamefont
  {T.}~\bibnamefont {{Kisner}}}, \bibinfo {author} {\bibfnamefont
  {N.}~\bibnamefont {{Krachmalnicoff}}}, \bibinfo {author} {\bibfnamefont
  {A.}~\bibnamefont {{Kusaka}}}, \bibinfo {author} {\bibfnamefont
  {M.}~\bibnamefont {{Le Jeune}}}, \bibinfo {author} {\bibfnamefont {A.~T.}\
  \bibnamefont {{Lee}}}, \bibinfo {author} {\bibfnamefont {E.~M.}\ \bibnamefont
  {{Leitch}}}, \bibinfo {author} {\bibfnamefont {D.}~\bibnamefont {{Leon}}},
  \bibinfo {author} {\bibfnamefont {E.}~\bibnamefont {{Linder}}}, \bibinfo
  {author} {\bibfnamefont {L.}~\bibnamefont {{Lowry}}}, \bibinfo {author}
  {\bibfnamefont {F.}~\bibnamefont {{Matsuda}}}, \bibinfo {author}
  {\bibfnamefont {T.}~\bibnamefont {{Matsumura}}}, \bibinfo {author}
  {\bibfnamefont {Y.}~\bibnamefont {{Minami}}}, \bibinfo {author}
  {\bibfnamefont {J.}~\bibnamefont {{Montgomery}}}, \bibinfo {author}
  {\bibfnamefont {M.}~\bibnamefont {{Navaroli}}}, \bibinfo {author}
  {\bibfnamefont {H.}~\bibnamefont {{Nishino}}}, \bibinfo {author}
  {\bibfnamefont {H.}~\bibnamefont {{Paar}}}, \bibinfo {author} {\bibfnamefont
  {J.}~\bibnamefont {{Peloton}}}, \bibinfo {author} {\bibfnamefont {A.~T.~P.}\
  \bibnamefont {{Pham}}}, \bibinfo {author} {\bibfnamefont {D.}~\bibnamefont
  {{Poletti}}}, \bibinfo {author} {\bibfnamefont {G.}~\bibnamefont
  {{Puglisi}}}, \bibinfo {author} {\bibfnamefont {C.~L.}\ \bibnamefont
  {{Reichardt}}}, \bibinfo {author} {\bibfnamefont {P.~L.}\ \bibnamefont
  {{Richards}}}, \bibinfo {author} {\bibfnamefont {C.}~\bibnamefont {{Ross}}},
  \bibinfo {author} {\bibfnamefont {Y.}~\bibnamefont {{Segawa}}}, \bibinfo
  {author} {\bibfnamefont {B.~D.}\ \bibnamefont {{Sherwin}}}, \bibinfo {author}
  {\bibfnamefont {M.}~\bibnamefont {{Silva-Feaver}}}, \bibinfo {author}
  {\bibfnamefont {P.}~\bibnamefont {{Siritanasak}}}, \bibinfo {author}
  {\bibfnamefont {N.}~\bibnamefont {{Stebor}}}, \bibinfo {author}
  {\bibfnamefont {R.}~\bibnamefont {{Stompor}}}, \bibinfo {author}
  {\bibfnamefont {A.}~\bibnamefont {{Suzuki}}}, \bibinfo {author}
  {\bibfnamefont {O.}~\bibnamefont {{Tajima}}}, \bibinfo {author}
  {\bibfnamefont {S.}~\bibnamefont {{Takakura}}}, \bibinfo {author}
  {\bibfnamefont {S.}~\bibnamefont {{Takatori}}}, \bibinfo {author}
  {\bibfnamefont {D.}~\bibnamefont {{Tanabe}}}, \bibinfo {author}
  {\bibfnamefont {G.~P.}\ \bibnamefont {{Teply}}}, \bibinfo {author}
  {\bibfnamefont {T.}~\bibnamefont {{Tomaru}}}, \bibinfo {author}
  {\bibfnamefont {C.}~\bibnamefont {{Tucker}}}, \bibinfo {author}
  {\bibfnamefont {N.}~\bibnamefont {{Whitehorn}}}, \ and\ \bibinfo {author}
  {\bibfnamefont {A.}~\bibnamefont {{Zahn}}},\ }\href {\doibase
  10.3847/1538-4357/aa8e9f} {\bibfield  {journal} {\bibinfo  {journal} {\apj}\
  }\textbf {\bibinfo {volume} {848}},\ \bibinfo {eid} {121} (\bibinfo {year}
  {2017})},\ \Eprint {http://arxiv.org/abs/1705.02907} {arXiv:1705.02907}
  \BibitemShut {NoStop}%
\bibitem [{\citenamefont {{Kusaka}}\ \emph {et~al.}(2018)\citenamefont
  {{Kusaka}}, \citenamefont {{Appel}}, \citenamefont {{Essinger-Hileman}},
  \citenamefont {{Beall}}, \citenamefont {{Campusano}}, \citenamefont {{Cho}},
  \citenamefont {{Choi}}, \citenamefont {{Crowley}}, \citenamefont {{Fowler}},
  \citenamefont {{Gallardo}}, \citenamefont {{Hasselfield}}, \citenamefont
  {{Hilton}}, \citenamefont {{Ho}}, \citenamefont {{Irwin}}, \citenamefont
  {{Jarosik}}, \citenamefont {{Niemack}}, \citenamefont {{Nixon}},
  \citenamefont {{\~{}Nolta}}, \citenamefont {{Page}}, \citenamefont {{Palma}},
  \citenamefont {{Parker}}, \citenamefont {{Raghunathan}}, \citenamefont
  {{Reintsema}}, \citenamefont {{Sievers}}, \citenamefont {{Simon}},
  \citenamefont {{Staggs}}, \citenamefont {{Visnjic}},\ and\ \citenamefont
  {{Yoon}}}]{abs18}%
  \BibitemOpen
  \bibfield  {author} {\bibinfo {author} {\bibfnamefont {A.}~\bibnamefont
  {{Kusaka}}}, \bibinfo {author} {\bibfnamefont {J.}~\bibnamefont {{Appel}}},
  \bibinfo {author} {\bibfnamefont {T.}~\bibnamefont {{Essinger-Hileman}}},
  \bibinfo {author} {\bibfnamefont {J.~A.}\ \bibnamefont {{Beall}}}, \bibinfo
  {author} {\bibfnamefont {L.~E.}\ \bibnamefont {{Campusano}}}, \bibinfo
  {author} {\bibfnamefont {H.-M.}\ \bibnamefont {{Cho}}}, \bibinfo {author}
  {\bibfnamefont {S.~K.}\ \bibnamefont {{Choi}}}, \bibinfo {author}
  {\bibfnamefont {K.}~\bibnamefont {{Crowley}}}, \bibinfo {author}
  {\bibfnamefont {J.~W.}\ \bibnamefont {{Fowler}}}, \bibinfo {author}
  {\bibfnamefont {P.}~\bibnamefont {{Gallardo}}}, \bibinfo {author}
  {\bibfnamefont {M.}~\bibnamefont {{Hasselfield}}}, \bibinfo {author}
  {\bibfnamefont {G.}~\bibnamefont {{Hilton}}}, \bibinfo {author}
  {\bibfnamefont {S.-P.~P.}\ \bibnamefont {{Ho}}}, \bibinfo {author}
  {\bibfnamefont {K.}~\bibnamefont {{Irwin}}}, \bibinfo {author} {\bibfnamefont
  {N.}~\bibnamefont {{Jarosik}}}, \bibinfo {author} {\bibfnamefont {M.~D.}\
  \bibnamefont {{Niemack}}}, \bibinfo {author} {\bibfnamefont {G.~W.}\
  \bibnamefont {{Nixon}}}, \bibinfo {author} {\bibfnamefont {M.}~\bibnamefont
  {{\~{}Nolta}}}, \bibinfo {author} {\bibfnamefont {L.~A.}\ \bibnamefont
  {{Page}}, \bibfnamefont {Jr.}}, \bibinfo {author} {\bibfnamefont {G.~A.}\
  \bibnamefont {{Palma}}}, \bibinfo {author} {\bibfnamefont {L.}~\bibnamefont
  {{Parker}}}, \bibinfo {author} {\bibfnamefont {S.}~\bibnamefont
  {{Raghunathan}}}, \bibinfo {author} {\bibfnamefont {C.~D.}\ \bibnamefont
  {{Reintsema}}}, \bibinfo {author} {\bibfnamefont {J.}~\bibnamefont
  {{Sievers}}}, \bibinfo {author} {\bibfnamefont {S.~M.}\ \bibnamefont
  {{Simon}}}, \bibinfo {author} {\bibfnamefont {S.~T.}\ \bibnamefont
  {{Staggs}}}, \bibinfo {author} {\bibfnamefont {K.}~\bibnamefont {{Visnjic}}},
  \ and\ \bibinfo {author} {\bibfnamefont {K.-W.}\ \bibnamefont {{Yoon}}},\
  }\href {\doibase 10.1088/1475-7516/2018/09/005} {\bibfield  {journal}
  {\bibinfo  {journal} {\jcap}\ }\textbf {\bibinfo {volume} {9}},\ \bibinfo
  {eid} {005} (\bibinfo {year} {2018})},\ \Eprint
  {http://arxiv.org/abs/1801.01218} {arXiv:1801.01218} \BibitemShut {NoStop}%
\bibitem [{\citenamefont {{Seljak}}(1997)}]{seljak97b}%
  \BibitemOpen
  \bibfield  {author} {\bibinfo {author} {\bibfnamefont {U.}~\bibnamefont
  {{Seljak}}},\ }\href {\doibase 10.1086/304123} {\bibfield  {journal}
  {\bibinfo  {journal} {\apj}\ }\textbf {\bibinfo {volume} {482}},\ \bibinfo
  {pages} {6} (\bibinfo {year} {1997})},\ \Eprint
  {http://arxiv.org/abs/astro-ph/9608131} {astro-ph/9608131} \BibitemShut
  {NoStop}%
\bibitem [{\citenamefont {{Kamionkowski}}\ \emph {et~al.}(1997)\citenamefont
  {{Kamionkowski}}, \citenamefont {{Kosowsky}},\ and\ \citenamefont
  {{Stebbins}}}]{kamionkowski97}%
  \BibitemOpen
  \bibfield  {author} {\bibinfo {author} {\bibfnamefont {M.}~\bibnamefont
  {{Kamionkowski}}}, \bibinfo {author} {\bibfnamefont {A.}~\bibnamefont
  {{Kosowsky}}}, \ and\ \bibinfo {author} {\bibfnamefont {A.}~\bibnamefont
  {{Stebbins}}},\ }\href {\doibase 10.1103/PhysRevLett.78.2058} {\bibfield
  {journal} {\bibinfo  {journal} {\prl}\ }\textbf {\bibinfo {volume} {78}},\
  \bibinfo {pages} {2058} (\bibinfo {year} {1997})},\ \Eprint
  {http://arxiv.org/abs/astro-ph/9609132} {astro-ph/9609132} \BibitemShut
  {NoStop}%
\bibitem [{\citenamefont {{Seljak}}\ and\ \citenamefont
  {{Zaldarriaga}}(1997)}]{seljak97a}%
  \BibitemOpen
  \bibfield  {author} {\bibinfo {author} {\bibfnamefont {U.}~\bibnamefont
  {{Seljak}}}\ and\ \bibinfo {author} {\bibfnamefont {M.}~\bibnamefont
  {{Zaldarriaga}}},\ }\href {\doibase 10.1103/PhysRevLett.78.2054} {\bibfield
  {journal} {\bibinfo  {journal} {\prl}\ }\textbf {\bibinfo {volume} {78}},\
  \bibinfo {pages} {2054} (\bibinfo {year} {1997})},\ \Eprint
  {http://arxiv.org/abs/astro-ph/9609169} {astro-ph/9609169} \BibitemShut
  {NoStop}%
\bibitem [{\citenamefont {{Zaldarriaga}}\ and\ \citenamefont
  {{Seljak}}(1998)}]{zaldarriaga98}%
  \BibitemOpen
  \bibfield  {author} {\bibinfo {author} {\bibfnamefont {M.}~\bibnamefont
  {{Zaldarriaga}}}\ and\ \bibinfo {author} {\bibfnamefont {U.}~\bibnamefont
  {{Seljak}}},\ }\href {\doibase 10.1103/PhysRevD.58.023003} {\bibfield
  {journal} {\bibinfo  {journal} {\prd}\ }\textbf {\bibinfo {volume} {58}},\
  \bibinfo {eid} {023003} (\bibinfo {year} {1998})},\ \Eprint
  {http://arxiv.org/abs/astro-ph/9803150} {astro-ph/9803150} \BibitemShut
  {NoStop}%
\bibitem [{\citenamefont {{\biceptwo\ Collaboration I}}(2014)}]{biceptwoI}%
  \BibitemOpen
  \bibfield  {author} {\bibinfo {author} {\bibnamefont {{\biceptwo\
  Collaboration I}}},\ }\href {\doibase 10.1103/PhysRevLett.112.241101}
  {\bibfield  {journal} {\bibinfo  {journal} {Physical Review Letters}\
  }\textbf {\bibinfo {volume} {112}},\ \bibinfo {eid} {241101} (\bibinfo {year}
  {2014})},\ \Eprint {http://arxiv.org/abs/1403.3985} {arXiv:1403.3985}
  \BibitemShut {NoStop}%
\bibitem [{\citenamefont {{\biceptwo/\keck\ and \planck\
  Collaborations}}(2015)}]{bkp}%
  \BibitemOpen
  \bibfield  {author} {\bibinfo {author} {\bibnamefont {{\biceptwo/\keck\ and
  \planck\ Collaborations}}},\ }\href {\doibase 10.1103/PhysRevLett.114.101301}
  {\bibfield  {journal} {\bibinfo  {journal} {Physical Review Letters}\
  }\textbf {\bibinfo {volume} {114}},\ \bibinfo {eid} {101301} (\bibinfo {year}
  {2015})},\ \Eprint {http://arxiv.org/abs/1502.00612} {arXiv:1502.00612}
  \BibitemShut {NoStop}%
\bibitem [{\citenamefont {{\keckarray\ and \biceptwo\ Collaborations
  VI}}(2016)}]{biceptwoVI}%
  \BibitemOpen
  \bibfield  {author} {\bibinfo {author} {\bibnamefont {{\keckarray\ and
  \biceptwo\ Collaborations VI}}},\ }\href {\doibase
  10.1103/PhysRevLett.116.031302} {\bibfield  {journal} {\bibinfo  {journal}
  {Physical Review Letters}\ }\textbf {\bibinfo {volume} {116}},\ \bibinfo
  {eid} {031302} (\bibinfo {year} {2016})},\ \Eprint
  {http://arxiv.org/abs/1510.09217} {arXiv:1510.09217} \BibitemShut {NoStop}%
\bibitem [{\citenamefont {{\biceptwo\ Collaboration II}}(2014)}]{biceptwoII}%
  \BibitemOpen
  \bibfield  {author} {\bibinfo {author} {\bibnamefont {{\biceptwo\
  Collaboration II}}},\ }\href {\doibase 10.1088/0004-637X/792/1/62} {\bibfield
   {journal} {\bibinfo  {journal} {\apj}\ }\textbf {\bibinfo {volume} {792}},\
  \bibinfo {eid} {62} (\bibinfo {year} {2014})},\ \Eprint
  {http://arxiv.org/abs/1403.4302} {arXiv:1403.4302} \BibitemShut {NoStop}%
\bibitem [{\citenamefont {{\keckarray\ and \biceptwo\ Collaborations
  V}}(2015)}]{biceptwoV}%
  \BibitemOpen
  \bibfield  {author} {\bibinfo {author} {\bibnamefont {{\keckarray\ and
  \biceptwo\ Collaborations V}}},\ }\href {\doibase
  10.1088/0004-637X/811/2/126} {\bibfield  {journal} {\bibinfo  {journal}
  {\apj}\ }\textbf {\bibinfo {volume} {811}},\ \bibinfo {eid} {126} (\bibinfo
  {year} {2015})},\ \Eprint {http://arxiv.org/abs/1502.00643}
  {arXiv:1502.00643} \BibitemShut {NoStop}%
\bibitem [{\citenamefont {{\biceptwo/\keck\ and Spider
  Collaborations}}(2015)}]{bkdets}%
  \BibitemOpen
  \bibfield  {author} {\bibinfo {author} {\bibnamefont {{\biceptwo/\keck\ and
  Spider Collaborations}}},\ }\href {\doibase 10.1088/0004-637X/812/2/176}
  {\bibfield  {journal} {\bibinfo  {journal} {\apj}\ }\textbf {\bibinfo
  {volume} {812}},\ \bibinfo {eid} {176} (\bibinfo {year} {2015})},\ \Eprint
  {http://arxiv.org/abs/1502.00619} {arXiv:1502.00619} \BibitemShut {NoStop}%
\bibitem [{\citenamefont {{\keckarray\ and \biceptwo\ Collaobrations
  VII}}(2016)}]{biceptwoVII}%
  \BibitemOpen
  \bibfield  {author} {\bibinfo {author} {\bibnamefont {{\keckarray\ and
  \biceptwo\ Collaobrations VII}}},\ }\href {\doibase
  10.3847/0004-637X/825/1/66} {\bibfield  {journal} {\bibinfo  {journal}
  {\apj}\ }\textbf {\bibinfo {volume} {825}},\ \bibinfo {eid} {66} (\bibinfo
  {year} {2016})},\ \Eprint {http://arxiv.org/abs/1603.05976}
  {arXiv:1603.05976} \BibitemShut {NoStop}%
\bibitem [{Note1()}]{Note1}%
  \BibitemOpen
  \bibinfo {note} {See \protect \url
  {http://lambda.gsfc.nasa.gov/product/map/dr5/m_products.cfm}}\BibitemShut
  {NoStop}%
\bibitem [{\citenamefont {{Bennett}}\ \emph {et~al.}(2013)\citenamefont
  {{Bennett}}, \citenamefont {{Larson}}, \citenamefont {{Weiland}},
  \citenamefont {{Jarosik}}, \citenamefont {{Hinshaw}}, \citenamefont
  {{Odegard}}, \citenamefont {{Smith}}, \citenamefont {{Hill}}, \citenamefont
  {{Gold}}, \citenamefont {{Halpern}}, \citenamefont {{Komatsu}}, \citenamefont
  {{Nolta}}, \citenamefont {{Page}}, \citenamefont {{Spergel}}, \citenamefont
  {{Wollack}}, \citenamefont {{Dunkley}}, \citenamefont {{Kogut}},
  \citenamefont {{Limon}}, \citenamefont {{Meyer}}, \citenamefont {{Tucker}},\
  and\ \citenamefont {{Wright}}}]{bennett13}%
  \BibitemOpen
  \bibfield  {author} {\bibinfo {author} {\bibfnamefont {C.~L.}\ \bibnamefont
  {{Bennett}}}, \bibinfo {author} {\bibfnamefont {D.}~\bibnamefont {{Larson}}},
  \bibinfo {author} {\bibfnamefont {J.~L.}\ \bibnamefont {{Weiland}}}, \bibinfo
  {author} {\bibfnamefont {N.}~\bibnamefont {{Jarosik}}}, \bibinfo {author}
  {\bibfnamefont {G.}~\bibnamefont {{Hinshaw}}}, \bibinfo {author}
  {\bibfnamefont {N.}~\bibnamefont {{Odegard}}}, \bibinfo {author}
  {\bibfnamefont {K.~M.}\ \bibnamefont {{Smith}}}, \bibinfo {author}
  {\bibfnamefont {R.~S.}\ \bibnamefont {{Hill}}}, \bibinfo {author}
  {\bibfnamefont {B.}~\bibnamefont {{Gold}}}, \bibinfo {author} {\bibfnamefont
  {M.}~\bibnamefont {{Halpern}}}, \bibinfo {author} {\bibfnamefont
  {E.}~\bibnamefont {{Komatsu}}}, \bibinfo {author} {\bibfnamefont {M.~R.}\
  \bibnamefont {{Nolta}}}, \bibinfo {author} {\bibfnamefont {L.}~\bibnamefont
  {{Page}}}, \bibinfo {author} {\bibfnamefont {D.~N.}\ \bibnamefont
  {{Spergel}}}, \bibinfo {author} {\bibfnamefont {E.}~\bibnamefont
  {{Wollack}}}, \bibinfo {author} {\bibfnamefont {J.}~\bibnamefont
  {{Dunkley}}}, \bibinfo {author} {\bibfnamefont {A.}~\bibnamefont {{Kogut}}},
  \bibinfo {author} {\bibfnamefont {M.}~\bibnamefont {{Limon}}}, \bibinfo
  {author} {\bibfnamefont {S.~S.}\ \bibnamefont {{Meyer}}}, \bibinfo {author}
  {\bibfnamefont {G.~S.}\ \bibnamefont {{Tucker}}}, \ and\ \bibinfo {author}
  {\bibfnamefont {E.~L.}\ \bibnamefont {{Wright}}},\ }\href {\doibase
  10.1088/0067-0049/208/2/20} {\bibfield  {journal} {\bibinfo  {journal}
  {\apjs}\ }\textbf {\bibinfo {volume} {208}},\ \bibinfo {eid} {20} (\bibinfo
  {year} {2013})},\ \Eprint {http://arxiv.org/abs/1212.5225} {arXiv:1212.5225}
  \BibitemShut {NoStop}%
\bibitem [{Note2()}]{Note2}%
  \BibitemOpen
  \bibinfo {note} {Public Release 2 ``full mission'' maps as available at
  \protect \url {http://www.cosmos.esa.int/web/planck/pla}. We will update to
  PR3 in our next analysis.}\BibitemShut {Stop}%
\bibitem [{\citenamefont {{Planck Collaboration 2015 I}}(2016)}]{planck2015I}%
  \BibitemOpen
  \bibfield  {author} {\bibinfo {author} {\bibnamefont {{Planck Collaboration
  2015 I}}},\ }\href {\doibase 10.1051/0004-6361/201527101} {\bibfield
  {journal} {\bibinfo  {journal} {\aap}\ }\textbf {\bibinfo {volume} {594}},\
  \bibinfo {eid} {A1} (\bibinfo {year} {2016})},\ \Eprint
  {http://arxiv.org/abs/1502.01582} {arXiv:1502.01582} \BibitemShut {NoStop}%
\bibitem [{\citenamefont {{Planck Collaboration Int.\
  XXX}}(2016)}]{planckiXXX}%
  \BibitemOpen
  \bibfield  {author} {\bibinfo {author} {\bibnamefont {{Planck Collaboration
  Int.\ XXX}}},\ }\href {\doibase 10.1051/0004-6361/201425034} {\bibfield
  {journal} {\bibinfo  {journal} {\aap}\ }\textbf {\bibinfo {volume} {586}},\
  \bibinfo {eid} {A133} (\bibinfo {year} {2016})},\ \Eprint
  {http://arxiv.org/abs/1409.5738} {arXiv:1409.5738} \BibitemShut {NoStop}%
\bibitem [{\citenamefont {{Planck Collaboration 2018 XI}}(2018)}]{planckiLIV}%
  \BibitemOpen
  \bibfield  {author} {\bibinfo {author} {\bibnamefont {{Planck Collaboration
  2018 XI}}},\ }\href@noop {} {\bibfield  {journal} {\bibinfo  {journal} {ArXiv
  e-prints}\ } (\bibinfo {year} {2018})},\ \Eprint
  {http://arxiv.org/abs/1801.04945} {arXiv:1801.04945} \BibitemShut {NoStop}%
\bibitem [{\citenamefont {{van Engelen}}\ \emph {et~al.}(2012)\citenamefont
  {{van Engelen}}, \citenamefont {{Keisler}}, \citenamefont {{Zahn}},
  \citenamefont {{Aird}}, \citenamefont {{Benson}}, \citenamefont {{Bleem}},
  \citenamefont {{Carlstrom}}, \citenamefont {{Chang}}, \citenamefont {{Cho}},
  \citenamefont {{Crawford}}, \citenamefont {{Crites}}, \citenamefont {{de
  Haan}}, \citenamefont {{Dobbs}}, \citenamefont {{Dudley}}, \citenamefont
  {{George}}, \citenamefont {{Halverson}}, \citenamefont {{Holder}},
  \citenamefont {{Holzapfel}}, \citenamefont {{Hoover}}, \citenamefont {{Hou}}
  \emph {et~al.}}]{vanengelen12}%
  \BibitemOpen
  \bibfield  {author} {\bibinfo {author} {\bibfnamefont {A.}~\bibnamefont {{van
  Engelen}}}, \bibinfo {author} {\bibfnamefont {R.}~\bibnamefont {{Keisler}}},
  \bibinfo {author} {\bibfnamefont {O.}~\bibnamefont {{Zahn}}}, \bibinfo
  {author} {\bibfnamefont {K.~A.}\ \bibnamefont {{Aird}}}, \bibinfo {author}
  {\bibfnamefont {B.~A.}\ \bibnamefont {{Benson}}}, \bibinfo {author}
  {\bibfnamefont {L.~E.}\ \bibnamefont {{Bleem}}}, \bibinfo {author}
  {\bibfnamefont {J.~E.}\ \bibnamefont {{Carlstrom}}}, \bibinfo {author}
  {\bibfnamefont {C.~L.}\ \bibnamefont {{Chang}}}, \bibinfo {author}
  {\bibfnamefont {H.~M.}\ \bibnamefont {{Cho}}}, \bibinfo {author}
  {\bibfnamefont {T.~M.}\ \bibnamefont {{Crawford}}}, \bibinfo {author}
  {\bibfnamefont {A.~T.}\ \bibnamefont {{Crites}}}, \bibinfo {author}
  {\bibfnamefont {T.}~\bibnamefont {{de Haan}}}, \bibinfo {author}
  {\bibfnamefont {M.~A.}\ \bibnamefont {{Dobbs}}}, \bibinfo {author}
  {\bibfnamefont {J.}~\bibnamefont {{Dudley}}}, \bibinfo {author}
  {\bibfnamefont {E.~M.}\ \bibnamefont {{George}}}, \bibinfo {author}
  {\bibfnamefont {N.~W.}\ \bibnamefont {{Halverson}}}, \bibinfo {author}
  {\bibfnamefont {G.~P.}\ \bibnamefont {{Holder}}}, \bibinfo {author}
  {\bibfnamefont {W.~L.}\ \bibnamefont {{Holzapfel}}}, \bibinfo {author}
  {\bibfnamefont {S.}~\bibnamefont {{Hoover}}}, \bibinfo {author}
  {\bibfnamefont {Z.}~\bibnamefont {{Hou}}},  \emph {et~al.},\ }\href {\doibase
  10.1088/0004-637X/756/2/142} {\bibfield  {journal} {\bibinfo  {journal}
  {\apj}\ }\textbf {\bibinfo {volume} {756}},\ \bibinfo {eid} {142} (\bibinfo
  {year} {2012})},\ \Eprint {http://arxiv.org/abs/1202.0546} {arXiv:1202.0546}
  \BibitemShut {NoStop}%
\bibitem [{\citenamefont {{Hamimeche}}\ and\ \citenamefont
  {{Lewis}}(2008)}]{hamimeche08}%
  \BibitemOpen
  \bibfield  {author} {\bibinfo {author} {\bibfnamefont {S.}~\bibnamefont
  {{Hamimeche}}}\ and\ \bibinfo {author} {\bibfnamefont {A.}~\bibnamefont
  {{Lewis}}},\ }\href {\doibase 10.1103/PhysRevD.77.103013} {\bibfield
  {journal} {\bibinfo  {journal} {\prd}\ }\textbf {\bibinfo {volume} {77}},\
  \bibinfo {eid} {103013} (\bibinfo {year} {2008})},\ \Eprint
  {http://arxiv.org/abs/0801.0554} {arXiv:0801.0554} \BibitemShut {NoStop}%
\bibitem [{\citenamefont {{Lewis}}\ and\ \citenamefont
  {{Bridle}}(2002)}]{cosmomc}%
  \BibitemOpen
  \bibfield  {author} {\bibinfo {author} {\bibfnamefont {A.}~\bibnamefont
  {{Lewis}}}\ and\ \bibinfo {author} {\bibfnamefont {S.}~\bibnamefont
  {{Bridle}}},\ }\href {\doibase 10.1103/PhysRevD.66.103511} {\bibfield
  {journal} {\bibinfo  {journal} {\prd}\ }\textbf {\bibinfo {volume} {66}},\
  \bibinfo {eid} {103511} (\bibinfo {year} {2002})},\ \Eprint
  {http://arxiv.org/abs/astro-ph/0205436} {astro-ph/0205436} \BibitemShut
  {NoStop}%
\bibitem [{\citenamefont {{Planck Collaboration Int.\
  XXII}}(2015)}]{planckiXXII}%
  \BibitemOpen
  \bibfield  {author} {\bibinfo {author} {\bibnamefont {{Planck Collaboration
  Int.\ XXII}}},\ }\href {\doibase 10.1051/0004-6361/201424088} {\bibfield
  {journal} {\bibinfo  {journal} {\aap}\ }\textbf {\bibinfo {volume} {576}},\
  \bibinfo {eid} {A107} (\bibinfo {year} {2015})},\ \Eprint
  {http://arxiv.org/abs/1405.0874} {arXiv:1405.0874} \BibitemShut {NoStop}%
\bibitem [{\citenamefont {{Fuskeland}}\ \emph {et~al.}(2014)\citenamefont
  {{Fuskeland}}, \citenamefont {{Wehus}}, \citenamefont {{Eriksen}},\ and\
  \citenamefont {{N{\ae}ss}}}]{fuskeland14}%
  \BibitemOpen
  \bibfield  {author} {\bibinfo {author} {\bibfnamefont {U.}~\bibnamefont
  {{Fuskeland}}}, \bibinfo {author} {\bibfnamefont {I.~K.}\ \bibnamefont
  {{Wehus}}}, \bibinfo {author} {\bibfnamefont {H.~K.}\ \bibnamefont
  {{Eriksen}}}, \ and\ \bibinfo {author} {\bibfnamefont {S.~K.}\ \bibnamefont
  {{N{\ae}ss}}},\ }\href {\doibase 10.1088/0004-637X/790/2/104} {\bibfield
  {journal} {\bibinfo  {journal} {\apj}\ }\textbf {\bibinfo {volume} {790}},\
  \bibinfo {eid} {104} (\bibinfo {year} {2014})},\ \Eprint
  {http://arxiv.org/abs/1404.5323} {arXiv:1404.5323} \BibitemShut {NoStop}%
\bibitem [{\citenamefont {{Krachmalnicoff}}\ \emph {et~al.}(2018)\citenamefont
  {{Krachmalnicoff}}, \citenamefont {{Carretti}}, \citenamefont
  {{Baccigalupi}}, \citenamefont {{Bernardi}}, \citenamefont {{Brown}},
  \citenamefont {{Gaensler}}, \citenamefont {{Haverkorn}}, \citenamefont
  {{Kesteven}}, \citenamefont {{Perrotta}}, \citenamefont {{Poppi}},\ and\
  \citenamefont {{Staveley-Smith}}}]{krachmalnicoff18}%
  \BibitemOpen
  \bibfield  {author} {\bibinfo {author} {\bibfnamefont {N.}~\bibnamefont
  {{Krachmalnicoff}}}, \bibinfo {author} {\bibfnamefont {E.}~\bibnamefont
  {{Carretti}}}, \bibinfo {author} {\bibfnamefont {C.}~\bibnamefont
  {{Baccigalupi}}}, \bibinfo {author} {\bibfnamefont {G.}~\bibnamefont
  {{Bernardi}}}, \bibinfo {author} {\bibfnamefont {S.}~\bibnamefont {{Brown}}},
  \bibinfo {author} {\bibfnamefont {B.~M.}\ \bibnamefont {{Gaensler}}},
  \bibinfo {author} {\bibfnamefont {M.}~\bibnamefont {{Haverkorn}}}, \bibinfo
  {author} {\bibfnamefont {M.}~\bibnamefont {{Kesteven}}}, \bibinfo {author}
  {\bibfnamefont {F.}~\bibnamefont {{Perrotta}}}, \bibinfo {author}
  {\bibfnamefont {S.}~\bibnamefont {{Poppi}}}, \ and\ \bibinfo {author}
  {\bibfnamefont {L.}~\bibnamefont {{Staveley-Smith}}},\ }\href@noop {}
  {\bibfield  {journal} {\bibinfo  {journal} {ArXiv e-prints}\ } (\bibinfo
  {year} {2018})},\ \Eprint {http://arxiv.org/abs/1802.01145}
  {arXiv:1802.01145} \BibitemShut {NoStop}%
\bibitem [{\citenamefont {{Dunkley}}\ \emph {et~al.}(2009)\citenamefont
  {{Dunkley}}, \citenamefont {{Amblard}}, \citenamefont {{Baccigalupi}},
  \citenamefont {{Betoule}}, \citenamefont {{Chuss}}, \citenamefont {{Cooray}},
  \citenamefont {{Delabrouille}}, \citenamefont {{Dickinson}}, \citenamefont
  {{Dobler}}, \citenamefont {{Dotson}}, \citenamefont {{Eriksen}},
  \citenamefont {{Finkbeiner}}, \citenamefont {{Fixsen}}, \citenamefont
  {{Fosalba}}, \citenamefont {{Fraisse}}, \citenamefont {{Hirata}},
  \citenamefont {{Kogut}}, \citenamefont {{Kristiansen}}, \citenamefont
  {{Lawrence}}, \citenamefont {{Magalh\~{a}es}}, \citenamefont
  {{Miville-Deschenes}} \emph {et~al.}}]{dunkley08}%
  \BibitemOpen
  \bibfield  {author} {\bibinfo {author} {\bibfnamefont {J.}~\bibnamefont
  {{Dunkley}}}, \bibinfo {author} {\bibfnamefont {A.}~\bibnamefont
  {{Amblard}}}, \bibinfo {author} {\bibfnamefont {C.}~\bibnamefont
  {{Baccigalupi}}}, \bibinfo {author} {\bibfnamefont {M.}~\bibnamefont
  {{Betoule}}}, \bibinfo {author} {\bibfnamefont {D.}~\bibnamefont {{Chuss}}},
  \bibinfo {author} {\bibfnamefont {A.}~\bibnamefont {{Cooray}}}, \bibinfo
  {author} {\bibfnamefont {J.}~\bibnamefont {{Delabrouille}}}, \bibinfo
  {author} {\bibfnamefont {C.}~\bibnamefont {{Dickinson}}}, \bibinfo {author}
  {\bibfnamefont {G.}~\bibnamefont {{Dobler}}}, \bibinfo {author}
  {\bibfnamefont {J.}~\bibnamefont {{Dotson}}}, \bibinfo {author}
  {\bibfnamefont {H.~K.}\ \bibnamefont {{Eriksen}}}, \bibinfo {author}
  {\bibfnamefont {D.}~\bibnamefont {{Finkbeiner}}}, \bibinfo {author}
  {\bibfnamefont {D.}~\bibnamefont {{Fixsen}}}, \bibinfo {author}
  {\bibfnamefont {P.}~\bibnamefont {{Fosalba}}}, \bibinfo {author}
  {\bibfnamefont {A.}~\bibnamefont {{Fraisse}}}, \bibinfo {author}
  {\bibfnamefont {C.}~\bibnamefont {{Hirata}}}, \bibinfo {author}
  {\bibfnamefont {A.}~\bibnamefont {{Kogut}}}, \bibinfo {author} {\bibfnamefont
  {J.}~\bibnamefont {{Kristiansen}}}, \bibinfo {author} {\bibfnamefont
  {C.}~\bibnamefont {{Lawrence}}}, \bibinfo {author} {\bibfnamefont {A.~M.}\
  \bibnamefont {{Magalh\~{a}es}}}, \bibinfo {author} {\bibfnamefont {M.~A.}\
  \bibnamefont {{Miville-Deschenes}}},  \emph {et~al.},\ }\href {\doibase
  10.1063/1.3160888} {\bibfield  {journal} {\bibinfo  {journal} {AIP Conf.\
  Proc.}\ }\textbf {\bibinfo {volume} {1141}},\ \bibinfo {pages} {222}
  (\bibinfo {year} {2009})},\ \Eprint {http://arxiv.org/abs/0811.3915}
  {arXiv:0811.3915} \BibitemShut {NoStop}%
\bibitem [{\citenamefont {{Choi}}\ and\ \citenamefont {{Page}}(2015)}]{choi15}%
  \BibitemOpen
  \bibfield  {author} {\bibinfo {author} {\bibfnamefont {S.~K.}\ \bibnamefont
  {{Choi}}}\ and\ \bibinfo {author} {\bibfnamefont {L.~A.}\ \bibnamefont
  {{Page}}},\ }\href {\doibase 10.1088/1475-7516/2015/12/020} {\bibfield
  {journal} {\bibinfo  {journal} {\jcap}\ }\textbf {\bibinfo {volume} {12}},\
  \bibinfo {eid} {020} (\bibinfo {year} {2015})},\ \Eprint
  {http://arxiv.org/abs/1509.05934} {arXiv:1509.05934} \BibitemShut {NoStop}%
\bibitem [{\citenamefont {{Planck Collaboration 2015
  II}}(2016)}]{planck2015II}%
  \BibitemOpen
  \bibfield  {author} {\bibinfo {author} {\bibnamefont {{Planck Collaboration
  2015 II}}},\ }\href {\doibase 10.1051/0004-6361/201525818} {\bibfield
  {journal} {\bibinfo  {journal} {\aap}\ }\textbf {\bibinfo {volume} {594}},\
  \bibinfo {eid} {A2} (\bibinfo {year} {2016})},\ \Eprint
  {http://arxiv.org/abs/1502.01583} {arXiv:1502.01583 [astro-ph.IM]}
  \BibitemShut {NoStop}%
\bibitem [{\citenamefont {{Thorne}}\ \emph {et~al.}(2017)\citenamefont
  {{Thorne}}, \citenamefont {{Dunkley}}, \citenamefont {{Alonso}},\ and\
  \citenamefont {{N{\ae}ss}}}]{thorne17}%
  \BibitemOpen
  \bibfield  {author} {\bibinfo {author} {\bibfnamefont {B.}~\bibnamefont
  {{Thorne}}}, \bibinfo {author} {\bibfnamefont {J.}~\bibnamefont {{Dunkley}}},
  \bibinfo {author} {\bibfnamefont {D.}~\bibnamefont {{Alonso}}}, \ and\
  \bibinfo {author} {\bibfnamefont {S.}~\bibnamefont {{N{\ae}ss}}},\ }\href
  {\doibase 10.1093/mnras/stx949} {\bibfield  {journal} {\bibinfo  {journal}
  {\mnras}\ }\textbf {\bibinfo {volume} {469}},\ \bibinfo {pages} {2821}
  (\bibinfo {year} {2017})},\ \Eprint {http://arxiv.org/abs/1608.02841}
  {arXiv:1608.02841} \BibitemShut {NoStop}%
\bibitem [{\citenamefont {Hensley}(2015)}]{hensley2015}%
  \BibitemOpen
  \bibfield  {author} {\bibinfo {author} {\bibfnamefont {B.}~\bibnamefont
  {Hensley}},\ }\emph {\bibinfo {title} {On the nature of interstellar
  grains}},\ \href@noop {} {Ph.D. thesis},\ \bibinfo  {school} {Princeton
  University} (\bibinfo {year} {2015})\BibitemShut {NoStop}%
\bibitem [{\citenamefont {{Kritsuk}}\ \emph {et~al.}(2017)\citenamefont
  {{Kritsuk}}, \citenamefont {{Ustyugov}},\ and\ \citenamefont
  {{Norman}}}]{kritsuk17}%
  \BibitemOpen
  \bibfield  {author} {\bibinfo {author} {\bibfnamefont {A.~G.}\ \bibnamefont
  {{Kritsuk}}}, \bibinfo {author} {\bibfnamefont {S.~D.}\ \bibnamefont
  {{Ustyugov}}}, \ and\ \bibinfo {author} {\bibfnamefont {M.~L.}\ \bibnamefont
  {{Norman}}},\ }\href {\doibase 10.1088/1367-2630/aa7156} {\bibfield
  {journal} {\bibinfo  {journal} {New Journal of Physics}\ }\textbf {\bibinfo
  {volume} {19}},\ \bibinfo {eid} {065003} (\bibinfo {year} {2017})},\ \Eprint
  {http://arxiv.org/abs/1705.01912} {arXiv:1705.01912} \BibitemShut {NoStop}%
\bibitem [{\citenamefont {{Planck Collaboration Int.\ L}}(2017)}]{planckiL}%
  \BibitemOpen
  \bibfield  {author} {\bibinfo {author} {\bibnamefont {{Planck Collaboration
  Int.\ L}}},\ }\href {\doibase 10.1051/0004-6361/201629164} {\bibfield
  {journal} {\bibinfo  {journal} {\aap}\ }\textbf {\bibinfo {volume} {599}},\
  \bibinfo {eid} {A51} (\bibinfo {year} {2017})},\ \Eprint
  {http://arxiv.org/abs/1606.07335} {arXiv:1606.07335} \BibitemShut {NoStop}%
\bibitem [{\citenamefont {{Sheehy}}\ and\ \citenamefont
  {{Slosar}}(2018)}]{sheehy17}%
  \BibitemOpen
  \bibfield  {author} {\bibinfo {author} {\bibfnamefont {C.}~\bibnamefont
  {{Sheehy}}}\ and\ \bibinfo {author} {\bibfnamefont {A.}~\bibnamefont
  {{Slosar}}},\ }\href {\doibase 10.1103/PhysRevD.97.043522} {\bibfield
  {journal} {\bibinfo  {journal} {\prd}\ }\textbf {\bibinfo {volume} {97}},\
  \bibinfo {eid} {043522} (\bibinfo {year} {2018})},\ \Eprint
  {http://arxiv.org/abs/1709.09729} {arXiv:1709.09729} \BibitemShut {NoStop}%
\bibitem [{\citenamefont {{Planck Collaboration XLVI}}(2016)}]{planck2016XLVI}%
  \BibitemOpen
  \bibfield  {author} {\bibinfo {author} {\bibnamefont {{Planck Collaboration
  XLVI}}},\ }\href {\doibase 10.1051/0004-6361/201628890} {\bibfield  {journal}
  {\bibinfo  {journal} {\aap}\ }\textbf {\bibinfo {volume} {596}},\ \bibinfo
  {eid} {A107} (\bibinfo {year} {2016})},\ \Eprint
  {http://arxiv.org/abs/1605.02985v2} {arXiv:1605.02985v2} \BibitemShut
  {NoStop}%
\bibitem [{\citenamefont {{\keckarray\ and \biceptwo\ Collaborations
  XI}}(shed)}]{biceptwoXI}%
  \BibitemOpen
  \bibfield  {author} {\bibinfo {author} {\bibnamefont {{\keckarray\ and
  \biceptwo\ Collaborations XI}}},\ }\href@noop {} {\  (\bibinfo {year} {to be
  published})}\BibitemShut {NoStop}%
\bibitem [{\citenamefont {{Planck Collaboration 2015 X}}(2016)}]{planck2015X}%
  \BibitemOpen
  \bibfield  {author} {\bibinfo {author} {\bibnamefont {{Planck Collaboration
  2015 X}}},\ }\href {\doibase 10.1051/0004-6361/201525967} {\bibfield
  {journal} {\bibinfo  {journal} {\aap}\ }\textbf {\bibinfo {volume} {594}},\
  \bibinfo {eid} {A10} (\bibinfo {year} {2016})},\ \Eprint
  {http://arxiv.org/abs/1502.01588} {arXiv:1502.01588} \BibitemShut {NoStop}%
\bibitem [{\citenamefont {{Finkbeiner}}\ \emph {et~al.}(1999)\citenamefont
  {{Finkbeiner}}, \citenamefont {{Davis}},\ and\ \citenamefont
  {{Schlegel}}}]{finkbeiner99}%
  \BibitemOpen
  \bibfield  {author} {\bibinfo {author} {\bibfnamefont {D.~P.}\ \bibnamefont
  {{Finkbeiner}}}, \bibinfo {author} {\bibfnamefont {M.}~\bibnamefont
  {{Davis}}}, \ and\ \bibinfo {author} {\bibfnamefont {D.~J.}\ \bibnamefont
  {{Schlegel}}},\ }\href {\doibase 10.1086/307852} {\bibfield  {journal}
  {\bibinfo  {journal} {\apj}\ }\textbf {\bibinfo {volume} {524}},\ \bibinfo
  {pages} {867} (\bibinfo {year} {1999})},\ \Eprint
  {http://arxiv.org/abs/astro-ph/9905128} {astro-ph/9905128} \BibitemShut
  {NoStop}%
\bibitem [{\citenamefont {{Vansyngel}}\ \emph {et~al.}(2017)\citenamefont
  {{Vansyngel}}, \citenamefont {{Boulanger}}, \citenamefont {{Ghosh}},
  \citenamefont {{Wandelt}}, \citenamefont {{Aumont}}, \citenamefont
  {{Bracco}}, \citenamefont {{Levrier}}, \citenamefont {{Martin}},\ and\
  \citenamefont {{Montier}}}]{vansyngel16}%
  \BibitemOpen
  \bibfield  {author} {\bibinfo {author} {\bibfnamefont {F.}~\bibnamefont
  {{Vansyngel}}}, \bibinfo {author} {\bibfnamefont {F.}~\bibnamefont
  {{Boulanger}}}, \bibinfo {author} {\bibfnamefont {T.}~\bibnamefont
  {{Ghosh}}}, \bibinfo {author} {\bibfnamefont {B.~D.}\ \bibnamefont
  {{Wandelt}}}, \bibinfo {author} {\bibfnamefont {J.}~\bibnamefont {{Aumont}}},
  \bibinfo {author} {\bibfnamefont {A.}~\bibnamefont {{Bracco}}}, \bibinfo
  {author} {\bibfnamefont {F.}~\bibnamefont {{Levrier}}}, \bibinfo {author}
  {\bibfnamefont {P.~G.}\ \bibnamefont {{Martin}}}, \ and\ \bibinfo {author}
  {\bibfnamefont {L.}~\bibnamefont {{Montier}}},\ }\href {\doibase
  10.1051/0004-6361/201629992} {\bibfield  {journal} {\bibinfo  {journal}
  {\aap}\ }\textbf {\bibinfo {volume} {603}},\ \bibinfo {eid} {A62} (\bibinfo
  {year} {2017})},\ \Eprint {http://arxiv.org/abs/1611.02577}
  {arXiv:1611.02577} \BibitemShut {NoStop}%
\bibitem [{\citenamefont {{Planck Collaboration IX}}(2014)}]{planck2013IX}%
  \BibitemOpen
  \bibfield  {author} {\bibinfo {author} {\bibnamefont {{Planck Collaboration
  IX}}},\ }\href@noop {} {\bibfield  {journal} {\bibinfo  {journal} {\aap}\
  }\textbf {\bibinfo {volume} {571}},\ \bibinfo {pages} {A9} (\bibinfo {year}
  {2014})}\BibitemShut {NoStop}%
\bibitem [{\citenamefont {{\keckarray\ and \biceptwo\ Collaborations
  VIII}}(2016)}]{biceptwoVIII}%
  \BibitemOpen
  \bibfield  {author} {\bibinfo {author} {\bibnamefont {{\keckarray\ and
  \biceptwo\ Collaborations VIII}}},\ }\href {\doibase
  10.3847/1538-4357/833/2/228} {\bibfield  {journal} {\bibinfo  {journal}
  {\apj}\ }\textbf {\bibinfo {volume} {833}},\ \bibinfo {eid} {228} (\bibinfo
  {year} {2016})},\ \Eprint {http://arxiv.org/abs/1606.01968}
  {arXiv:1606.01968} \BibitemShut {NoStop}%
\bibitem [{\citenamefont {Ade}\ \emph {et~al.}(2016)\citenamefont {Ade} \emph
  {et~al.}}]{planckXV}%
  \BibitemOpen
  \bibfield  {author} {\bibinfo {author} {\bibfnamefont {P.~A.~R.}\
  \bibnamefont {Ade}} \emph {et~al.} (\bibinfo {collaboration} {Planck}),\
  }\href {\doibase 10.1051/0004-6361/201525941} {\bibfield  {journal} {\bibinfo
   {journal} {Astron. Astrophys.}\ }\textbf {\bibinfo {volume} {594}},\
  \bibinfo {pages} {A15} (\bibinfo {year} {2016})},\ \Eprint
  {http://arxiv.org/abs/1502.01591} {arXiv:1502.01591} \BibitemShut {NoStop}%
%%CITATION = ARXIV:1502.01591;%%
\end{thebibliography}%

\clearpage

\begin{appendix}

\section{Maps}
\label{app:maps}

\begin{figure*}
\resizebox{\textwidth}{!}{\includegraphics{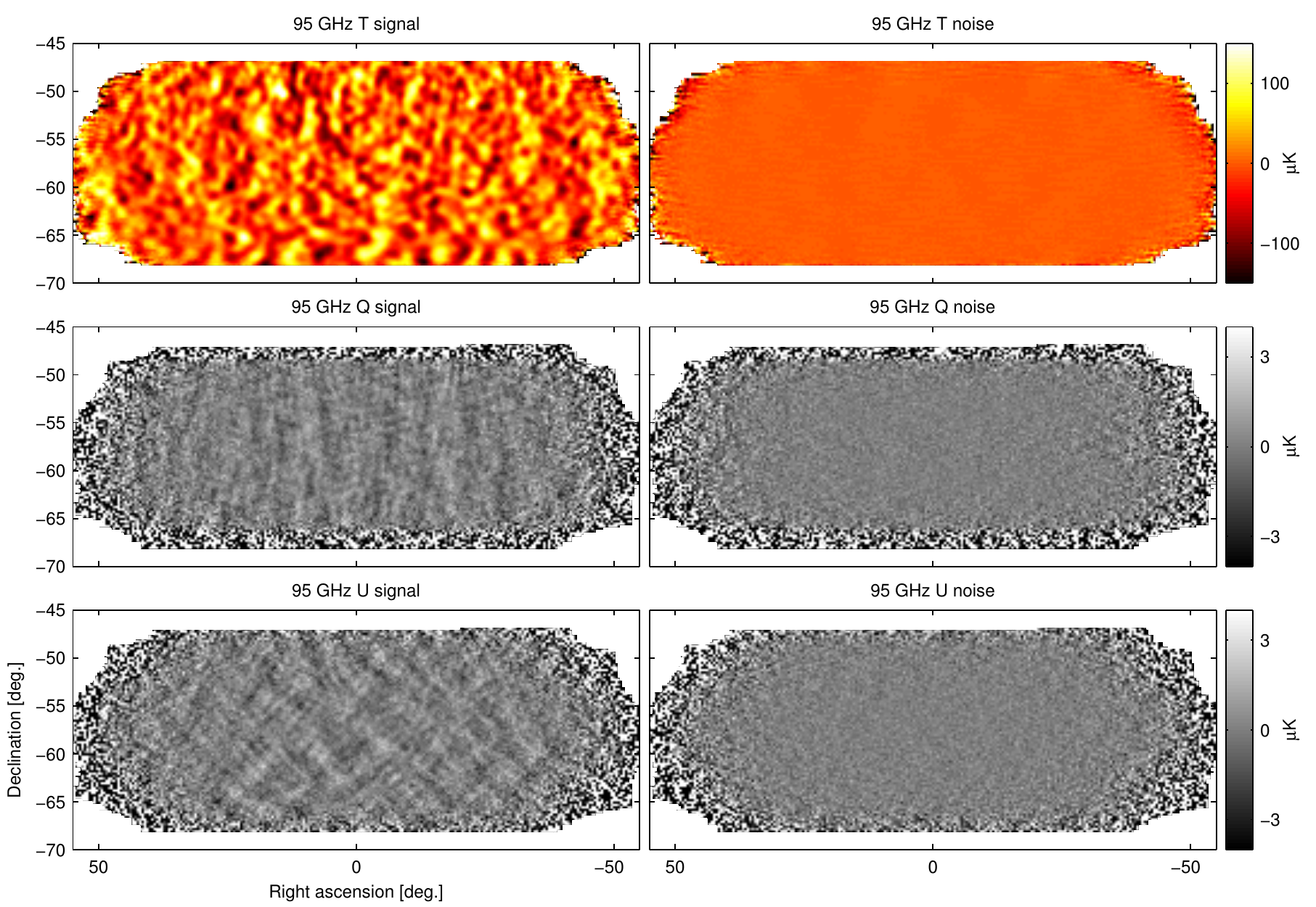}}
\caption{$T$, $Q$, $U$ maps at 95\,GHz
using data taken by two receivers of \keckarray\ during the 2014 \& 2015
seasons---we refer to these maps as BK15$_{95}$.
The left column shows the real data maps with $0.25\deg$ pixelization
as output by the reduction pipeline.
The right column shows a noise realization made by randomly assigning
positive and negative signs while coadding the data.
These maps are filtered by the instrument beam
(FWHM 43~arcmin \cite{biceptwoXI}),
timestream processing, and (for $Q$~\&~$U$) deprojection of
beam systematics.
Note that the horizontal/vertical and $45\deg$ structures seen in the
$Q$ and $U$ signal maps are expected for an \emode\ dominated sky.}
\label{fig:tqu_maps_95}
\end{figure*}

\begin{figure*}
\resizebox{\textwidth}{!}{\includegraphics{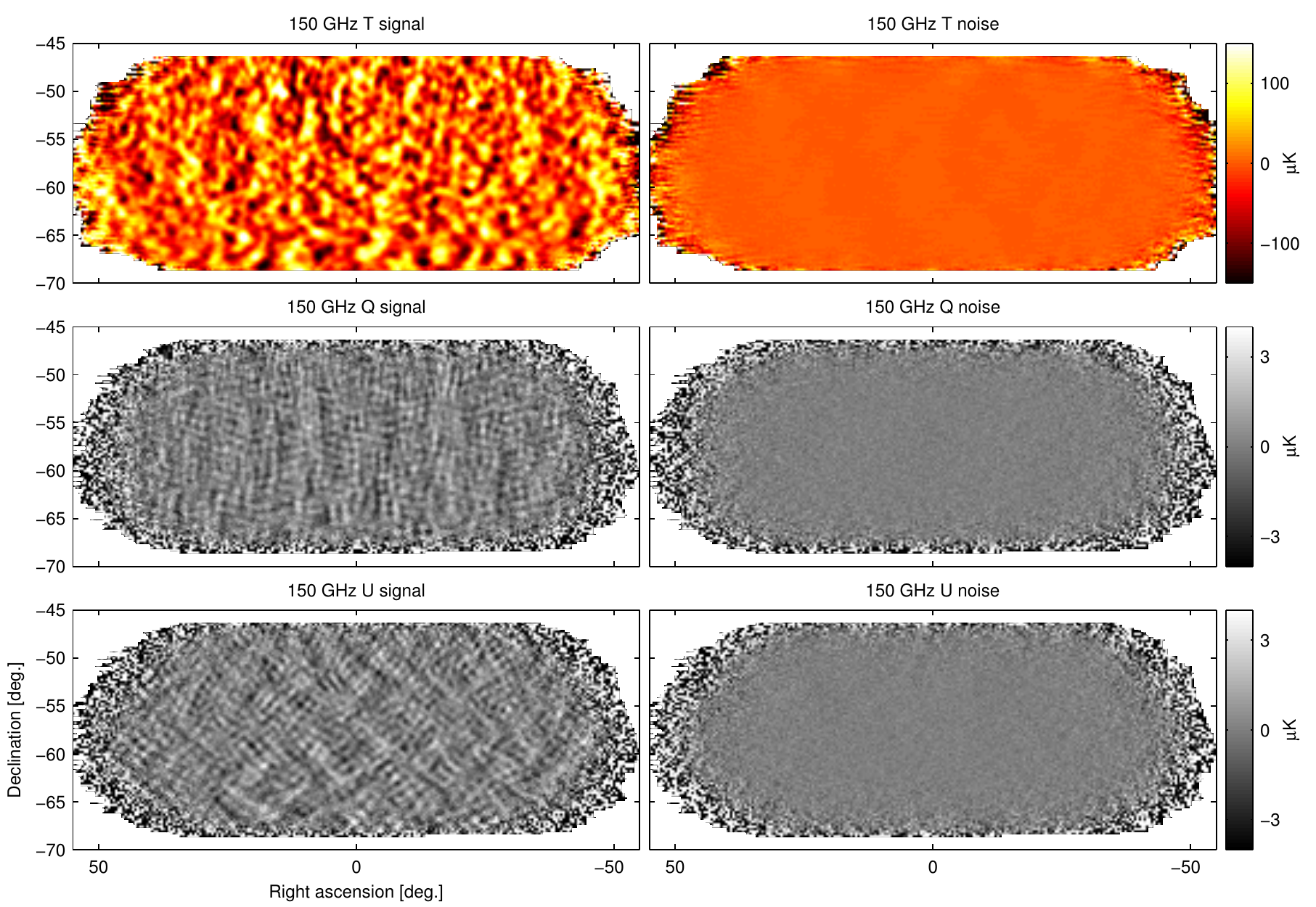}}
\caption{$T$, $Q$, $U$ maps at 150\,GHz
using all \biceptwo/\keck\ data up to and including that taken
during the 2015 observing season---we refer to these maps as BK15$_{150}$.
These maps are directly analogous to the 95\,GHz maps shown in
Fig.~\ref{fig:tqu_maps_95} except that the
instrument beam filtering is in this case 30~arcmin FWHM
\cite{biceptwoXI}.}
\label{fig:tqu_maps_150}
\end{figure*}

\begin{figure*}
\resizebox{\textwidth}{!}{\includegraphics{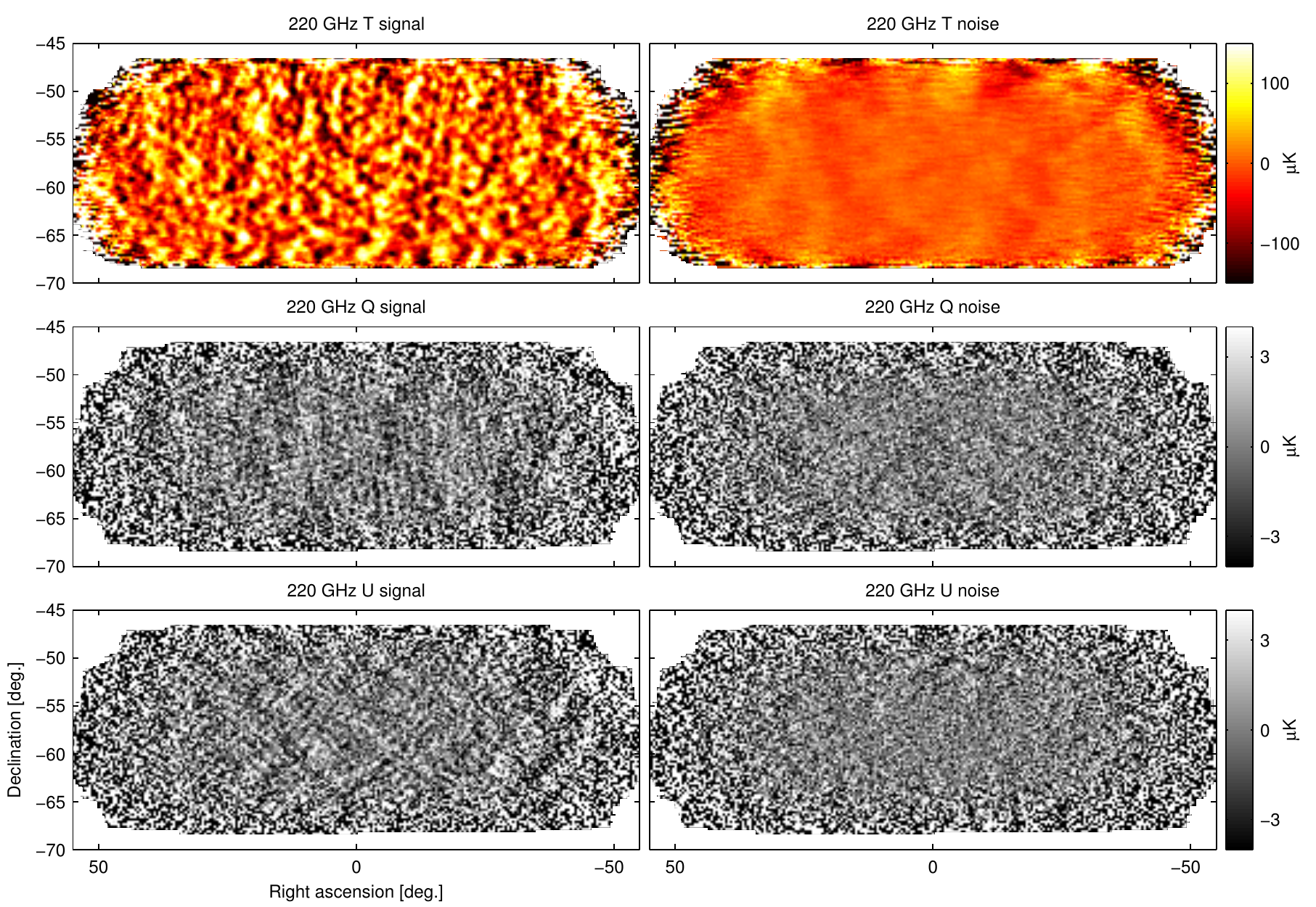}}
\caption{$T$, $Q$, $U$ maps at 220\,GHz
using data taken by two receivers of \keckarray\ during the 2015
season---we refer to these maps as BK15$_{220}$.
These maps are directly analogous to the 95\,GHz maps shown in
Fig.~\ref{fig:tqu_maps_95} except that the
instrument beam filtering is in this case 20~arcmin FWHM
\cite{biceptwoXI}.}
\label{fig:tqu_maps_220}
\end{figure*}

Figures~\ref{fig:tqu_maps_95},~\ref{fig:tqu_maps_150}~\&~\ref{fig:tqu_maps_220}
show the full sets of BK15 $T$/$Q$/$U$ maps at 95, 150 \& 220~GHz.
The right side of each figure shows realizations of noise
created by randomly flipping the sign of data subsets while
coadding the map---see Sec.~V.B of Ref.~\cite{biceptwoI} for
further details.

\section{95\,GHz and 220\,GHz Internal Consistency Tests}
\label{app:mapjack}

A powerful internal consistency test are data split difference tests
which we refer to as ``jackknives''.
As well as the full coadd signal maps we also form many pairs
of split maps where the splits are chosen such that one might
expect different systematic contamination in the two halves
of the split.
The split halves are differenced and the power spectra
taken.
We then take the deviations of these from the mean of
signal+noise simulations and
form $\chi^2$ and $\chi$ (sum of deviations) statistics.
In this section we perform tests of the 95\,GHz
and 220\,GHz data sets which are exactly analogous to the tests
of the 150\,GHz data sets performed
in Sec.~VII.C of Ref.~\cite{biceptwoI} and Sec.~6.3 of Ref.~\cite{biceptwoV}.
(Since going from 8 to 9 receiver-years of \keck\ 150\,GHz barely % REFB
shifts the results we omit those tests for brevity.)

\begin{figure}[htb]
\begin{center}
\resizebox{0.7\columnwidth}{!}{\includegraphics{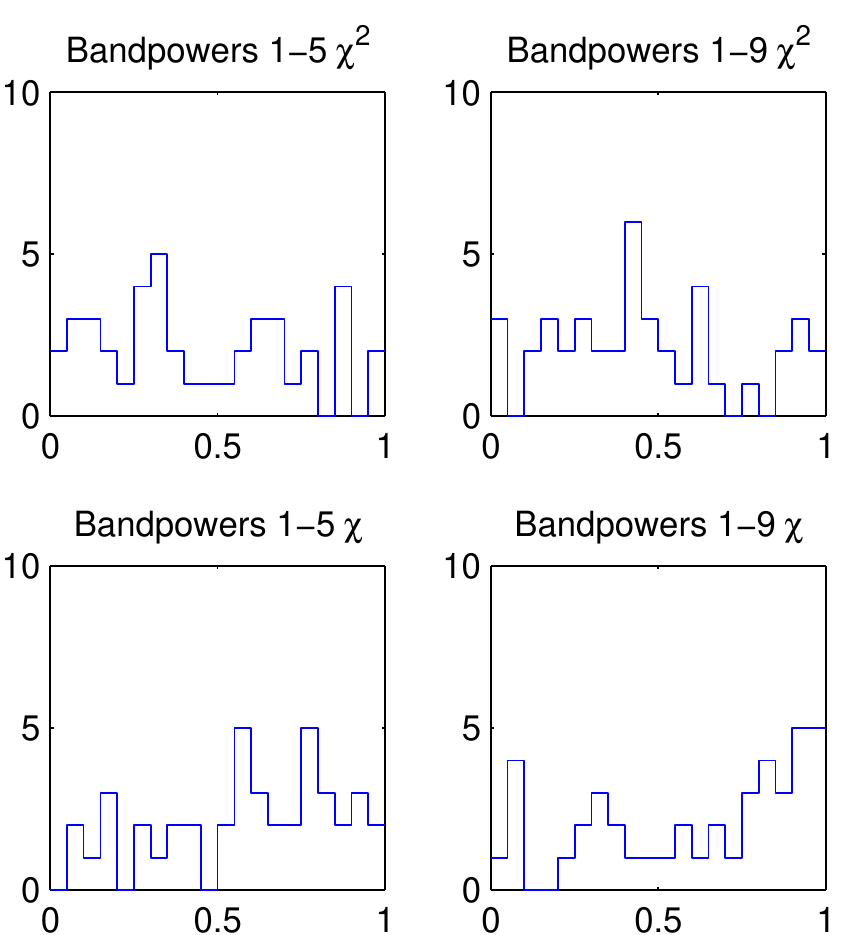}}
\end{center}
\caption{Distributions of the jackknife $\chi^2$ and $\chi$ PTE
values for the \keckarray\ 2014 \& 2015 95\,GHz data over the tests and
spectra given in Table~\ref{tab:ptes_95}.
This figure is analogous to Fig.~12 of Ref.~\cite{biceptwoVI}.}
\label{fig:ptedist_95}
\end{figure}

\begin{figure}[htb]
\begin{center}
\resizebox{0.7\columnwidth}{!}{\includegraphics{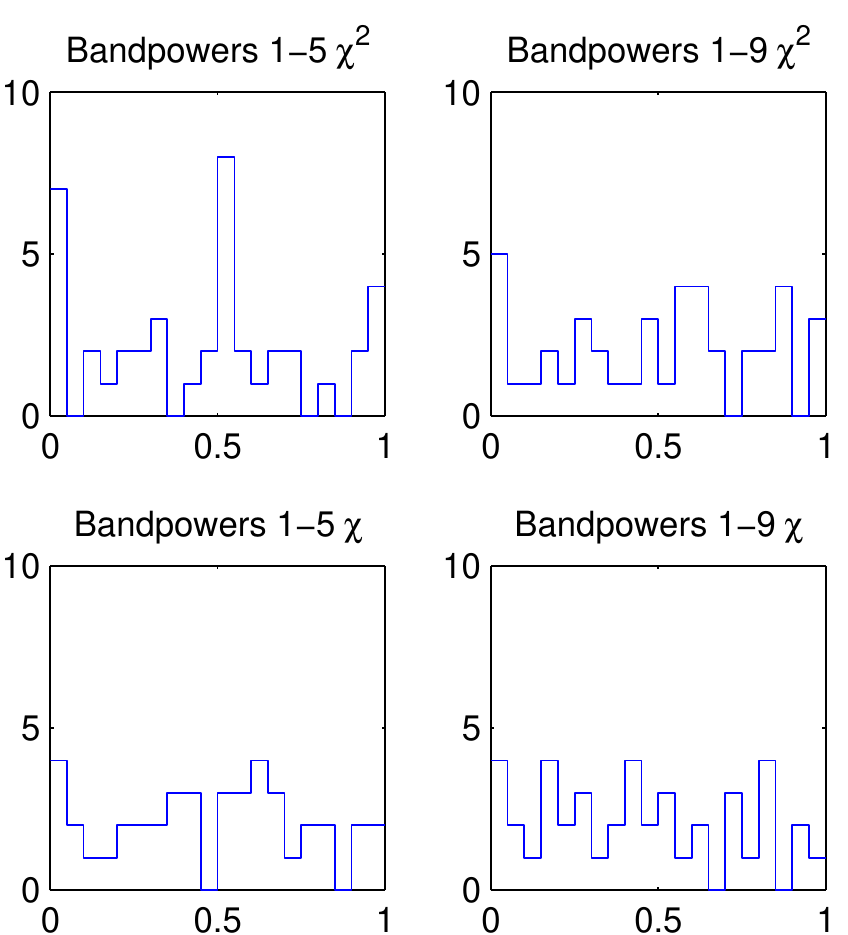}}
\end{center}
\caption{Distributions of the jackknife $\chi^2$ and $\chi$ PTE
values for the \keckarray\ 2015 220\,GHz data over the tests and
spectra given in Table~\ref{tab:ptes_220}.}
\label{fig:ptedist_220}
\end{figure}

Tables~\ref{tab:ptes_95} and~\ref{tab:ptes_220} show the
$\chi^2$ and $\chi$ statistics for the 95\,GHz and 220\,GHz
jackknife tests respectively, while
Figures~\ref{fig:ptedist_95}~\&~\ref{fig:ptedist_220} present the
same results in graphical form.
Note that these values are partially correlated---particularly
the 1--5 and 1--9 versions of each statistic.
We conclude that there is no evidence for corruption of the
data at a level exceeding the noise.

\begingroup
\squeezetable
\begin{table}[pht]
\caption{\label{tab:ptes_95} Jackknife PTE values from $\chi^2$ and $\chi$
(sum of deviations) tests 
for \keckarray\ 95\,GHz data taken in 2014 and 2015.
This table is analogous to Table~I of Ref.~\cite{biceptwoVI} but extended to two seasons
of data.}
\begin{ruledtabular}
\begin{tabular}{l c c c c }
Jackknife & Band powers & Band powers & Band powers & Band powers\\
& 1--5 $\chi^2$ & 1--9 $\chi^2$ & 1--5 $\chi$ & 1--9 $\chi$ \\
\hline
\\ 
\multicolumn{5}{l}{Deck jackknife} \\ 
EE & 0.042 & 0.176 & 0.421 & 0.501 \\ 
BB & 0.132 & 0.186 & 0.852 & 0.952 \\ 
EB & 0.705 & 0.922 & 0.196 & 0.361 \\ 
\multicolumn{5}{l}{Scan Dir jackknife} \\ 
EE & 0.281 & 0.136 & 0.553 & 0.920 \\ 
BB & 0.154 & 0.100 & 0.980 & 0.968 \\ 
EB & 0.269 & 0.263 & 0.096 & 0.050 \\ 
\multicolumn{5}{l}{Tag Split jackknife} \\ 
EE & 0.194 & 0.377 & 0.743 & 0.930 \\ 
BB & 0.084 & 0.160 & 0.920 & 0.898 \\ 
EB & 0.685 & 0.870 & 0.259 & 0.319 \\ 
\multicolumn{5}{l}{Tile jackknife} \\ 
EE & 0.321 & 0.517 & 0.800 & 0.916 \\ 
BB & 0.862 & 0.978 & 0.832 & 0.792 \\ 
EB & 0.363 & 0.279 & 0.758 & 0.711 \\ 
\multicolumn{5}{l}{Phase jackknife} \\ 
EE & 0.858 & 0.800 & 0.627 & 0.621 \\ 
BB & 0.010 & 0.048 & 0.186 & 0.200 \\ 
EB & 0.337 & 0.423 & 0.721 & 0.758 \\ 
\multicolumn{5}{l}{Mux Col jackknife} \\ 
EE & 0.778 & 0.912 & 0.904 & 0.804 \\ 
BB & 0.651 & 0.497 & 0.419 & 0.880 \\ 
EB & 0.343 & 0.224 & 0.569 & 0.253 \\ 
\multicolumn{5}{l}{Alt Deck jackknife} \\ 
EE & 0.110 & 0.409 & 0.399 & 0.483 \\ 
BB & 0.335 & 0.487 & 0.569 & 0.677 \\ 
EB & 0.643 & 0.347 & 0.517 & 0.950 \\ 
\multicolumn{5}{l}{Mux Row jackknife} \\ 
EE & 0.459 & 0.557 & 0.599 & 0.896 \\ 
BB & 0.784 & 0.447 & 0.665 & 0.832 \\ 
EB & 0.697 & 0.621 & 0.132 & 0.042 \\ 
\multicolumn{5}{l}{Tile/Deck jackknife} \\ 
EE & 0.393 & 0.693 & 0.812 & 0.691 \\ 
BB & 0.267 & 0.309 & 0.303 & 0.333 \\ 
EB & 0.579 & 0.355 & 0.760 & 0.934 \\ 
\multicolumn{5}{l}{Focal Plane inner/outer jackknife} \\ 
EE & 0.617 & 0.419 & 0.906 & 0.992 \\ 
BB & 0.132 & 0.226 & 0.892 & 0.972 \\ 
EB & 0.984 & 0.629 & 0.683 & 0.806 \\ 
\multicolumn{5}{l}{Tile top/bottom jackknife} \\ 
EE & 0.595 & 0.020 & 0.593 & 0.407 \\ 
BB & 0.954 & 0.990 & 0.615 & 0.357 \\ 
EB & 0.289 & 0.505 & 0.954 & 0.840 \\ 
\multicolumn{5}{l}{Tile inner/outer jackknife} \\ 
EE & 0.305 & 0.605 & 0.158 & 0.090 \\ 
BB & 0.509 & 0.601 & 0.527 & 0.567 \\ 
EB & 0.449 & 0.447 & 0.375 & 0.096 \\ 
\multicolumn{5}{l}{Moon jackknife} \\ 
EE & 0.086 & 0.299 & 0.066 & 0.086 \\ 
BB & 0.900 & 0.852 & 0.291 & 0.325 \\ 
EB & 0.200 & 0.477 & 0.782 & 0.796 \\ 
\multicolumn{5}{l}{A/B offset best/worst} \\ 
EE & 0.090 & 0.034 & 0.766 & 0.295 \\ 
BB & 0.882 & 0.435 & 0.806 & 0.970 \\ 
EB & 0.613 & 0.902 & 0.611 & 0.561 \\ 

\end{tabular}
\end{ruledtabular}
\end{table}
\endgroup

\begingroup
\squeezetable
\begin{table}[pht]
\caption{\label{tab:ptes_220} Jackknife PTE values from $\chi^2$ and $\chi$
(sum of deviations) tests 
for \keckarray\ 220\,GHz data taken in 2015.}
\begin{ruledtabular}
\begin{tabular}{l c c c c }
Jackknife & Band powers & Band powers & Band powers & Band powers\\
& 1--5 $\chi^2$ & 1--9 $\chi^2$ & 1--5 $\chi$ & 1--9 $\chi$ \\
\hline
\\ 
\multicolumn{5}{l}{Deck jackknife} \\ 
EE & 0.515 & 0.198 & 0.918 & 0.365 \\ 
BB & 0.024 & 0.028 & 0.008 & 0.178 \\ 
EB & 0.343 & 0.551 & 0.359 & 0.383 \\ 
\multicolumn{5}{l}{Scan Dir jackknife} \\ 
EE & 0.962 & 0.968 & 0.643 & 0.579 \\ 
BB & 0.154 & 0.261 & 0.579 & 0.754 \\ 
EB & 0.713 & 0.896 & 0.631 & 0.447 \\ 
\multicolumn{5}{l}{Tag Split jackknife} \\ 
EE & 0.030 & 0.014 & 0.715 & 0.976 \\ 
BB & 0.327 & 0.587 & 0.966 & 0.948 \\ 
EB & 0.483 & 0.840 & 0.234 & 0.431 \\ 
\multicolumn{5}{l}{Tile jackknife} \\ 
EE & 0.008 & 0.026 & 0.228 & 0.208 \\ 
BB & 0.242 & 0.469 & 0.846 & 0.850 \\ 
EB & 0.138 & 0.377 & 0.597 & 0.643 \\ 
\multicolumn{5}{l}{Phase jackknife} \\ 
EE & 0.549 & 0.858 & 0.966 & 0.928 \\ 
BB & 0.343 & 0.281 & 0.768 & 0.479 \\ 
EB & 0.447 & 0.271 & 0.669 & 0.727 \\ 
\multicolumn{5}{l}{Mux Col jackknife} \\ 
EE & 0.263 & 0.647 & 0.257 & 0.166 \\ 
BB & 0.567 & 0.693 & 0.116 & 0.257 \\ 
EB & 0.936 & 0.752 & 0.509 & 0.719 \\ 
\multicolumn{5}{l}{Alt Deck jackknife} \\ 
EE & 0.968 & 0.844 & 0.573 & 0.824 \\ 
BB & 0.030 & 0.172 & 0.409 & 0.539 \\ 
EB & 0.517 & 0.425 & 0.331 & 0.106 \\ 
\multicolumn{5}{l}{Mux Row jackknife} \\ 
EE & 0.695 & 0.611 & 0.166 & 0.094 \\ 
BB & 0.840 & 0.609 & 0.649 & 0.168 \\ 
EB & 0.509 & 0.311 & 0.605 & 0.347 \\ 
\multicolumn{5}{l}{Tile/Deck jackknife} \\ 
EE & 0.675 & 0.220 & 0.768 & 0.182 \\ 
BB & 0.968 & 0.990 & 0.681 & 0.834 \\ 
EB & 0.972 & 0.994 & 0.363 & 0.246 \\ 
\multicolumn{5}{l}{Focal Plane inner/outer jackknife} \\ 
EE & 0.020 & 0.038 & 0.010 & 0.016 \\ 
BB & 0.108 & 0.313 & 0.032 & 0.026 \\ 
EB & 0.012 & 0.040 & 0.509 & 0.433 \\ 
\multicolumn{5}{l}{Tile top/bottom jackknife} \\ 
EE & 0.210 & 0.108 & 0.076 & 0.028 \\ 
BB & 0.030 & 0.096 & 0.010 & 0.006 \\ 
EB & 0.709 & 0.581 & 0.685 & 0.549 \\ 
\multicolumn{5}{l}{Tile inner/outer jackknife} \\ 
EE & 0.503 & 0.637 & 0.503 & 0.828 \\ 
BB & 0.531 & 0.549 & 0.317 & 0.465 \\ 
EB & 0.477 & 0.471 & 0.826 & 0.723 \\ 
\multicolumn{5}{l}{Moon jackknife} \\ 
EE & 0.507 & 0.671 & 0.910 & 0.649 \\ 
BB & 0.942 & 0.894 & 0.281 & 0.267 \\ 
EB & 0.639 & 0.756 & 0.389 & 0.539 \\ 
\multicolumn{5}{l}{A/B offset best/worst} \\ 
EE & 0.561 & 0.854 & 0.066 & 0.082 \\ 
BB & 0.273 & 0.457 & 0.443 & 0.257 \\ 
EB & 0.531 & 0.569 & 0.425 & 0.441 \\ 

\end{tabular}
\end{ruledtabular}
\end{table}
\endgroup

\section{95\,GHz Spectral Stability}
\label{app:specjack}

\begin{figure}[htb]
\begin{center}
\resizebox{\columnwidth}{!}{\includegraphics{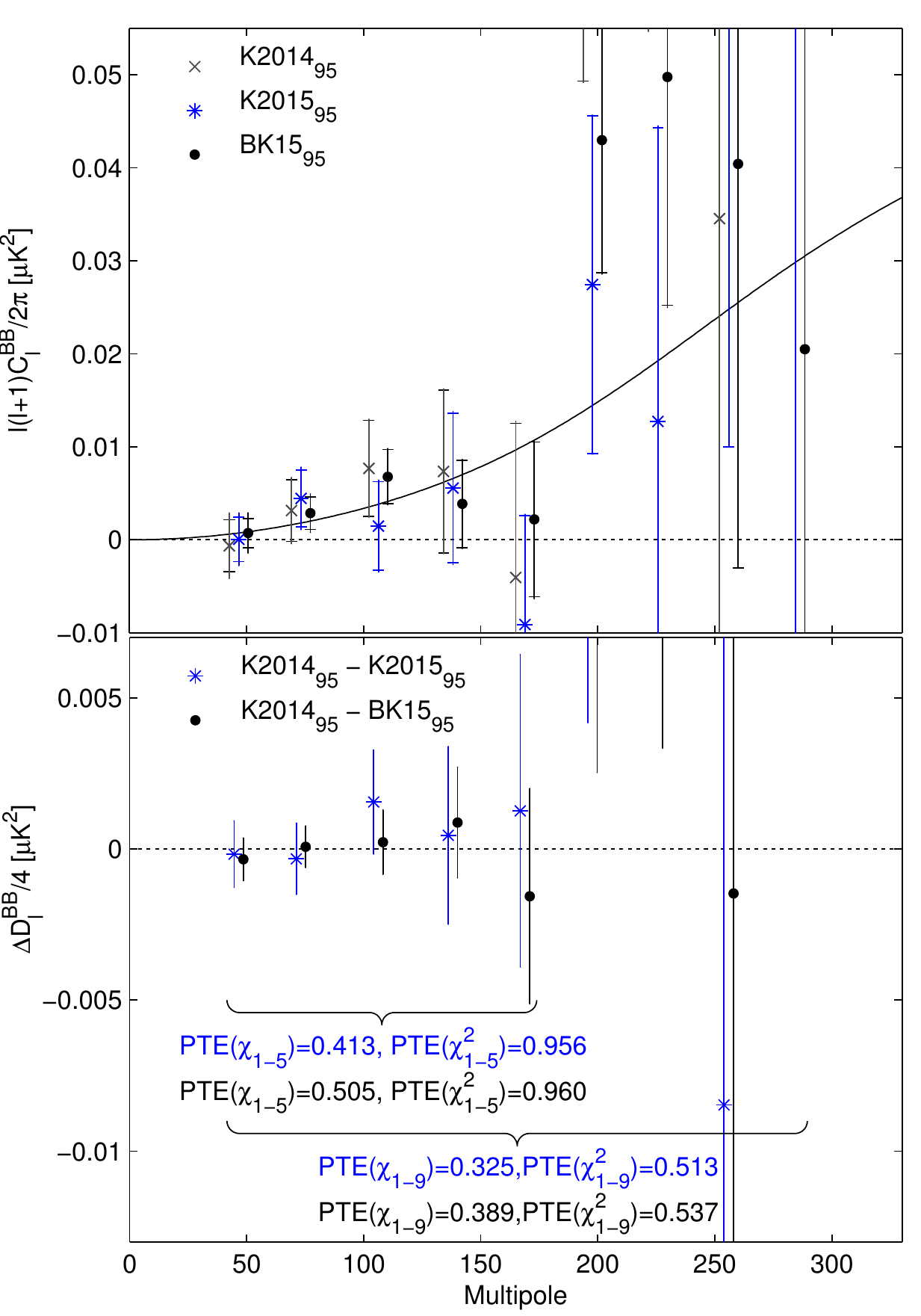}}
\end{center}
\caption{{\it Upper:} Comparison of the 95\,GHz $BB$ auto-spectrum
as previously published (K2014$_{95}$), for 2015 alone (K2015$_{95}$),
and for the combination of the two (BK15$_{95}$).
The inner error bars are the standard deviation of the lensed-\lcdm+noise 
simulations, while the outer error bars include the additional
fluctuation induced by the dust signal.
Note that neither of these uncertainties are appropriate for comparison
of the band power values---for this see the lower panel.
(For clarity the sets of points are offset horizontally.)
{\it Lower:} The difference of the pairs of spectra shown 
in the upper panel divided by a factor of four.
The error bars are the standard deviation of the pairwise
differences of signal+noise simulations
which share common input skies (the simulations used to
derive the outer error bars in the upper panel).
Comparison of these points with null is an appropriate test
of the compatibility of the spectra, and the PTE of $\chi$
and $\chi^2$ are shown.
This figure is similar to Fig.~13 of Ref.~\cite{biceptwoVI}.}
\label{fig:specjack_95}
\end{figure}

We next test the mutual compatibility of the 2014
and 2015 95\,GHz spectra.
We compare the differences of the real spectra
to the differences of simulations which share the same underlying
input skies.
We perform the test in two ways: firstly by differencing
the single season spectra (K2014$_{95}$ and K2015$_{95}$),
and secondly by differencing the 2014 single season from the
2014+2015 season combined spectrum.
Fig.~\ref{fig:specjack_95} shows the results---the differences
are seen to be consistent with noise fluctuation.

\section{Additional Spectra}
\label{app:allspec}

\begin{figure*}
\begin{center}
\resizebox{\textwidth}{!}{\includegraphics{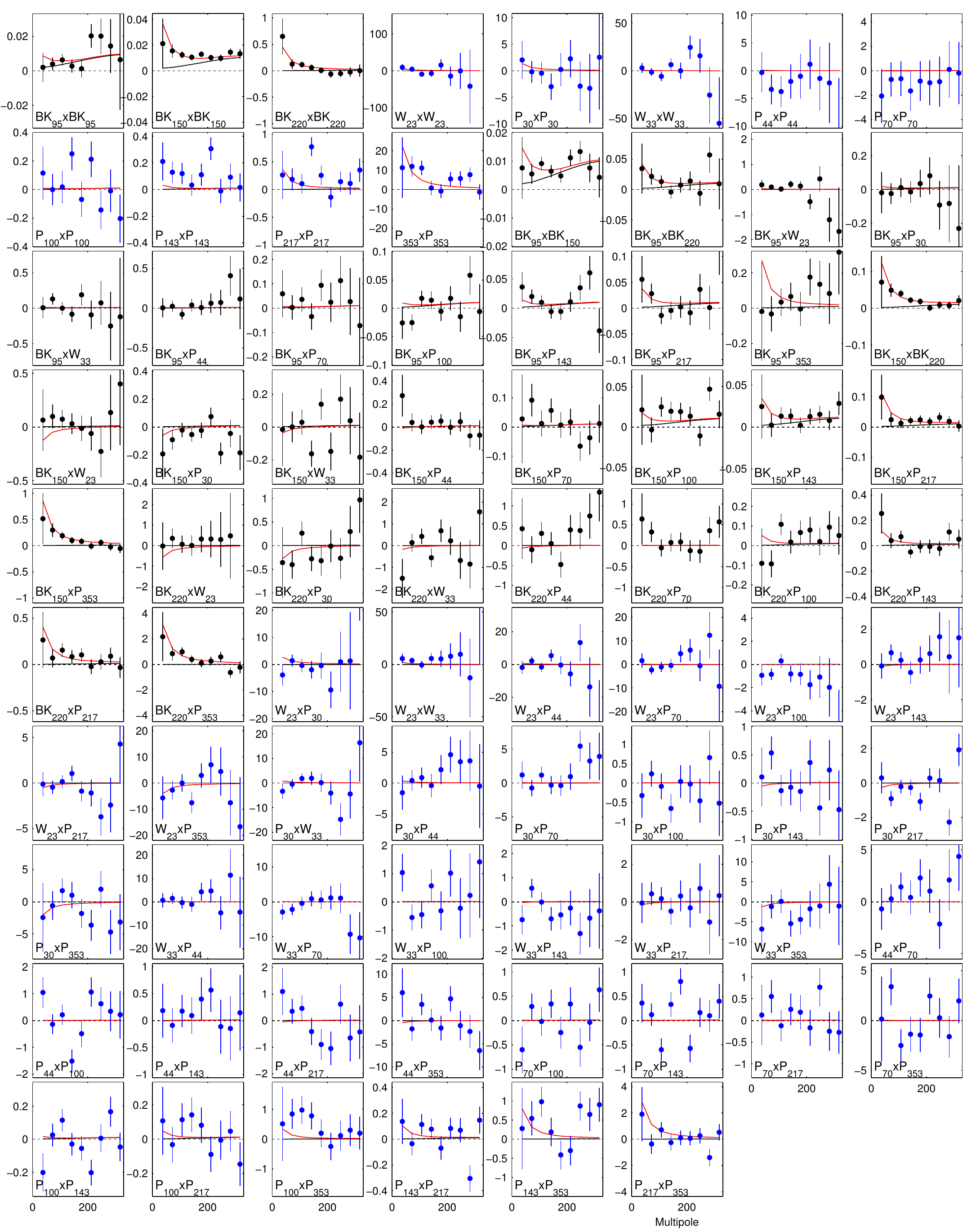}}
\end{center}
\caption{The full set of $BB$ auto- and cross-spectra
from which the joint model likelihood is derived.
In all cases the quantity plotted is $100 \ell C_\ell / 2 \pi$ (\uksq).
Spectra involving \biceptwo/\keck\ data are shown as black points
while those using only \wmap/\planck\ data are shown as blue
points.
The black lines show the expectation values
for lensed-\lcdm, while the 
red lines show the expectation values of the
maximum likelihood lensed-\lcdm+$r$+dust+synchrotron model
($r=\rmlm$, $\Adf=\Admlm$\,\uksq, $\Bd=\Bdmlm$, $\ad=\admlm$,
 $\Asf=\Asmlm$\,\uksq, $\Bs=\Bsmlm$, $\as=\asmlm$, $\epsilon=\emlm$),
and the error bars are scaled to that model.}
\label{fig:powspec_all}
\end{figure*}

Fig.~\ref{fig:powspecres_bkbands}
shows only a small subset of the spectra which are used in the
likelihood analysis and included in the \texttt{COSMOMC}
input file.
We are using three \biceptwo/\keck\ bands, two \wmap\ bands,
and seven \planck\ bands resulting in 12 auto- and 66
cross-spectra.
In Fig.~\ref{fig:powspec_all} we show all of these
together with the maximum likelihood model from the baseline analysis
whose parameters were quoted above.
Most of the spectra not already shown in Fig.~\ref{fig:powspecres_bkbands}
have low signal-to-noise, although a few of them carry interesting
additional information on the possible level of synchrotron
as will be noted later.

The HL likelihood~\citep{hamimeche08} we use for the primary analysis
accounts for the full joint PDF of auto- and cross-spectral
bandpowers which are derived from maps which are a
combination of (correlated) signal and (mostly uncorrelated)
noise.
We choose to quantify the absolute goodness-of-fit of the data
to the maximum likelihood model using a simple $\chi^2$ statistic which 
assumes that the bandpowers are normally distributed
about their expectation values.
We find that the distribution of this $\chi^2$ statistic
for the standard (499) lensed-\lcdm+dust+noise simulations versus
their input model is significantly broader than the nominal
theoretical distribution---presumably because of the non-normal
distribution of the bandpowers.
It is therefore most appropriate to compare the real data
value to the simulated distribution.

For the $9\times78=702$ bandpowers shown in Fig.~\ref{fig:powspec_all},
$\chi^2=(d-m)^T C^{-1}(d-m)=\chitwo$, where $d$ are the
bandpower values, $m$ are the model expectation values,
and $C$ is the bandpower covariance matrix.
This has a nominal theory PTE of $\chitwoptenom$ but a PTE versus the
simulations of $\chitwoptesim$.
If instead we take the sum of the normalized deviations
($\sum{((d-m)/e)^2}$ where $e$ is the square-root of the
diagonal of $C$) we find that the PTE versus the simulations
is \chioneptesim.
We conclude that the parametric model which we have chosen---in
combination with the approximation of Gaussian fluctuation
of the dust (and synchrotron) sky patterns---is an adequate
description of the presently available data.

\begin{figure}
\begin{center}
\resizebox{\columnwidth}{!}{\includegraphics{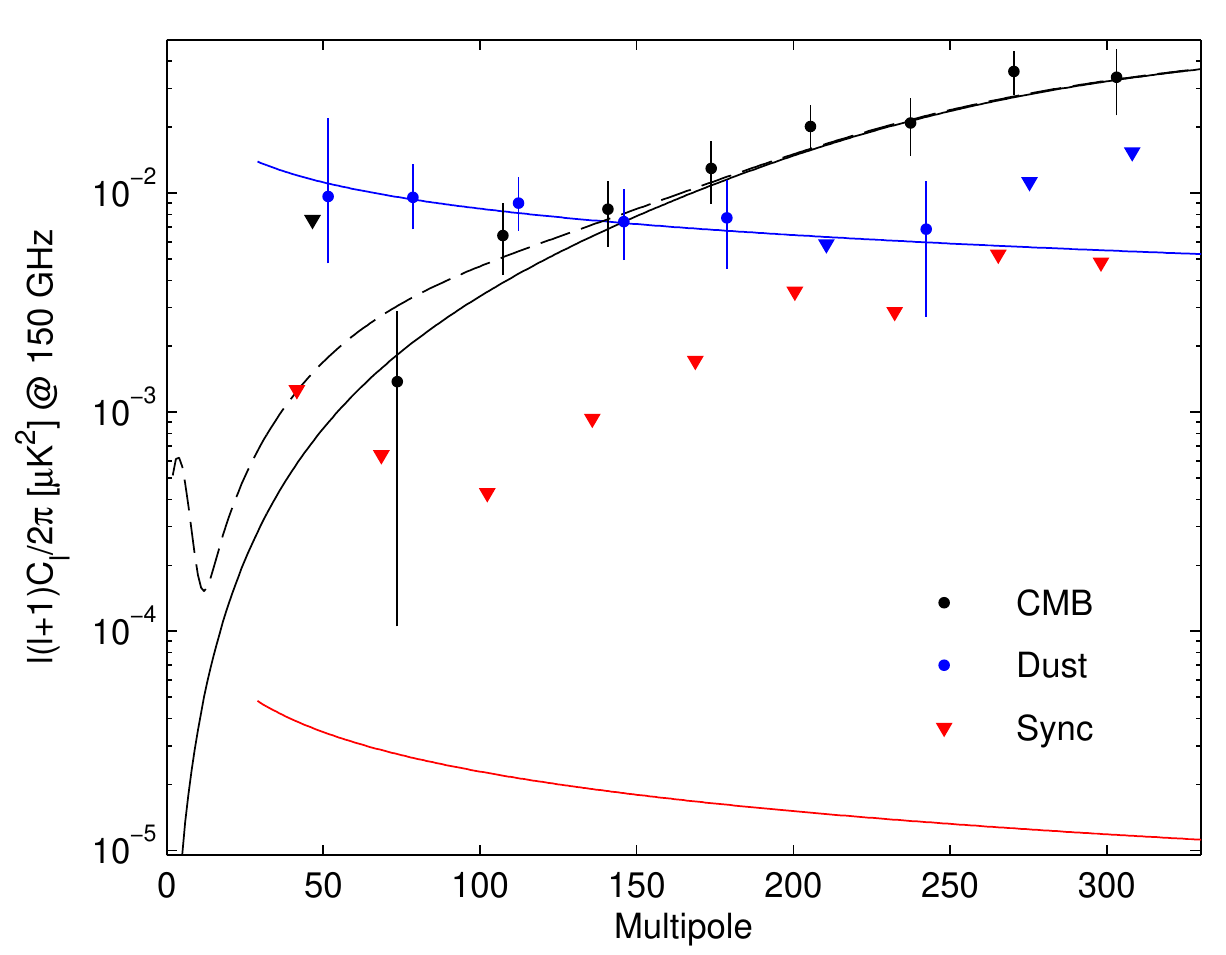}}
\end{center}
\caption{Spectral decomposition of the $BB$ data into synchrotron
(red), CMB (black) and dust (blue) components at 150\,GHz.
The decomposition is calculated independently in each bandpower,
marginalizing over $\Bd$, $\Bs$ and $\epsilon$ with
the same priors as the baseline analysis.
Error bars denote 68\protect\% credible intervals, with the point marking the most
probable value.
If the 68\protect\% interval includes zero, we also indicate the 95\protect\% upper
limit with a downward triangle.
(For clarity the sets of points are offset horizontally.)
The solid black line shows lensed-\lcdm\ with the dashed
line adding on top an $r_{0.05}=0.02$ tensor contribution.
The blue/red curves show sync/dust models consistent with the
baseline analysis
($\Adf=\Adcentval$\,\uksq, $\Bd=1.6$, $\ad=-0.4$ and
 $\Asf=\Ascentval$\,\uksq, $\Bs=-3.1$, $\as=-0.6$ respectively).
}
\label{fig:specdecomp}
\end{figure}

We also run a likelihood analysis
to find the CMB and foreground contributions on
a bandpower-by-bandpower basis.
The baseline analysis is a single fit to all 9 bandpowers
across 78 spectra with 8 parameters.
Instead we now perform 9 separate
fits---one for each bandpower---across the 78 spectra,
with 6 parameters in each fit.
These 6 parameters are the amplitudes of CMB,
dust and synchrotron plus $\Bd$, $\Bs$ and $\epsilon$ with
identical priors to the baseline analysis. 
The results are shown in Fig.~\ref{fig:specdecomp}---the
resulting CMB values are consistent with lensed-\lcdm\
while the dust values are consistent with the level
of dust found in the baseline analysis.
Synchrotron is tightly limited in all the multipole ranges,
and not detected in any of them.

\section{Likelihood Variation and Validation}
\label{app:likeevolvar}

\subsection{Likelihood Evolution}
\label{app:likeevol}

We make only one model change versus the BK14 baseline analysis---we 
extend the range over which the sync/dust correlation parameter
is marginalized from $0<\epsilon<1$ to the full possible range
$-1<\epsilon<1$.
This change was motivated by noting that the likelihood
of this parameter peaked at zero in the BK14 analysis and
following the philosophy of ``allowing the data to select
the model it prefers so long as this does not result in bias on $r$.''
While we are not aware of any theoretical motivation to consider
negative values, anti-correlation is presumably physically
possible.
Empirical evidence is sparse;
Ref.~\cite{choi15} reports a correlation of 0.2 for $30<\ell<200$,
but the most recent \planck\ analysis detects (positive) sync/dust correlation
only for $\ell < 50$~\cite{planckiLIV}.

\begin{figure*}[htb]
\begin{center}
\resizebox{1.0\textwidth}{!}{\includegraphics{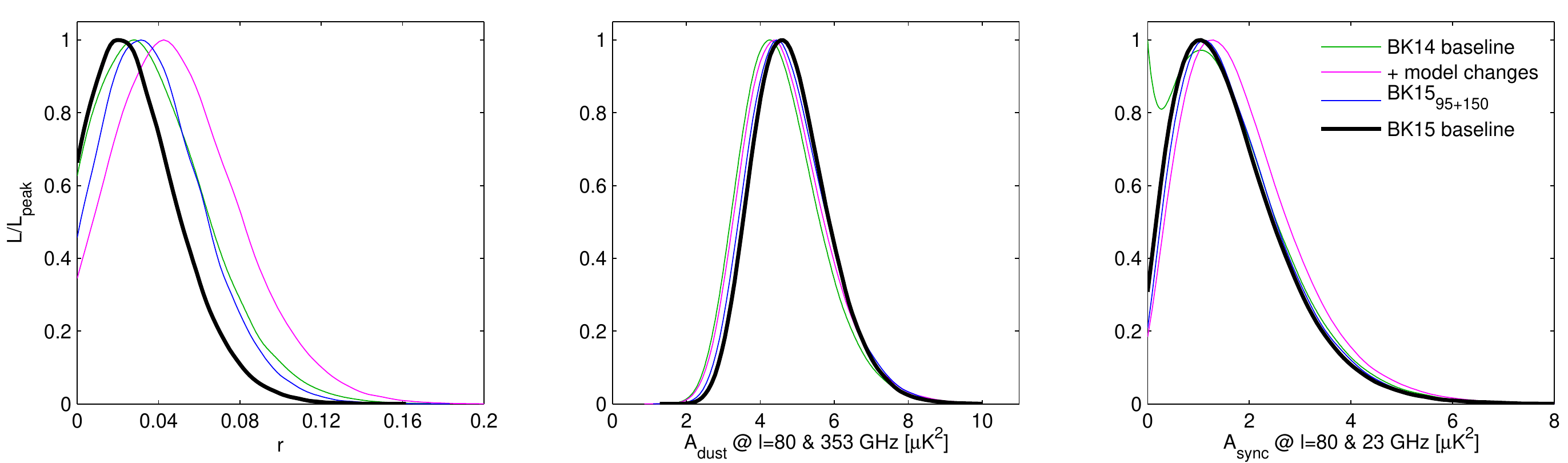}}
\end{center}
\caption{Evolution of the BK14 analysis to the ``baseline'' analysis as
defined in this paper---see Appendix~\ref{app:likeevol} for details.}
\label{fig:likeevol}
\end{figure*}

Fig.~\ref{fig:likeevol} shows the sequence of steps from
the BK14 baseline analysis to the new baseline.
Changing the $\epsilon$ marginalization range results in the
change from green to magenta.
Adding the 2015 data at 95 \& 150\,GHz causes the change
from magenta to blue.
Finally adding the new 220\,GHz band results in the change
from blue to black.
The net result is a narrowing of the $r$ likelihood curve
and a slight downward shift in the peak position.
Note that we made the choice to change the $\epsilon$
prior based on the considerations above, and
before looking at these real data results.

\subsection{Likelihood Variation}
\label{app:likevar}

\begin{figure*}[htb]
\begin{center}
\resizebox{1.0\textwidth}{!}{\includegraphics{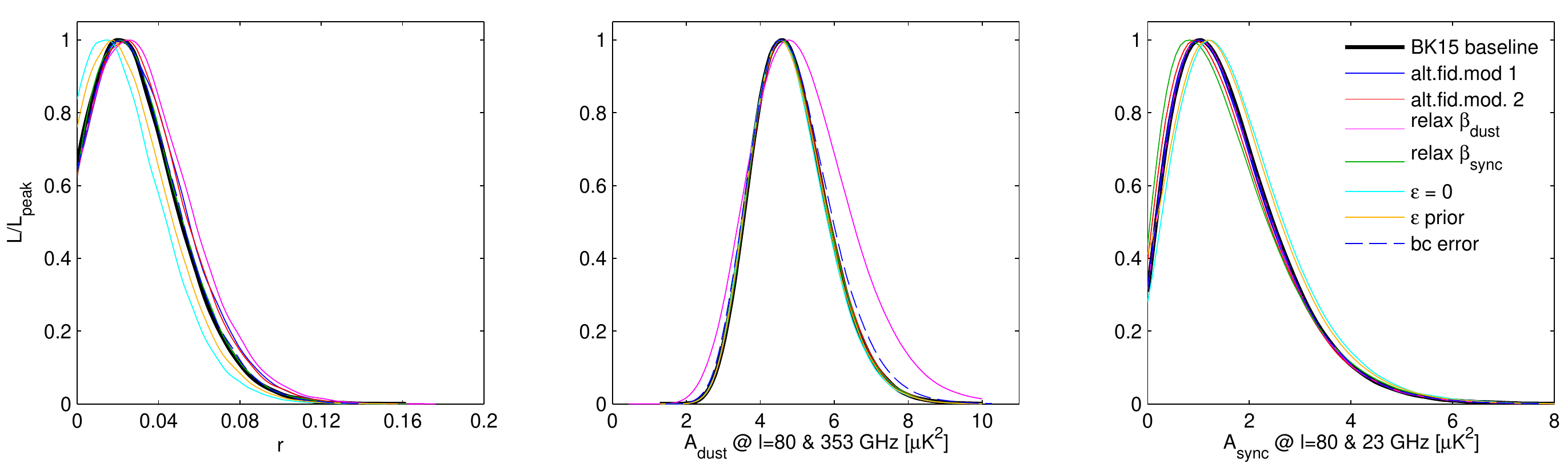}}
\end{center}
\caption{Likelihood results when varying the analysis choices---see
Appendix~\ref{app:likevar} for details.}
\label{fig:likevar}
\end{figure*}

Fig.~\ref{fig:likevar} shows some variations from the baseline
analysis choices.
The HL likelihood~\citep{hamimeche08} requires that one provide a
``fiducial model'', but it is not supposed to matter very much what this
model is so long as it is reasonably close to reality.
Since the BKP paper we have used $\Adf=3.6$\,\uksq, $\As=0$, $r=0$.
Switching to $\Adf=5$\,\uksq, $\As=0$, $r=0.05$ (blue) or
$\Adf=5$\,\uksq, $\Asf=2$\,\uksq, $r=0.05$ (red) makes little
difference.

Since BKP our baseline analysis has used a prior on the frequency spectral index
of dust of $\Bd=1.59 \pm 0.11$, using a Gaussian prior
with the given $1\sigma$ width.
These numbers are based on external information from
\planck~\cite{planckiXXII,bkp} derived from other regions of the sky.
In BK14 removing this prior resulted in a significant upshift in
the $r$ constraint curve and a shift and broadening of the $\Ad$
curve.
However, with the addition of the new \keck\ 220\,GHz data we are
now able to constrain $\Bd$ sufficiently well that changes
when removing this prior are small (black to magenta).
The $\Bd$ constraint curve (not shown) is close to Gaussian in
shape with mean/$\sigma$ of 1.65/0.20.
With further improvements in the data in the future we will no longer
need the $\Bd$ prior and hence will be able to remove the uncertainty
that comes from assuming that dust behavior in our sky patch is the
same as the average behavior over larger regions of sky.

Our baseline prior on the frequency spectral index of synchrotron
is $\Bs=-3.1\pm0.3$~\citep{fuskeland14}, with a Gaussian shape
with the given $1\sigma$ width.
Relaxing to a uniform prior over the range $-4.5<\Bs<-2.0$ produces
no significant changes (black to green).
The data has little preference for the value of this parameter
within the allowed range, which is not surprising since non-zero
synchrotron amplitude is only weakly preferred.

Tightening the prior on the dust/sync correlation
parameter from the baseline $-1<\epsilon<1$ to
$\epsilon=0$ produces a small downshift in the $r$ constraint
curve (black to cyan), as expected given what we already saw in Fig.~\ref{fig:likeevol}.
We show this case as we will invoke it when adding dust decorrelation
to the model in Appendix~\ref{app:decorr} below.
Putting a Gaussian prior on the dust/sync correlation with mean/$\sigma$
of 0.48/0.50~\cite{planckiLIV} produces a smaller downshift in $r$
than setting $\epsilon=0$ (comparing yellow and cyan).

We explore the effect of uncertainty in the measured bandpasses for
\bk\ 95, 150 and 220~GHz channels.
We expect such difference to be small and parameterize it as a fractional
shift in the band center.
We include one parameter for each frequency plus a correlated shift applied
to all three channels.
For each parameter, we use a Gaussian prior with mean/$\sigma$ of 0/0.02.
These potential bandcenter shifts have little effect on the likelihood
(black to dashed blue).

\begin{figure*}[htb]
\begin{center}
\resizebox{1.0\textwidth}{!}{\includegraphics{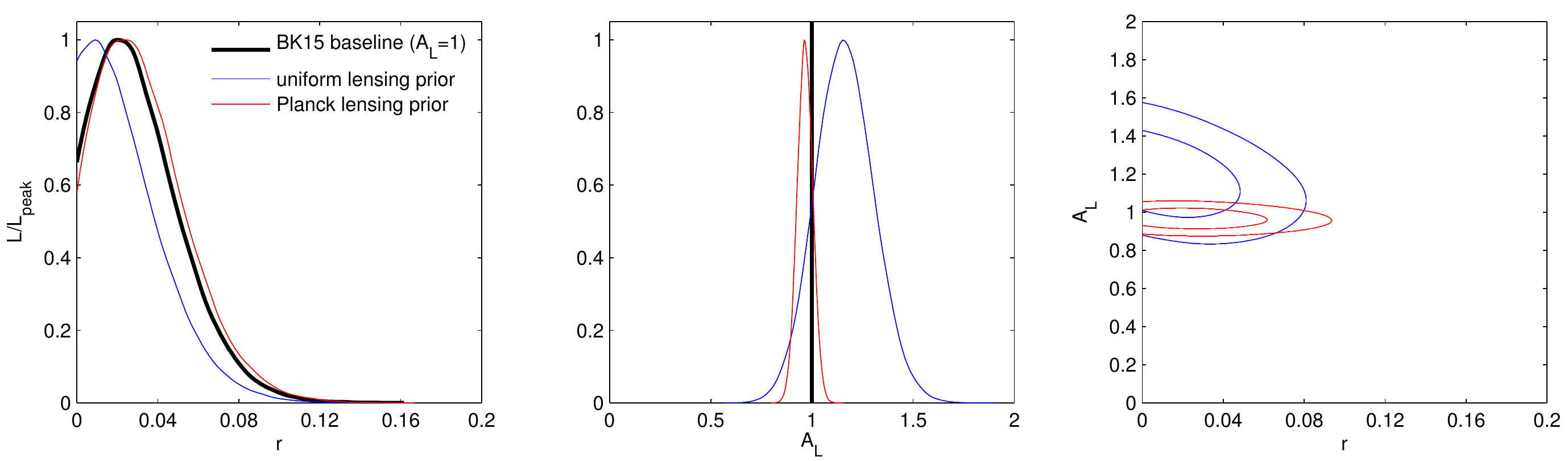}}
\end{center}
\caption{
Likelihood results when allowing the lensing amplitude
to be a free parameter---see Appendix~\ref{app:likevar} for details.}
\label{fig:freelens}
\end{figure*}

In the baseline analysis, the lensing amplitude is fixed to the \lcdm\ 
expected value ($A_{\rm L}^{\rm BB} =1$).
Relaxing this assumption we obtain the results shown in Fig.~\ref{fig:freelens}.
With a unifrom prior, and marginalizing over all other parameters,
we obtain $A_{\rm L}= 1.15^{+0.16}_{-0.14}$. 
The zero-to-peak likelihood ratio is $1.3 \times 10^{-17}$, 
and the probability to have a lower value is $5.8 \times 10^{-19}$, 
which corresponds to a $8.8 \sigma$ detection. 
This is the most significant detection of lensing using B-mode polarization to date. 
Due to the degeneracy between $r$ and $A_{\rm L}$, the $r$ likelihood curve shifts down. 
If we impose a prior from Planck, $A_{\rm L}=0.95\pm0.04$~\citep{planck2015XIII},
the recovered $r$ likelihood curve is almost indistinguishable from the baseline case. 

\begin{figure*}[htb]
\begin{center}
\resizebox{1.0\textwidth}{!}{\includegraphics{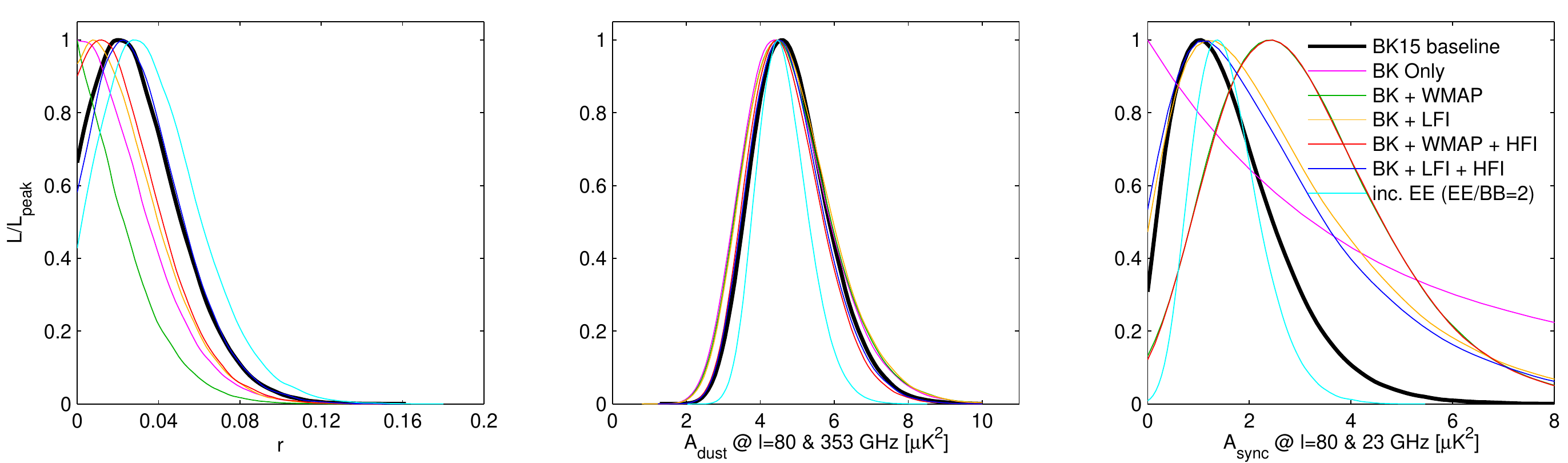}}
\end{center}
\caption{
Likelihood results when varying the data set selection---see
Appendix~\ref{app:likevar} for details.}
\label{fig:datavar}
\end{figure*}

Fig.~\ref{fig:datavar} shows some variations of the data set selection.
If we use the \biceptwo/\keck\ data only (magenta) the $r$ constraint
curve shifts down to peak at zero, while the $\Ad$ curve broadens
slightly, and much larger values of $\As$ become allowed.
Bringing back \wmap\ (green) produces an even stronger downshift in $r$,
and $\As$ becomes better constrained.
Switching LFI for \wmap\ (green to yellow) brings $r$ back up a bit and $\As$ down
(note the internal consistency problems of the LFI maps~\cite{planck2015II}). % REFB
Adding HFI to \bicep/\keck+\wmap\ (green to red) brings
$r$ up and leaves $\As$ unchanged.
\bicep/\keck+\planck\ (blue) has almost exactly the same $r$ curve
as the baseline but a considerably wider $\As$ curve.
We can understand the behaviors in the $\As$ curves,
at least in part, by noting that in Fig.~\ref{fig:powspec_all} the
BK$_{95}\times$W$_{23}$ bandpowers are positive while the
BK$_{95}\times$P$_{30}$ bandpowers are negative.

One additional variation which we explore is to include the
$EE$ spectra (and hence also $EB$) in the fit under the
assumption that $EE/BB=2$ for dust and synchrotron, as is shown to
be close to the case in Refs~\cite{planckiLIV} and~\cite{krachmalnicoff18}.
(While we have not included the $EE$ jackknife tests in this, or previous, % REFB
papers they also produce distributions of $\chi$ and $\chi^2$ PTE values
which are consistent with uniform.)
As we can see in Fig.~\ref{fig:powspecres_bkbands}, $EE$ spectra
such as BK$_{220}\times$P$_{353}$ and P$_{353}\times$P$_{353}$
certainly carry information on the amplitude of the dust
emission and can presumably help indirectly to constrain $r$.
In Fig.~\ref{fig:datavar} adding $EE$ results
in a small upshift in $r$ and significant tightening
of the constraints on $\Ad$ and $\As$.
We will consider adding $EE$ to the baseline in future
analyses, marginalizing over some range in the $EE/BB$ ratios.

At first glance it may appear surprising how large the shifts
in the $r$ constraint are under the variations of the data
selection shown in Fig.~\ref{fig:datavar}, and that many of
the shifts are downward.
However, when viewing the equivalent plots for the standard
lensed-\lcdm+dust+noise simulation realizations---which contain
no tension between the data sets---the qualitative impression
in many cases is similar.
Note that while we verify in the next section that the
baseline $r$ constraint is unbiased, we have not tested this
for the data set variations explored here.

\subsection{Likelihood Validation}
\label{app:likevalid}

\begin{figure}
\resizebox{\columnwidth}{!}{\includegraphics{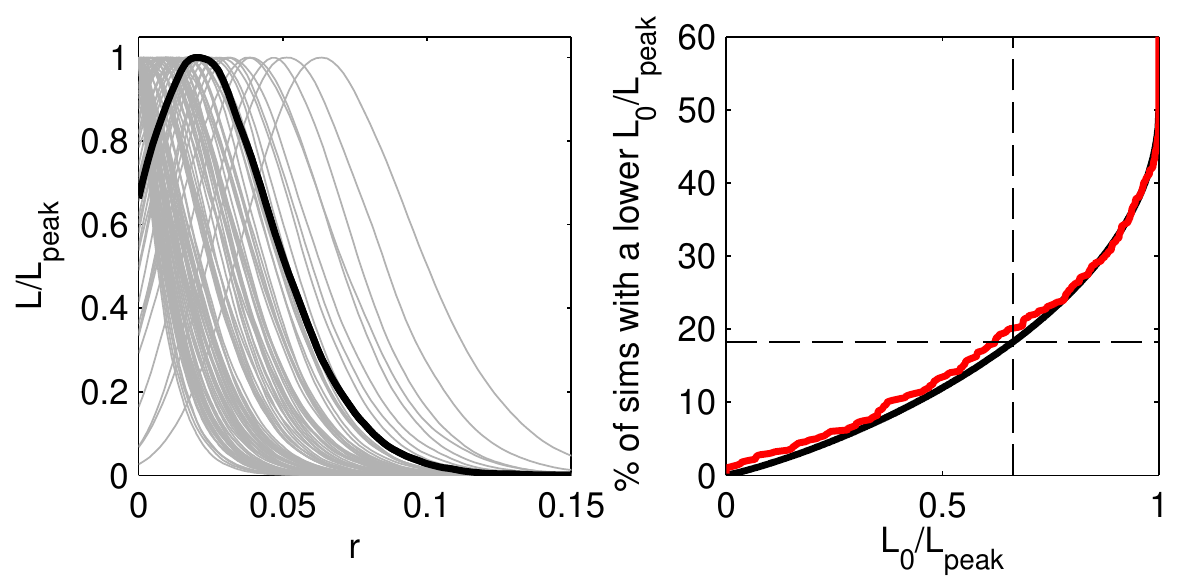}}
\caption{
{\it Left:} Likelihood curves for $r$ when running the baseline
analysis on each of the lensed-\lcdm+dust+noise simulations---half of
them peak at zero.
The real data curve is shown overplotted in heavy black.
{\it Right:} The CDF of the zero-to-peak ratio (red) of the curves shown
at right as compared to the simple analytic ansatz (solid black)
$\frac{1}{2} \left( 1-f \left( -2\log{L_0/L_{\rm peak}} \right) \right)$
where $f$ is the $\chi^2$ CDF (for one degree of freedom).
About one fifth of the simulations offer more evidence for non-zero
$r$ than the real data when the true value is actually zero (dashed black).}
\label{fig:simcons}
\end{figure}

The interpretation of $r$ likelihood curves such as the one
shown in the upper left panel of Fig.~\ref{fig:likebase}
is not necessarily straightforward.
Since the parameters are restricted to, and marginalized over,
physical values only, biases can result.
For instance, in a scenario where two parameters are fully
degenerate, power will be assigned on average equally
between them, and both will be biased low, with the curves for
greater than 50\% of realizations peaking at zero when the
true values are zero.
To investigate we make full \texttt{COSMOMC} runs on the
ensemble of lensed-\lcdm+dust+noise simulations.
The left panel of Fig.~\ref{fig:simcons} shows the resulting $r$ constraint
curves, while the right panel shows that 
the CDF of the zero-to-peak likelihood ratios
closely follows the simple analytic ansatz
$\frac{1}{2} \left( 1-f \left( -2\log{L_0/L_{\rm peak}} \right) \right)$
where $f$ is the $\chi^2$ CDF (for one degree of freedom).
We find that 53\% of the simulations peak at zero,
and 19\% have a lower zero-to-peak ratio than the real data---i.e.\ 
show more evidence for $r$ when the true value is in fact zero.
This study provides powerful empirical evidence that the real
data $r$ constraint curve can be taken at face value, provided
the assumed foreground parameterization is an adequate
description of reality.

\begin{figure*}[htb]
\begin{center}
\resizebox{0.8\textwidth}{!}{\includegraphics{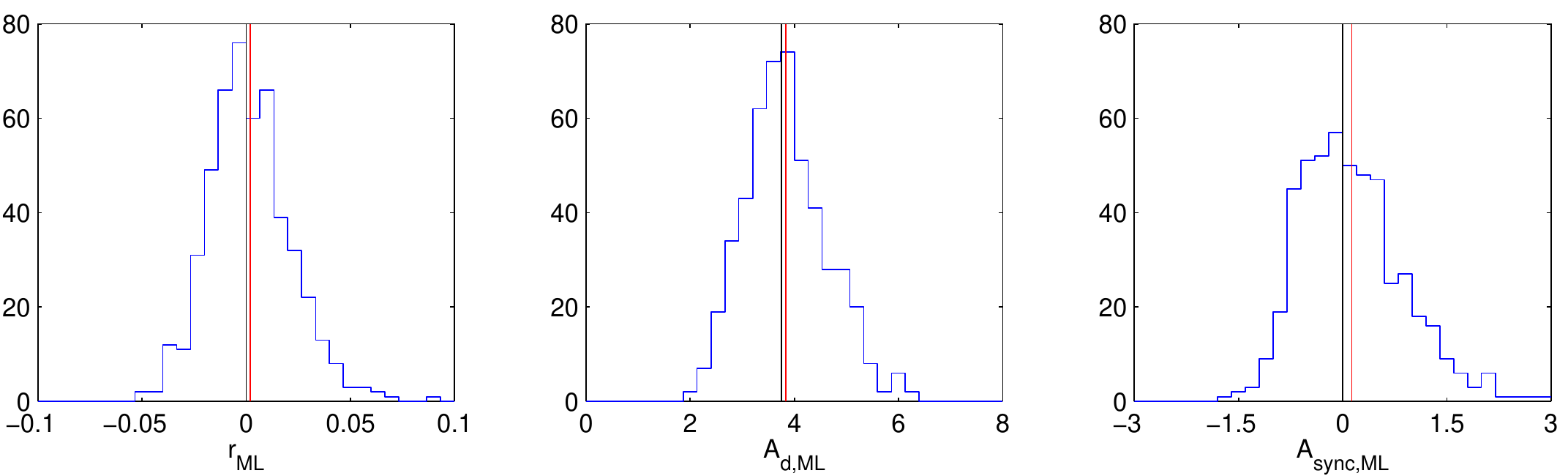}}
\end{center}
\caption{
Results of a validation test running maximum likelihood
search on simulations of a lensed-\lcdm+dust+noise model
with no synchrotron
($\Adf=3.75$\,\uksq, $\Bd=1.6$, $\ad=-0.4$, $\As=0$).
The baseline priors are applied on $\Bd$, $\Bs$, $\ad$, $\as$ and $\epsilon$.
The blue histograms are the recovered maximum likelihood values with the
red lines marking their means and the black lines showing
the input values.
In the left panel $\sigma(r)=0.020$.
See Appendix~\ref{app:likevalid} for details.}
\label{fig:likevalid}
\end{figure*}

An alternate (and much faster) likelihood validation
exercise is to run maximum likelihood searches,
with non-physical parameter values allowed (such as negative $r$).
When running on simulations generated
according to the model being re-fit, we then
have an a priori expectation that the input parameter values
should be recovered in the mean.
Fig.~\ref{fig:likevalid} shows the results when running
on the standard lensed-\lcdm+dust+noise simulations,
with the same priors as for the baseline analysis---the input values
are recovered in the mean.
The first row of Table~\ref{tab:altfgmod}
summarizes: $\sigma(r)=0.020$, and bias in the mean value
is small as compared to the noise.
We prefer this $\sigma(r)$ measure of the intrinsic constraining
power of the experiment since it is independent of the particular
noise fluctuation that is present in the real data.

\subsection{Exploration of Alternate Foreground Models}
\label{app:altfore}

We now extend the maximum likelihood validation study to
simulations using third-party foreground models.
These models do not necessarily conform to our foreground
parameterization and therefore when fit to it may potentially
produce bias in $r$ at levels relevant compared to the noise.
The second and subsequent rows of Table~\ref{tab:altfgmod} summarize
the results.
The third-party models provide only a single realization
of the foreground sky, and we add it on top of
each of the lensed-\lcdm+noise realizations that are used in
the standard simulations.

\newlength{\x}
\settowidth{\x}{+}
\begingroup
\squeezetable
\begin{table}[pht]
\caption{\label{tab:altfgmod}
Uncertainty and bias on $r$
in simulations using Gaussian and 3rd party foreground models.
(The numbers in parentheses suffer from disagreement
between the priors and the model so bias is expected---see text for details.)}
\begin{ruledtabular}
\begin{tabular}{l r r c c c}
      & \multicolumn{1}{c}{$\overline{A_d}$} & \multicolumn{1}{c}{$\overline{A_s}$} & \multicolumn{3}{c}{$\sigma(r)$, $\overline{r}/\sigma(r)$} \\
Model & (\uksq) & (\uksq) & $\Bd$ prior & $\Bd$ free & with decorr. \\
\hline
Gaussian    &  3.8 & 0.1 &  0.020, \makebox[\x] {+}0.1$\sigma$  &   0.023, \makebox[\x] { }0.0$\sigma$  &  0.021, \makebox[\x] {+}0.0$\sigma$  \\
PySM 1      & 10.9 & 1.1 &  0.026, \makebox[\x] {+}0.2$\sigma$  &   0.028, \makebox[\x] {+}0.2$\sigma$  &  0.028, \makebox[\x] {+}0.1$\sigma$  \\
PySM 2      & 24.2 & 0.9 &  0.028, \makebox[\x] {+}0.1$\sigma$  &   0.029, \makebox[\x] {+}0.1$\sigma$  &  0.032, \makebox[\x] {+}0.1$\sigma$  \\
PySM 3      & 12.1 & 1.1 & (0.030, \makebox[\x] {+}0.4$\sigma$) &   0.031, \makebox[\x] {+}0.1$\sigma$  & (0.032, \makebox[\x] {+}0.2$\sigma$) \\
MHDv2       &  2.9 & 5.6 &  0.020, \makebox[\x] {+}0.2$\sigma$  &   0.027, \makebox[\x]{--}0.2$\sigma$  &  0.021, \makebox[\x]{--}0.1$\sigma$  \\
G.\ Decorr. &  4.6 & 0.1 & (0.023, \makebox[\x] {+}1.5$\sigma$) &  (0.026, \makebox[\x] {+}1.3$\sigma$) &  0.022, \makebox[\x] {+}0.0$\sigma$  \\
\end{tabular}
\end{ruledtabular}
\end{table}
\endgroup

The PySM models 1, 2 and 3 are a1d1f1s1, a2d4f1s3 and a2d7f1s3
respectively, with the letters indicating AME (a), dust (d),
free-free (f) \& synchrotron (s), and the numbers referring to the various
models of each as described in the PySM paper~\citep{thorne17}.
The a1 and f1 models are unpolarized and hence not relevant.
The a2 model uses a \planck\ {\tt Commander}~\cite{planck2015X} derived template
and (dust) polarization
angles together with a conservative 2\% polarization fraction.
No account for AME is made in our
parametric model so this could potentially result in bias.
The d1 model again uses \planck\ {\tt Commander} derived templates
for both the 353\,GHz $Q/U$ patterns and the $T_d$ and $\Bd$
spectral parameters.
The dust SED thus varies spatially, and this model therefore
implements decorrelation of the dust pattern at some level (which
in practice is found to be very small).
Model d4 generalizes model d1 to the two temperature FDS
model~\citep{finkbeiner99}.
Model d7 is a sophisticated physical model of dust grains as described
in Ref.~\cite{hensley2015} which does not necessarily conform
to the modified blackbody SED.
The s1 model takes the \wmap\ 23\,GHz $Q/U$ maps and rescales
them according to a power law using a spectral index map,
and the s3 model adds on top of this a (spatially uniform) curvature
of the synchrotron SED.
The \wmap\ and \planck\ polarization templates are all
noise dominated at smaller angular scales, so PySM
filters out this noise and fills back in Gaussian
realizations of foreground structure according to the recipe
described in Sec. 3.1 of the PySM paper~\citep{thorne17}.

We see in Table~\ref{tab:altfgmod} that the PySM models
predict considerably higher dust power in the \bk\ field
than is actually observed and that this pushes up $\sigma(r)$
somewhat as compared to the Gaussian results.
The dust amplitude is sufficiently high in these models that
$\Bd$ becomes well constrained for the noise levels
and frequency range of the BK15 data---the prior on
$\Bd$ can therefore be relaxed, and this is actually
necessary for the PySM~3 model where the value
of $\Bd$ preferred by the model is outside of the prior
range, and bias on $r$ results if the prior is not relaxed.

The model labeled ``MHDv2'' is based on simulations
of the Galactic magnetic field~\citep{kritsuk17} and
naturally produces correlated dust and synchrotron emission.
Since this model contains no explicit experimental data
there is no noise issue, and the generated structure
is non-Gaussian across the full range of $\ell$.
This model gives a higher
level of synchrotron than that which is preferred by the
\bk\ data ($\Asf=5.6$\,\uksq\ as compared to the maximum likelihood
value of $\Asmlm$\,\uksq\ and 95\% upper limit of $\Asf<\Asul$\,\uksq).
This model also produces bias in the mean value of $r$
that is small compared to the noise level.

We conclude that none of the considered models produces
relevant bias on $r$ when fitted to our foreground
parameterization for the current experimental noise levels.
These models span a variety of assumptions and methods and
in some cases predict levels of foreground contamination
much stronger than we actually observe in our field.
However, there is no guarantee that the real foregrounds
do not in fact produce greater bias than any of the
considered models.
We note that all of the above models produce dust decorrelation
that is negligibly small compared to the current noise
levels.

\section{Adding dust decorrelation}
\label{app:decorr}

The simplest possible model of a given component of the polarized
foreground emission (e.g.\ dust or synchrotron) is
that it presents a fixed spatial pattern on the sky which
scales with frequency according to a single SED.
In this case the cross-spectrum between any two given
frequencies is simply the geometric mean of the respective
auto-spectra.
In reality the morphology of the polarization pattern will
inevitably vary as a function of frequency at some level.
If $Q$ and $U$ at each given point on the sky deviate in sympathy away
from the mean SED then the polarized intensity map will evolve
as a function of frequency, but the polarization angles will
remain constant.
If $Q$ and $U$ deviate independently from the mean SED
then both polarization intensity and angle will be functions of
observing frequency.
In either case the cross-spectra will be suppressed with
respect to the geometric mean of the auto-spectra---a phenomenon
which we refer to as decorrelation.

\planck\ Intermediate Paper~XXX~\citep{planckiXXX}
looked for suppression of the cross-spectral amplitudes in
Figs.~6 \& E.1 and did not find any evidence for decorrelation.
However, that analysis was implicitly weighted towards lower $\ell$.
Later \planck\ Intermediate Paper~L~\citep[hereafter PIPL]{planckiL}
examined the cross-spectrum
between 220 \& 353\,GHz as a function of $\ell$ and found
evidence for a suppression effect which increased
with $\ell$ and also when going to cleaner regions of sky
(as determined by neutral hydrogen column density---see Fig.~3
of that paper).
More recently, Ref.~\citep{sheehy17} re-analyzed the now public
\planck\ data and found no evidence for a detection
of dust decorrelation.
Finally the \planck\ team revisited the issue again in
\planck\ Intermediate Paper~LIV~\citep[hereafter PIPLIV]{planckiLIV}
and this time state that
``We find no evidence for a loss of correlation.''

Decorrelation certainly exists at some level---the question
is whether that level is relevant as compared to the
current instrumental noise.
To search for evidence of decorrelation in the BK15 data
we add decorrelation of the dust pattern to our parametric model.
We define the correlation ratio of the dust
\begin{equation}
\dd =
\frac{\mathcal{D}_{80}(217\times 353)}{\sqrt{\mathcal{D}_{80}(217\times 217)\mathcal{D}_{80}(353\times 353)}},
\label{eqn:Delta}
\end{equation}
where $\mathcal{D}_{80}$ is the dust power at $\ell=80$.
This makes $\dd$ close to equivalent to $\mathcal{R}_{80}^{BB}$ as defined
by PIPL and PIPLIV.
We scale to other frequency combinations using the factor
\begin{equation}
f(\nu_1,\nu_2) = \frac{(\log (\nu_1/\nu_2))^2}{(\log(217/353))^2},
\label{eqn:nuscale}
\end{equation}
as suggested by PIPL.

Fig.~2 of PIPL suggests that decorrelation grows with increasing $\ell$,
although in Sec.~4 they assume flat with $\ell$.
In this paper we consider two possible scalings
\begin{equation}
g(\ell) = \left\{ \begin{array}{ll}
                   1             & \, \text{flat case} \\
                   (\ell / 80)   & \, \text{linear case} %\\
                   %(\ell / 80)^2 & \, \text{quadratic case}
                   \end{array} \right..
\label{eqn:ellscale}
\end{equation}
Since the $\ell$ range we are concerned with is not broad this
choice turns out to make little practical difference.

The above scalings can produce extreme, and non-physical,
behavior for widely separated frequencies and low/high $\ell$.
We therefore re-map the nominal value using the following function
\begin{equation}
\ddp(\nu_1,\nu_2,\ell) = \exp \left[ \log(\dd) \, f(\nu_1,\nu_2) \, g(\ell) \right],
\label{eqn:remap}
\end{equation}
such that $\ddp$ remains in the
range 0 to 1 for all values of $f$ and $g$.
We note that for the frequency scaling this becomes
the same as Eqn.~14 of Ref.~\citep{vansyngel16} which
is shown in that paper to correspond to a Gaussian spatial
variation in the foreground spectral index.
(This is also used in PIPLIV.)
For the moment we defer consideration of models which have
both decorrelation of the dust pattern and correlation
of the dust and synchrotron patterns simultaneously, setting
$\epsilon=0$ whenever we allow $\dd \neq 1$.
Note that in Fig.~\ref{fig:likevar} we see that 
setting $\epsilon=0$ produces only small changes
from the baseline analysis.

Fig.~\ref{fig:powspecres_bkbands} shows the power spectra
of the frequency bands which have the most power to constrain
the dust contribution to the model.
We can see visually that the (non-decorrelated) model from our
previous BK14 analysis which is plotted there appears to be a good
explanation of the observations (and in Appendix~\ref{app:allspec}
it was shown formally that the new BK15 maximum likelihood model
is compatible with the data).
PIPL states that the mean neutral hydrogen column density in the \biceptwo/\keck\
field is $\sim 1.6\times10^{20}$\,cm$^{-2}$ for which their Eqn.~6
gives a predicted correlation ratio value
$\mathcal{R}_{50-160}^{BB}(217,353)=0.83$.
To illustrate the effect of decorrelation in
Fig.~\ref{fig:powspecres_bkbands} we also re-plot the
BK14 model modified with $\dd=0.85$
as the dashed red lines---this leaves the auto-spectra unchanged while
suppressing the cross-spectra.
The $150\times353$ data appears to weakly disfavor the change
while the $95\times353$ weakly favors it.
The above is simply for the purposes of illustration---we
proceed below to include decorrelation and re-fit the model.

\begin{figure*}[htb]
\begin{center}
\resizebox{1.0\textwidth}{!}{\includegraphics{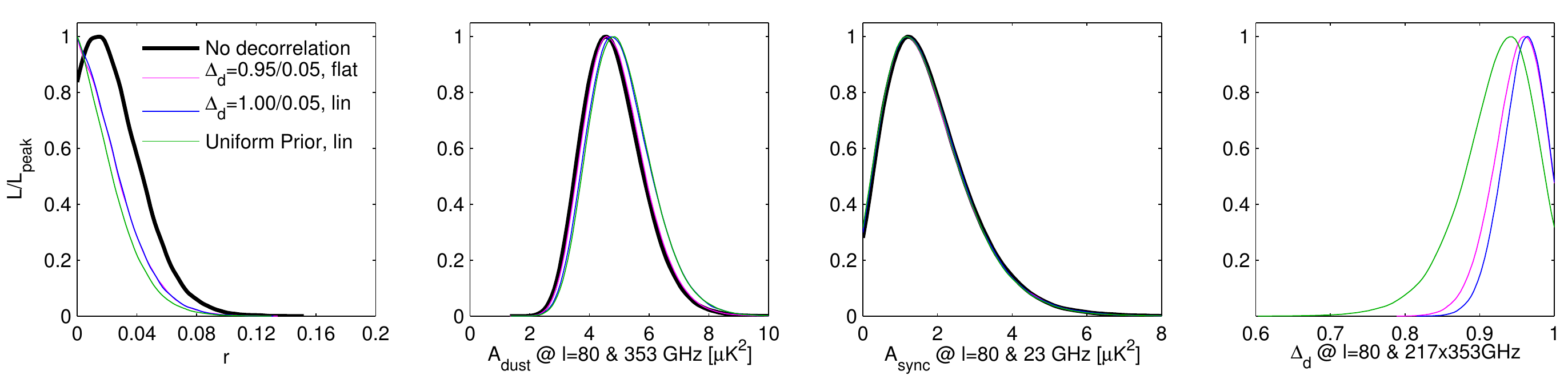}}
\end{center}
\caption{
Likelihood results when allowing dust decorrelation---see
Appendix~\ref{app:decorr} for details.}
\label{fig:likedecorr}
\end{figure*}

We expand the baseline likelihood analysis to include
decorrelation and show results in Fig.~\ref{fig:likedecorr}.
We consider several choices of prior on the 
$\dd$ parameter:
i) Based on Table~1 of Ref.~\citep{sheehy17} and Table~3 of Ref.~\cite{planckiLIV} we set a Gaussian
prior with mean/$\sigma$ of 0.95/0.05 (truncated above 1), flat with $\ell$. % REFB
ii) A Gaussian prior with mean/$\sigma$
of 1.00/0.05, linear with $\ell$.
iii) A uniform prior 0 to 1, linear with $\ell$.
All of these choices result in the $r$ likelihood curve
shifting down and peaking at zero.
However, note that introducing $\dd$ in a likelihood analysis
which marginalizes only over the physically meaningful
range $\dd \leq 1$ can result in a downward bias
on $r$ even in the absence of a real decorrelation effect.
For a given set of bandpowers it is possible to explain
observed power in cross-spectra such as $150\times353$
with a higher value of $A_d$ in combination with a lower
value of $\dd$.
The auto-spectra resist this preventing strong degeneracy,
but a net bias still results.
When we repeat the exercise of Fig.~\ref{fig:simcons} running
the full analysis on the standard lensed-\lcdm+dust+noise simulations
(which do not contain decorrelation), but include the decorrelation
parameter in the analysis, we find that 72\% of the $r$ curves
peak at zero, and many of the $\dd$ curves peak below 1.
We therefore choose not to include the decorrelation parameter
in our baseline analysis at this time.

\begin{figure*}[htb]
\begin{center}
\resizebox{\textwidth}{!}{\includegraphics{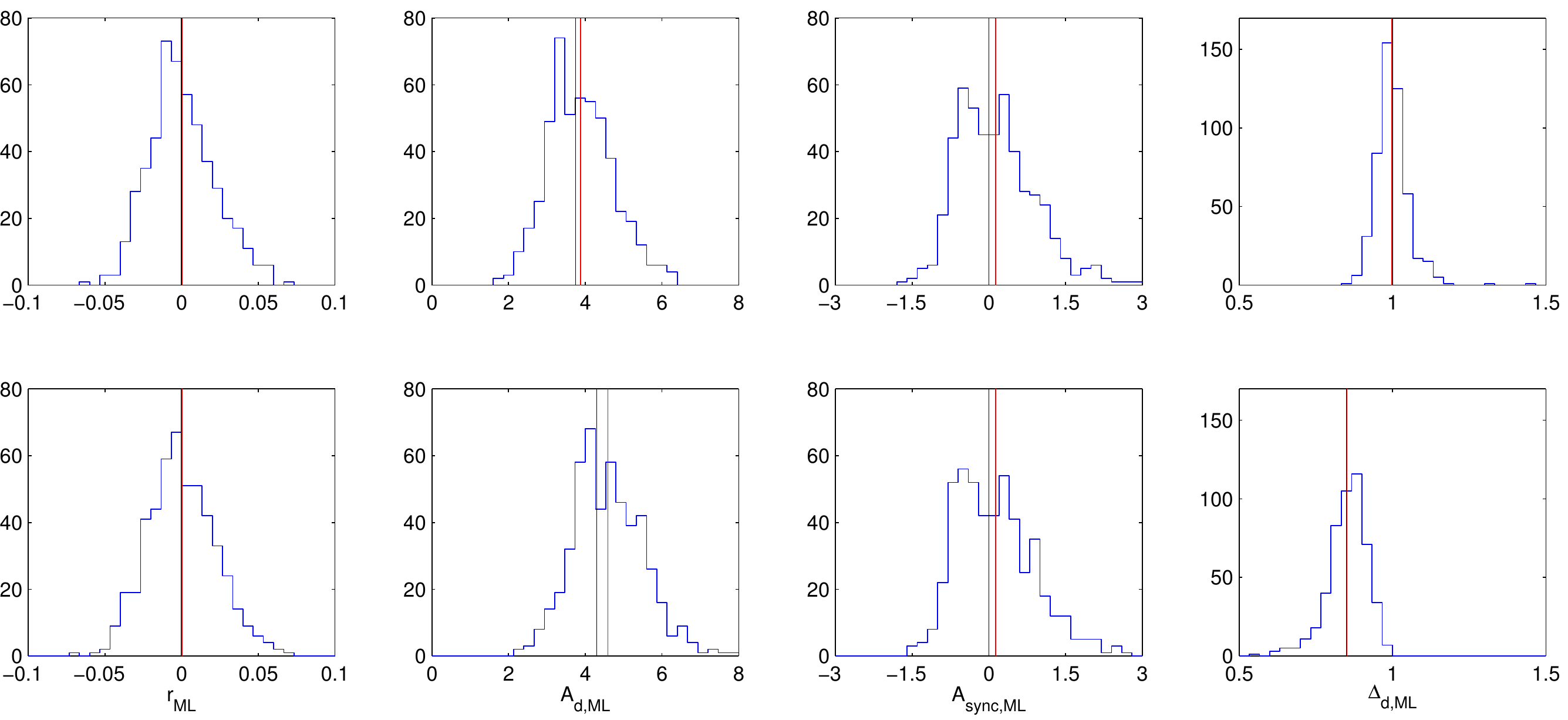}}
\end{center}
\caption{
Validation tests running the
likelihood with the dust decorrelation parameter
$\dd$ included.
{\it Upper row:} Results for the same lensed-\lcdm+dust+noise
simulations shown in Fig.~\ref{fig:likevalid}.
{\it Lower row:} Results for the toy highly decorrelated
dust model.
The blue histograms are the recovered maximum likelihood values with the
red lines marking their means and the black lines showing
the input values.
See Appendix~\ref{app:decorr} for details.}
\label{fig:likedecorrvalid}
\end{figure*}

To check that the machinery remains unbiased when running
maximum likelihood searches we repeat the exercise of Appendix~\ref{app:likevalid}
but this time including the decorrelation
parameter $\dd$ and allowing it to take values greater than
one.
To do this in a symmetrical manner we use
\begin{equation}
\ddp(\nu_1,\nu_2,\ell) = 2 - \exp \left[ \log(2-\dd) \, f(\nu_0,\nu_1) \, g(\ell) \right].
\end{equation}
In this exercise we take the linear $\ell$ scaling.
Fig.~\ref{fig:likedecorrvalid} shows the results for
the standard lensed-\lcdm+dust+noise simulations which
contain no decorrelation.
We see that $\dd=1$ is recovered, and $r$ remains
unbiased.
We also show results for a toy highly decorrelated model
which uses $\dd=0.85$ and linear scaling
with $\ell$, following Eqns.~\ref{eqn:nuscale}--\ref{eqn:remap}.
The input parameters of this model are also recovered
in the mean.
Finally we run the analysis with decorrelation
on the third-party foreground models and give results for
all the models in Table~\ref{tab:altfgmod}.
As expected we see that the decorrelated
simulations produce bias when re-analyzed without
allowing decorrelation in the model.

Running a maximum likelihood search including decorrelation
on the real data we obtain
$r_{0.05}=-0.012$, $\Adf=5.0$\,\uksq,
$\Asf=1.4$\,\uksq, 
$\Bd=1.6$, $\Bs=-3.1$,
$\ad=-0.38$, $\as=-0.52$, and $\dd=0.92$, i.e.\ $\Ad$
shifts up a bit, $r$ shifts down a bit, and $\dd$ is a little
less than one.
This model has a $\chi^2$ versus the data of \chitwodc\ to be compared
to the baseline model value of \chitwo---the data shows little
evidence for decorrelation of the dust pattern.
As the data improves in the future the ability to constrain decorrelation
while remaining unbiased on $r$ will improve.

\section{Definition of Multicomponent Model}
\label{app:multicomp}

The likelihood analysis uses a parametrized model to describe the bandpower
expectation values as a combination of cosmological and foreground signals.
The form of this model remains unchanged from BKP and \BKfourteen\ except
for the addition of foreground decorrelation (described in Appendix~\ref{app:decorr}). 
However, the choice of free parameters and priors has evolved over time
due to improved \bk\ data and new information from external sources.
The previous papers describe the important features of the model but
do not include a complete mathematical formulation, which we provide
here.

Equation~\ref{eq:BB_model} describes contributions to the $BB$ cross-spectrum
between maps at frequencies $\nu_1$ and $\nu_2$ (or auto-spectrum, if $\nu_1 = \nu_2$) from dust,
synchrotron, and the spatially-correlated component of dust and synchrotron.
Parameter $\Ad$ specifies the dust power in units of $\mu\mathrm{K}_\mathrm{\mbox{\tiny\sc cmb}}^2$ at
pivot frequency 353~GHz and angular scale $\ell = 80$.
Parameter $\As$ specifies synchrotron power similarly, except with a pivot
frequency of 23~GHz.
The dust and synchrotron components scale as power laws in $\ell$ with
slopes $\ad$ and $\as$, respectively.
Note that we define parameters $\ad$ and $\as$ as the $\ell$ scaling of
$\mathcal{D}_\ell \equiv \clstar$,
not \cl.

The level of spatial correlation between dust and synchrotron is set by
parameter $\epsilon$.
The correlated component scales with $\ell$ with a slope that is the average
of $\ad$ and $\as$, meaning that the correlation coefficient is
assumed to be constant across all $\ell$.

Parameter $\ddp$ accounts for decorrelation of the dust pattern between
$\nu_1$ and $\nu_2$ and is defined in equation \ref{eqn:remap}.
Note that if $\nu_1 = \nu_2$, then $\ddp = 1$ (perfect correlation).
Parameter $\Delta'_\mathrm{s}$ describes decorrelation of the synchrotron
pattern but is not currently used.
We currently do not include foreground decorrelation parameters in the
dust--synchrotron correlated component.
A complete foreground model would include the full set of correlations between
dust and synchrotron fields at $\nu_1$ and the dust and synchrotron fields
at $\nu_2$, but current data offer no guidance about the form of these correlations.
For the time being, we consider dust decorrelation only as an extension to
models with $\epsilon = 0$, as noted in Appendix~\ref{app:decorr}.

Additional coefficients $f_\mathrm{d}$ and $f_\mathrm{s}$ capture the scaling of dust and
synchrotron power from the pivot frequencies to the actual bandpasses of the
maps labeled $\nu_1$ and $\nu_2$.
This scaling includes the foreground SED as well as the conversion between
\ukcmb\ units at the pivot frequency and at the target map
bandpass.
The SED model used for dust is a blackbody with temperature $T_\mathrm{d} = 19.6 \mathrm{K}$
multiplied by a power law with emissivity spectral index $\Bd$ \cite{planckiXXII}.
The SED model used for synchrotron is a power law with spectral index $\Bs$
defined relative to a Rayleigh-Jeans spectrum.
When integrating the SED and unit conversion factors over a map bandpass
it is necessary to choose a bandpass convention.
We define our bandpass functions to be proportional to the response as a
function of frequency to a beam-filling source with uniform spectral
radiance (the same convention as used by \planck~\cite{planck2013IX}).

\begin{widetext}
  \begin{equation}
    \mathcal{D}_{\ell,BB}^{\nu_1 \times \nu_2} =
    \Ad \ddp f_\mathrm{d}^{\nu_1} f_\mathrm{d}^{\nu_2} \left( \frac{\ell}{80} \right)^{\ad} + 
    \As \Delta'_\mathrm{s} f_\mathrm{s}^{\nu_1} f_\mathrm{s}^{\nu_2} \left( \frac{\ell}{80} \right)^{\as} +
    \epsilon \sqrt{\Ad\As} (f_\mathrm{d}^{\nu_1} f_\mathrm{s}^{\nu_2} + f_\mathrm{s}^{\nu_1} f_\mathrm{d}^{\nu_2}) \left( \frac{\ell}{80} \right)^{(\ad + \as) / 2}
    \label{eq:BB_model}
  \end{equation}
\end{widetext}

The foreground contribution to $EE$ is similar, except that $\Ad$ and $\As$
are scaled by the $EE/BB$ ratios for dust and synchrotron, respectively,
which are both assumed to be equal to 2~\cite{planckiLIV,krachmalnicoff18}.
The model for the $EB$ spectrum is zero, since neither CMB nor foreground
signals are expected to break parity symmetry.
We do not model the unpolarized foregrounds, nor include $TT/TE/TB$ spectra
in the likelihood analysis.

\section{Summary of Simulations}
\label{app:simsum}

We interpret the single realization of real data through comparison to several
sets of simulations.
With the exception of the alternate foreground models mentioned in
Appendix~\ref{app:altfore} above these have all been described
and used in our previous papers~\cite{biceptwoI,bkp,biceptwoVI}.

We start by generating 499 pseudosimulations of noise by the sign-flip
technique~\cite{biceptwoI,vanengelen12}.
During the addition of multiple data subsets to form the final map
we randomly flip the signs to cancel out sky signal.
Each sequence is constructed to have equal weight in positives and
negatives, and since the sequences are $>10^4$ in length the resulting
noise realizations are found empirically to be uncorrelated.
The mean spectra of these noise simulations are used to debias
the real spectra (this being very important for the auto-spectra).

We also generate 499 realizations of lensed and unlensed \lcdm\
by resampling timestream from simulated input maps and passing
it through the full analysis pipeline (including
filtering etc.)~\cite{biceptwoI}.
The unlensed simulations are useful to empirically determine the
purity delivered by the matrix purification algorithm which is
used to extract the \bmode\ signal in the presence of a much
stronger \emode.

From the simulated signal-cross-signal, noise-cross-noise and signal-cross-noise
spectra we can construct the bandpower covariance matrix appropriate
for any model containing a set of signal components with given
SEDs~\cite{bkp}.
When we do this we set to zero any term which has an expectation
value of zero (under the assumption that signal and noise are uncorrelated)
to reduce the Monte Carlo error in the resulting covariance
matrix given the relatively modest number of realizations.
We also set to zero the covariance between bandpowers
that are separated by more than one bin in $\ell$, but, importantly,
preserve the covariance between the
the auto- and cross-spectra of the different frequency bands.
This covariance matrix construction is used for the HL likelihood, and also
to provide bandpower uncertainties shown, for example, in Fig.~\ref{fig:powspec_all}.

We also explicitly simulate simple dust input maps as power-law
Gaussian realizations (with amplitude set to the observed dust amplitude
in the \bicep/\keck\ patch) and pass these through the timestream sampling
and pipeline re-mapping operation.
They are then added to the lensed-LCDM and noise maps, and taken
through to power spectra.
We use these when it is important to match the fluctuations present
in the real data in detail.
One example is in the spectral stability tests shown in
Fig.~\ref{fig:specjack_95}.
Another example is when determining the PTE of the real data
$\chi^2$ value in Appendix~\ref{app:allspec}.

\section{Lensing analysis}
\label{app:lensing}

In Ref.~\cite{biceptwoVIII}, we showed a detection of 
the gravitational lensing signal using the BK14 $E$- and $B$-modes at 150\,GHz. 
We showed that the lensing signal is 
consistent with the standard $\Lambda$CDM model, and the BK14 $B$-mode 
spectrum at intermediate scales is dominated by lensing. 

At 150\,GHz, the sensitivity of BK15 to lensing is almost 
the same as that of BK14. 
Reconstructed lensing maps at 95\,GHz and 220\,GHz are still noisy.
However, reconstructing lensing signals from BK15 data is important 
to test consistency of the data and simulation. 

\begin{figure}[h]
\begin{center}
\vspace{12pt}
\resizebox{\columnwidth}{!}{\includegraphics{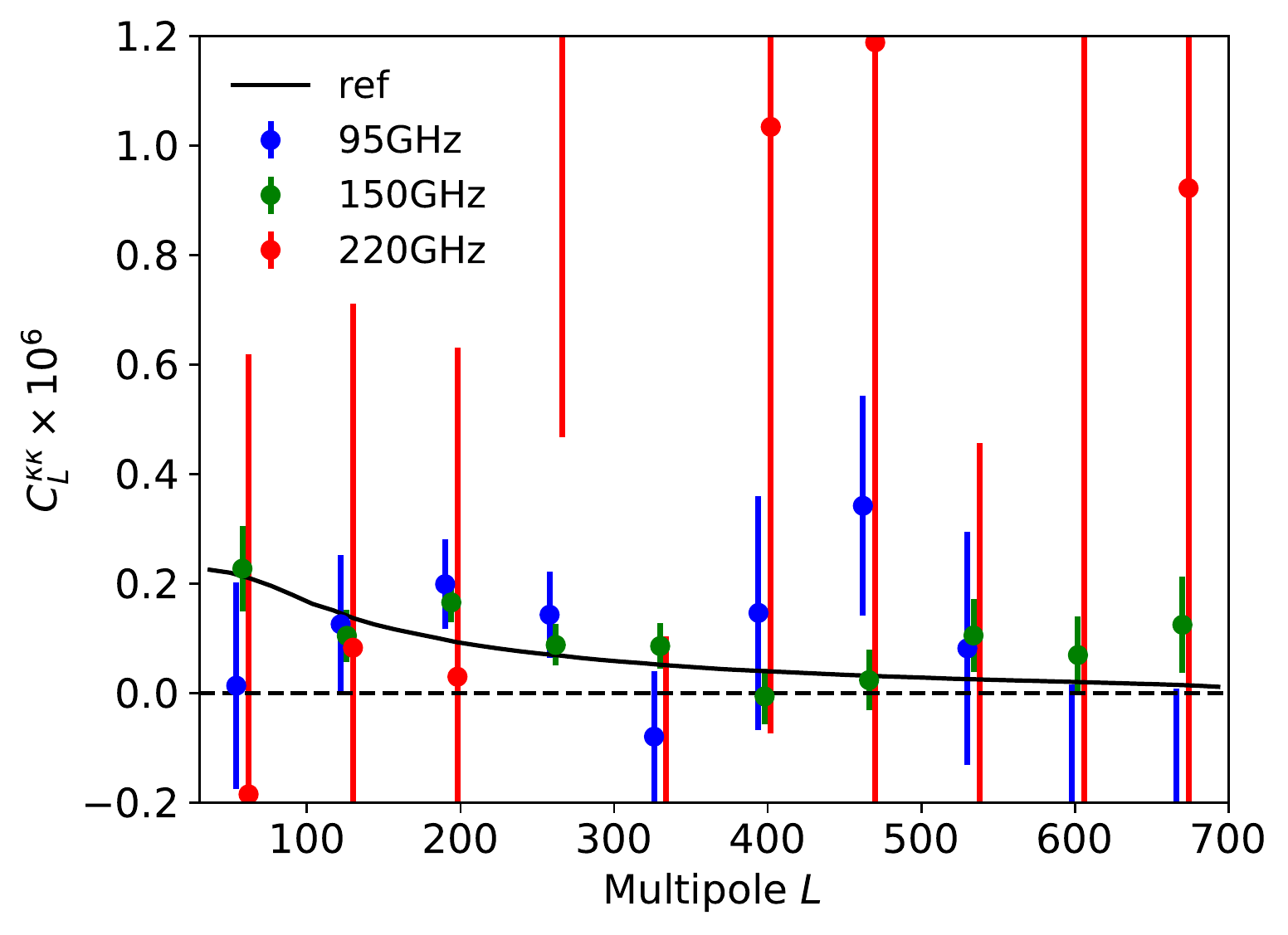}}
\end{center}
\caption{
Cross-correlation of the lensing reconstructions between \planck\ and BK15.
We show the spectra for reconstruction using the
BK15 95\,GHz, 150\,GHz and 220\,GHz bands.
The black solid line shows the theoretical lensing power spectrum. 
}
\label{fig:lensingpower}
\end{figure}

We reconstruct the lensing maps using BK15 data at 95\,GHz, 150\,GHz and 220\,GHz 
based on the method described in Ref.~\cite{biceptwoVIII}. 
Because the \planck\ lensing map has higher signal-to-noise than 
our reconstructed lensing maps, 
the BK15 lensing maps are then cross-correlated with the \planck\ lensing map 
provided by Ref.~\cite{planckXV}. 
Fig.~\ref{fig:lensingpower} shows the cross correlation of 
the reconstructed lensing signals between \planck\ 
and BK15 at each frequency. 
The amplitudes of the observed lensing spectra relative to 
the simulated spectra
are found to be $A_{\rm L}^{\phi\phi}=1.24\pm0.39$ (95\,GHz), $1.14\pm0.20$ (150\,GHz) 
and $-1.13\pm1.87$ (220\,GHz), respectively. 
The data are consistent with our baseline simulation,
and no spurious behavior is found in the lensing analysis. 

\end{appendix}

\end{document}